\pdfoutput=1
\documentclass[prd,twocolumn,aps,superscriptaddress,showpacs,preprintnumbers,nofootinbib]{revtex4-1}
\DeclareUnicodeCharacter{2212}{-}
\usepackage{latexsym,amsmath,amssymb}
\usepackage{graphicx}
\usepackage[table,usenames,dvipsnames]{xcolor}
\usepackage{hyperref}
\definecolor{azure(colorwheel)}{rgb}{0.0, 0.5, 1.0}
\definecolor{darkgreen}{rgb}{0.0, 0.5, 0.0}
\hypersetup{colorlinks=true,linkcolor=blue,citecolor=azure(colorwheel),pdfborder={0 0 0}}
\usepackage[utf8]{inputenc}
\usepackage{feynmp}
\usepackage{bbold}
\usepackage{fontawesome}
\newcolumntype{P}[1]{>{\centering\arraybackslash}p{#1}}

\makeatletter
\newcommand*{\rom}[1]{\expandafter\@slowromancap\romannumeral #1@}
\makeatother
\def\bea{\begin{eqnarray}}
\def\eea{\end{eqnarray}}
\def\beq{\begin{equation}}
\def\eeq{\end{equation}}

\begin{document}
\title{
The Heavy Dark Photon Handbook: Cosmological and Astrophysical Bounds
}

\author{Andrea Caputo}\email{andrea.caputo@cern.ch}
\affiliation{Theoretical Physics Department, CERN, 1211 Geneva 23, Switzerland}
\affiliation{Dipartimento di Fisica, ``Sapienza'' Universit\`a di Roma \& Sezione INFN Roma1, Piazzale Aldo Moro
5, 00185, Roma, Italy}
\affiliation{Department of Particle Physics and Astrophysics, Weizmann Institute of Science, Rehovot 7610001, Israel}

\author{Jaeyoung Park}\email{rsl958@snu.ac.kr}
\affiliation{Department of Physics and Astronomy and Center for Theoretical Physics, Seoul National University, Seoul 08826, Korea}
\affiliation{Theoretical Physics Department, CERN, 1211 Geneva 23, Switzerland}

\author{Seokhoon Yun}\email{seokhoon.yun@ibs.re.kr}
\affiliation{Center for Theoretical Physics of the Universe, Institute for Basic Science (IBS), Daejeon, 34126, Korea}

\date{\today}

\begin{abstract}

We investigate cosmological and astrophysical constraints on dark photons with masses $\sim 10^{-1}$-$10^3$ MeV. These dark photons can be copiously produced either in the early universe or during core-collapse supernovae, potentially leaving distinct observational signatures. First, we derive updated constraints from cosmological and astrophysical observables that rely on the thermal relic abundance of dark photons, including the CMB spectrum, primordial light element abundances, and galactic/extragalactic gamma-ray flux. We consider the minimal reheating temperature possible, $T_{\rm RH} = 6 \, \rm MeV$, such that our constraints are conservative, but unavoidable within the minimal dark photon model. Then, for supernova-sourced dark photons, we systematically examine all relevant observational bounds, revisit the standard cooling argument and derive limits from other arguments such as fireball formation, low energy supernovae and galactic positron injection. 
\end{abstract}


\preprint{CTPU-PTC-25-42}

\maketitle


\section{Introduction}
\label{sec:introduction}

New light and weakly coupled degrees of freedom appear ubiquitously in many well-motivated extensions of the Standard Model (SM). These hypothetical particles offer a fertile ground for probing physics beyond the SM through cosmological and astrophysical observations.
 
In this work, we explore one of the simplest and most theoretically motivated extensions of the SM: a new spin-one vector field identified as the gauge boson of an additional Abelian $U(1)^\prime$ gauge symmetry. We denote the corresponding quantum field by $A_\mu^\prime$, and refer to its particle excitation as the dark photon (DP), denoted by $\gamma^\prime$. More specifically, we investigate the constraints and prospects on the most minimal communication channel with the visible sector in the absence of other new particles at the low-energy scale probed by processes relevant to cosmology and astrophysics: the renormalizable Abelian kinetic mixing~\cite{Holdom:1985ag,Okun:1982xi, Galison:1983pa, Arkani-Hamed:2008hhe, Pospelov:2007mp}. 

\begin{figure}[t!]
\centering
\includegraphics[width=1.\linewidth]{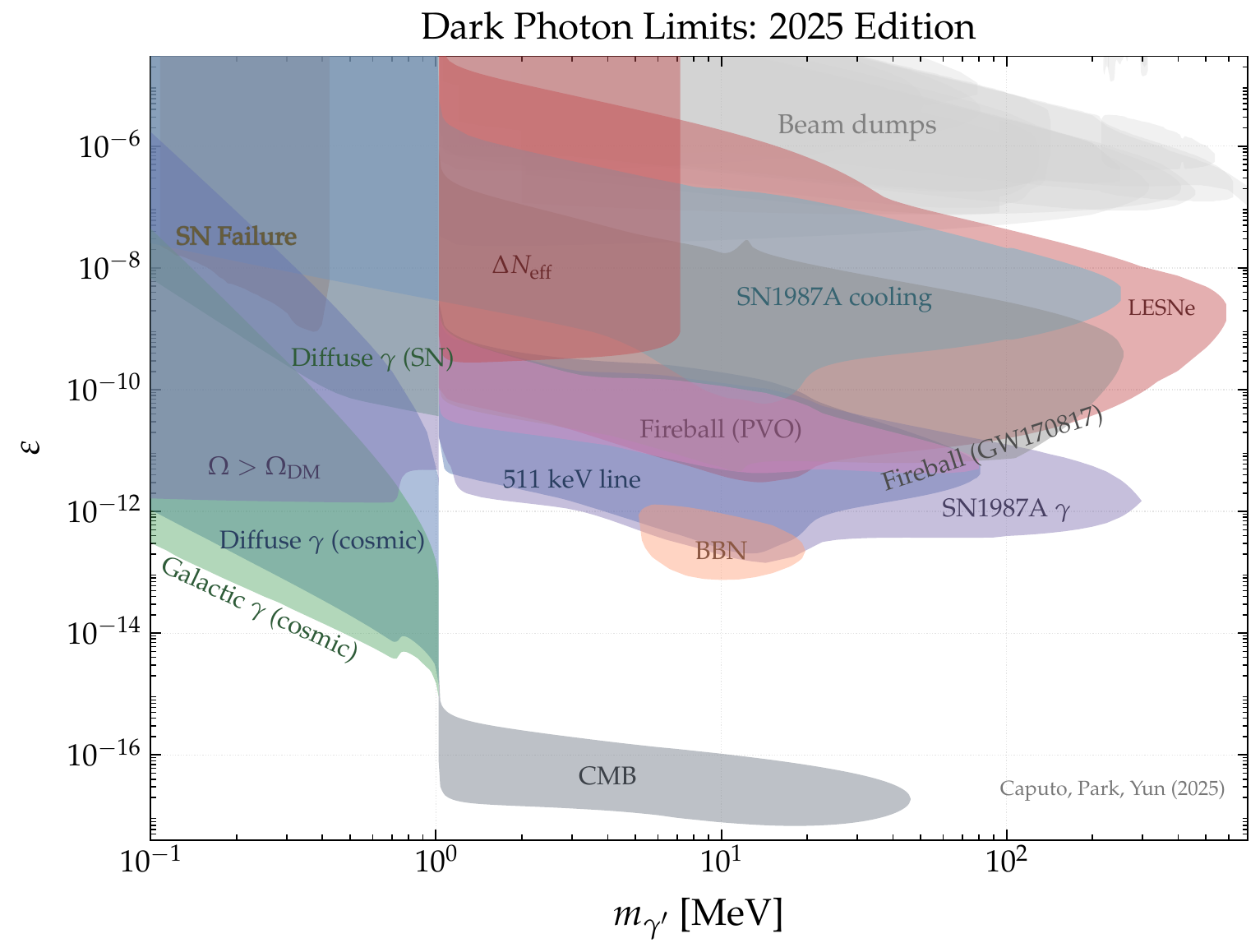}
 \caption{Astrophysical, cosmological and colliders constraints on the dark photon model in MeV mass range. In particular, we show constraints from SN~1987A cooling (dark-blue), failed supernovae (mustard), low-energy SNe (light red), BBN (salmon), $\Delta N_{\rm eff}$ (red), relic-density $\Omega_{\gamma'}$ (olive), 511\,keV line (violet-blue), diffuse supernova $\gamma$-rays, SN~1987A $\gamma$-ray constraints (lilac), Galactic and diffuse $\gamma$-ray limits from cosmic relic densities (green), CMB spectral distortions (dark gray), fireball formation both from both SN~1987A (magenta) and GW170817 (grey).
The light-grey region labelled ``Beam dumps'' shows the combined exclusion from fixed-target and collider experiments, obtained using \textsc{DarkCast}~\cite{DarkCast, Ilten:2018crw, Baruch:2022esd}. We stress that the combination of low-energy supernovae, fireball formation, supernova $\gamma$-ray signals, and Galactic positron injection is independent of the supernova explosion mechanism and, taken together, supersedes the standard cooling argument.
See the text for the details of each constraint.}
\label{fig:moneyplot}
\end{figure}

The DP can be probed through a wide variety of experimental and observational means.
Laboratory searches, including fixed-target~\cite{Bjorken:2009mm,Reece:2009un,Freytsis:2009bh, APEX:2011dww, Merkel:2014avp, A1:2011yso} and beam dump experiments~\cite{Batell:2009di,Essig:2010gu}, rare meson decay measurements~\cite{Pospelov:2008zw, Reece:2009un, KLOE-2:2011hhj, WASA-at-COSY:2013zom, HADES:2013nab, NA482:2015wmo, PHENIX:2014duq, Williams:2011qb, Blumlein:2011mv, Gninenko:2012eq, Blumlein:2013cua, NA64:2018lsq}, and collider signatures~\cite{Fayet:2007ua, Batell:2009psz, Essig:2009nc, Curtin:2014cca, BaBar:2014zli, LHCb:2017trq, BESIII:2017fwv, Anastasi:2015qla}, have excluded wide regions of the DP parameter space, particularly at higher kinetic mixing, where DPs are efficiently produced and decay promptly.
However, at lower values of the kinetic mixing, where the DP becomes long-lived on experimental timescales, terrestrial experiments loose sensitivity, and the role of astrophysical and cosmological probes becomes crucial.

In these regimes, DPs can be efficiently produced in extreme environments, such as the hot and dense interiors of stars, supernovae, and neutron star mergers.
Depending on its mass and mixing, DPs may escape the object, deposit energy outside via decay, or remain trapped and contribute to local energy transport, thereby modifying observable quantities such as neutrino emission, light curves, or post-merger signals.

In parallel, the cosmological implications of MeV-scale DPs offer complementary and powerful avenues of constraint.
In the early Universe, DPs produced through thermal processes can contribute to the expansion history.
Their radiative decays may inject entropy, alter the predicted light element abundances of Big Bang nucleosynthesis (BBN), or contribute to the effective number of relativistic species and distort the Cosmic Microwave Background (CMB), all of which are precisely measured by current observations.
At later times, decays of relic DPs into photons or electron-positron pairs can inject energy into the intergalactic medium, and be probed through cosmological observables such as the CMB anisotropy spectrum by affecting the ionization history.
In galactic and extragalactic environments, the decay products can lead to distinctive features in X-ray signals if DPs are long-lived and decay on astrophysical scales.
Constraints from the Milky Way, clusters, and nearby galaxies provide stringent bounds in this regime.

In this work, we provide a comprehensive survey of the phenomenology of MeV-scale DPs, bringing together a wide range of observational and experimental constraints (see Ref.~\cite{Caputo:2021eaa} instead for smaller DP masses).
Our analysis includes cosmological limits from early-Universe physics, and astrophysical probes from SNe-associated phenomena. In particular, we highlight the role of recent developments in supernova probes and neutron star merger simulations for probing long-lived DPs. We also collect updated colliders constraints in the regime of interest.

By combining all available constraints, we present an up-to-date map of the viable parameter space for MeV-scale DPs (see Fig.~\ref{fig:moneyplot}). This work serves as both a status report and a resource for future investigations, identifying open windows for discovery and clarifying the complementarity of different probes across cosmic and terrestrial frontiers.

\section{Dark Photon Theory, Interactions, and Production Rates}
\label{sec:DPcoupling}

This section defines the theoretical framework that forms the basis of our phenomenological analysis. We begin by introducing the low-energy effective DP interactions and discussing plausible UV origins for the two key parameters appearing on the Cartesian axes of Fig.~\ref{fig:moneyplot}. We then examine how Abelian kinetic mixing induces interactions between the DP and electrically charged SM particles. This requires identifying the physical propagating states, which we carry out both in vacuum and in a medium. Finally, we use these results to derive the master formula in Eq.~\eqref{eq:DPdifferentialrate}, which quantifies the number of DPs produced per unit time and volume. This expression will be repeatedly used in the following sections to study DP production in both the early universe and supernova environments.

The low-energy Lagrangian for the DP that captures the relevant phenomenology reads as follows
\beq
\begin{split}
\mathcal{L}_{\rm DP} = & -\frac{1}{4}F_{\mu\nu}F^{\mu\nu} -\frac{1}{4}F^{\prime}_{\mu\nu} F^{\prime \mu\nu}  +\frac{\varepsilon}{2}F_{\mu\nu}F^{\prime \mu\nu} \\
&   + \frac{m_{\gamma^\prime}}{2} A_{\mu}^\prime A^{\prime \mu} + eA_\mu J_{\rm EM}^{\mu} \, .
\end{split}
\label{eq:DPLagrangian}
\eeq
Here, $F_{\mu\nu}^{(\prime)} = \partial_\mu A_\nu^{(\prime)} - \partial_\nu A_\mu^{(\prime)}$ is the photon (DP) field strength tensor, $m_\gamma^\prime$ is the DP mass, and $J_{\rm EM}^\mu$ is the electromagnetic (EM) current. We notice how this Lagrangian contains the two parameters that define our framework: the kinetic mixing $\varepsilon$ and the DP mass $m_{\gamma^\prime}$. 

The first row of Eq.~\eqref{eq:DPLagrangian} contains the kinetic terms of the Lagrangian, including the operator that serves as the sole interaction channel between the visible and dark sectors: the kinetic mixing between the two Abelian field strengths.\footnote{If we were interested in a Lagrangian valid at arbitrarily high energies, the Abelian gauge group to which the DP couples would be the hypercharge $U(1)_Y$~\cite{Holdom:1985ag}. However, since the processes relevant to our analysis take place at energies well below the Fermi scale, we work with the SM in its broken electroweak phase and accordingly couple the DP to the electromagnetic gauge group; coupling to the neutral weak gauge group is suppressed by the Fermi scale as $(m_{\gamma^\prime}/m_Z)^2$~\cite{Baumgart:2009tn,Lee:2016ief,Bauer:2022nwt}.} This interaction is conventionally normalized by the dimensionless parameter $\varepsilon$. Because particles charged under both Abelian symmetries induce radiative corrections, the kinetic mixing parameter typically runs with the energy scale~\cite{Holdom:1985ag}. In any given UV complete theory, its value must be fixed at a specific energy scale, with the low-energy value determined accordingly. In this work, we adopt a practical approach and treat $\varepsilon$ as the kinetic mixing parameter evaluated at the low energies relevant for the constraints used in our analysis.

The mass term for the DP in the second row of Eq.~\eqref{eq:DPLagrangian} explicitly breaks the $U(1)^\prime$ gauge symmetry. However, there are several mechanisms that can generate such a mass at low energies while preserving gauge invariance at high energies. Two widely studied examples are the dark Higgs mechanism~\cite{Fayet:1990wx} and the Stueckelberg mechanism~\cite{Stueckelberg:1938hvi}. The former, analogous to the well-established mechanism for electroweak symmetry breaking, introduces a new scalar field that spontaneously breaks the Abelian symmetry, resulting in a massive gauge boson and an additional radial mode in the spectrum. In contrast, the Stueckelberg mechanism generates a gauge-invariant mass term without spontaneous symmetry breaking or additional physical scalars, by coupling the gauge field to a pseudoscalar degree of freedom. In this work, we remain agnostic about the origin of the DP mass. We assume that any additional states associated with the mass-generation mechanism are heavy enough to be phenomenologically irrelevant, and we treat $m_{\gamma^\prime}$ as a free parameter throughout our analysis.

Having introduced the effective parameters that define our framework and outlined our low-energy, pragmatic approach, we now turn to the physical implications of a non-zero kinetic mixing. As we will show below, this term induces a small effective coupling between the DP and electrically charged SM particles.

We start our discussion from the simplest case where physical processes happen in the vacuum and the SM photon field is massless. The diagonalization of the photon and DP fields in Eq.~\eqref{eq:DPLagrangian} can be achieved via the gauge-invariant shift of $A_\mu \rightarrow A_\mu + \varepsilon A_\mu^\prime$ at the leading order of $\varepsilon$ (higher order terms in $\varepsilon$ arise from the canonical field normalization)~\cite{Babu:1997st}.
Therefore, the interactions of DPs with the SM contents in the vacuum are simply written as~\cite{Babu:1997st}
\bea
\left.\mathcal{L}_{\rm DP}\right|_{\rm vacuum} \supset \varepsilon e A_\mu^\prime J_{\rm EM}^\mu \, .
\eea

The assumption that we are in vacuum is not always justified and, in some of the cases considered in this work, it is not valid. For this reason, we are also forced to consider situations in which the system is embedded in a dense medium. The effective DP couplings in this case are affected by plasma effects, which modifies photon dispersion. Considering the thermally corrected photon propagator in the form of the polarization tensor, these couplings can be described by the replacement of the kinetic mixing as~\cite{Redondo:2008ec,An:2013yfc} 
\bea
\varepsilon\rightarrow \varepsilon \frac{m_{\gamma^\prime}^2}{m_{\gamma^\prime}^2 - \pi_{\rm T,L}} \,,
\label{eq:EffDPcoupling}
\eea
where $\pi_{\rm T,L}$ quantifies the refractive properties of the intermediate photon propagator for transverse ($\rm T$) and longitudinal ($\rm L$) polarizations, respectively. Explicit expressions for the polarizations can be found in App.~\ref{app:plasma}. This expression indeed covers the case of the vacuum, where both polarizations $\pi_{\rm T,L}$ vanish. 

We conclude this section with a very useful expression that quantifies the DP production rate. The DP coupling structure, which is provided in Eq.~\eqref{eq:EffDPcoupling}, follows that of photons (i.e., EM interactions) up to an overall factor proportional to the kinetic mixing $\varepsilon$. Thus, we can express any interaction rates of DPs in terms of the total absorption rate of SM photons. To this purpose, it is useful to appreciate how the imaginary components of $\pi_{\rm T,L}$ rely on processes that bring photons into thermal equilibrium through scatterings, such as absorption and/or decay. When particles participating in such processes are in thermal equilibrium, detailed balance yields~\cite{Stodolsky:1986dx,Weldon:1983jn,Redondo:2013lna}
\bea
{\rm Im}\,\pi_{\rm T,L} = -\omega \left(1-e^{-\omega/T}\right) \Gamma_{\rm T,L} \, .
\label{eq:ImPi}
\eea 
Here, $\Gamma_{\rm T,L}$ represents the total absorption rate of photons with a modified dispersion relation that corresponds to an external DP with energy $\omega$ and momentum $\vec{k}$, or in other words satisfying the relation $\omega^2-k^2=m_{\gamma^\prime}^2$. Therefore, all the functions appearing in Eq.~\eqref{eq:ImPi} are determined by properties of the QED plasma populating the early universe.

In particular, the differential DP production rate per volume can be written as follows~\cite{Redondo:2008ec,Redondo:2013lna}
\bea
\frac{d \mathcal{N}_{\gamma^\prime}}{dV \, dt \, d^3\vec{k}} & = & \frac{\varepsilon^2}{\left(2\pi\right)^3} \sum_{i={\rm T,L}}g_ie^{-\omega/T}  \Gamma_i \left|\frac{m_{\gamma^\prime}^2}{m_{\gamma^\prime}^2 - \pi_i}\right|^2 \,\nonumber\\
&=& \frac{1}{\left(2\pi\right)^3 }\frac{\varepsilon^2 m_{\gamma^\prime}^4}{\omega} \frac{1}{e^{\omega/T}-1} \nonumber\\
&&\times \sum_{i={\rm T,L}}g_i   \frac{\left|{\rm Im}\,\pi_i\right|}{\left(m_{\gamma^\prime}^2 - {\rm Re}\,\pi_i\right)^2+\left|{\rm Im}\,\pi_i\right|^2}
\label{eq:DPdifferentialrate}
\eea
with $g_i = 2 \,(1)$ for the transverse (longitudinal) polarization. Under the typical condition of a narrow resonance width (i.e., ${\rm Re}\,\pi_i \gg {\rm Im}\,\pi_i$), the DP production during the resonance can be interpreted as the resonant conversion from background photons and become independent of the details of scattering processes as we will discuss.

\section{Cosmological Dark Photons: Thermal Production and Signals}
\label{sec:cosmo}

The hot plasma in the early universe serves as a natural source for the production of massive DPs.
In this paper, we focus on the primordial DP abundance with masses around the ${\rm MeV}$ scale, generated through thermal processes such as particle scatterings within the thermal bath.
Even if these thermally produced DPs contribute only a small fraction of the total dark matter density, their decay into the electromagnetic sector can leave observable imprints on astrophysical and cosmological phenomena. In the following sections, we examine sensitivities to probe a signal from the relic DP abundance via their decay by exploiting some observables, which align closely with predictions from the SM.

The cosmological evolution of the DP number density $n_{\gamma^\prime}$ is governed by the Boltzmann equation
\bea
\frac{d n_{\gamma^\prime}}{dt} + 3H n_{\gamma^\prime} =\gamma_{\gamma^\prime} \left(1-\frac{n_{\gamma^\prime}}{n_{\gamma^\prime}^{\rm eq}}\right) \, , 
\eea
where $H$ accounts for the Hubble rate, and $\gamma_{\gamma^\prime}$ is the DP interaction rate.
For the sake of convenience, we rewrite the Boltzmann equation in terms of the comoving number density $Y_{\gamma^\prime} = n_{\gamma^\prime}/s$
\bea
\frac{d Y_{\gamma^\prime}}{d\log T^{-1}} = \left(1-\frac{1}{3}\frac{d\log g_{*s}}{d\log T^{-1}}\right) \frac{\gamma_{\gamma^\prime}}{sH} \,, 
\label{eq:Boltzmann}
\eea
where $s = (2\pi^2/45)g_{*s}T^3$ denotes the entropy density, $T$ is the temperature, and $g_{*s}$ represents the effective entropic degrees of freedom.
Since DPs with MeV-scale masses are produced during the radiation-domanited era, the Hubble rate reads $H = 1.66\times g_*^{1/2} T^2/m_{\rm pl}$ with $m_{\rm pl} =G^{-1/2} = 1.22\times 10^{19}\,{\rm GeV}$ the Planck mass and $g_*$ the total number of relativistic degrees of freedom.

The interaction rate $\gamma_{\gamma^\prime}$ is proportional to $(n_{\gamma^\prime}^{\rm eq} - n_{\gamma^\prime})$, where $n_{\gamma^\prime}^{\rm eq}$ denotes the equilibrium number density of DPs.
If the DP coupling strength (i.e., the kinetic mixing $\varepsilon$) is sufficiently large, the DP number density follows thermal equilibrium and eventually freezes out.
On the other hand, for small kinetic mixing, DPs never reach thermal equilibrium but are gradually produced from the thermal bath via the so-called `freeze-in' production mechanism.
In this freeze-in regime, the DP production rate is predominantly contributed from collisions that generate DPs, approximately given by $\gamma_{\gamma^\prime} \propto n_{\gamma^\prime}^{\rm eq}$.



\begin{figure}[t!]
\centering
 \includegraphics[width=.9\linewidth]{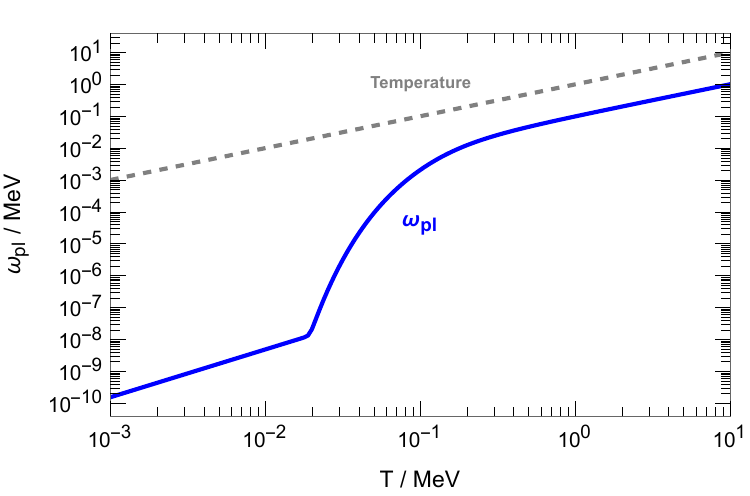}
 \caption{Cosmic plasma frequency as a function of temperature.} 
 \label{fig:plasmafrequency}
\end{figure}

Building upon the description in Sec.~\ref{sec:DPcoupling}, the cosmological DP production rate can be expressed as
\beq
\begin{split}
\gamma_{\gamma^\prime}  = &   \int d^3\vec{k}\frac{d N_{\gamma^\prime}}{dV \, dt \, d^3\vec{k}}  
\end{split}
\label{eq:ThermalDPrate}
\eeq
The plasma frequency $\omega_{\rm pl}$, which determines ${\rm Re}\,\pi_i$ as outlined in Eqs.~\eqref{eq:RePiT} and \eqref{eq:RePiL}, is depicted in Fig.~\ref{fig:plasmafrequency},
incorporating the baryon asymmetry $n_B/n_{\gamma}\simeq 6\times 10^{-10}$ with the proton abundance $n_p/n_B \simeq 0.878$ after BBN, leading to $n_{e^-} - n_{e^+} \simeq n_p$ due to the electric-charge neutrality condition.
The imaginary component ${\rm Im}\,\pi_i$ encode information on the absorption (and, correspondingly, production in the detailed balance relation) of photon states with the energy $\omega$ and the dispersion relation of $\omega^2 - k^2 = m_{\gamma^\prime}^2$.

It is noteworthy that cosmological DP production is IR-dominated.
This IR-sensitive feature stems from the renormalizable nature of DP couplings, governed by the dimensionless kinetic mixing $\varepsilon$.
Moreover, at high temperatures of $\omega_{\rm pl} > m_{\gamma^\prime}$, plasma effects further suppress the effective DP couplings and, consequently, the interaction rate.
Hence, the thermal production of cosmological DPs becomes efficient at lower temperatures when $\omega_{\rm pl} \leq m_{\gamma^\prime}$.

A notable contribution to the DP abundance can arise from the resonance point, ${\rm Re}\,\pi_i=m_{\gamma^\prime}^2$, similar to the case of supernovae discussed above.
Since the resonance width is narrow (i.e., ${\rm Im}\,\pi_i \ll {\rm Re}\,\pi_i$), the resonance contribution to the DP abundance becomes independent of the imaginary components ${\rm Im}\,\pi_i$, as discussed in Eq.~\eqref{eq:resonanceAPP}, can be understood as a non-adiabatic resonant conversion of photons into DPs.
Details of the analytic derivation can be found in Appendix.~\ref{app:CosDPres}.

We notice that the resonant production is dominated by the transverse components, which occur when $\omega_{\rm pl} \simeq m_{\gamma^\prime}$.
This is because the resonance of the longitudinal polarization occurs at the specific low energy of $\tilde{\omega} = \mathcal{O}(\omega_{\rm pl})$, considerably smaller than temperature.
Therefore, the photon number density available for resonant conversion into longitudinal DPs is suppressed by factors of $(\omega_{\rm pl}/T)^2$ and $(1/\pi)^2$ relative to the equilibrium number density $n_{\gamma}$; the second factor comes from the kinematic space integral.
There is a compensatory enhancement of $(T/\omega_{\rm pl})$ since the non-adiabatic resonant conversion probability is inversely proportional to $\omega$.
As a result, the resonance contribution to the longitudinal DP abundance is approximately two orders of magnitude lower than that of transverse ones.

The thermal production of DPs persists until $T\sim m_{\gamma^\prime}$.
The conventional method for evaluating production rates assumes that statistical factors for the final states are neglected, and it exploits the advantage of invariance under Lorentz transformations to the final states.
This approach allows the use of Mandelstam variables to describe scattering processes~\cite{Gondolo:1990dk} in the center-of-mass frame.
However, in principle, this approach is not directly applicable to calculating the DP production rate because the plasma effect on the DP couplings depends on polarization, which is defined in a specific frame characterized by thermodynamic properties such as temperature.

\begin{figure}[t!]
\centering 
\includegraphics[width=.9\linewidth]{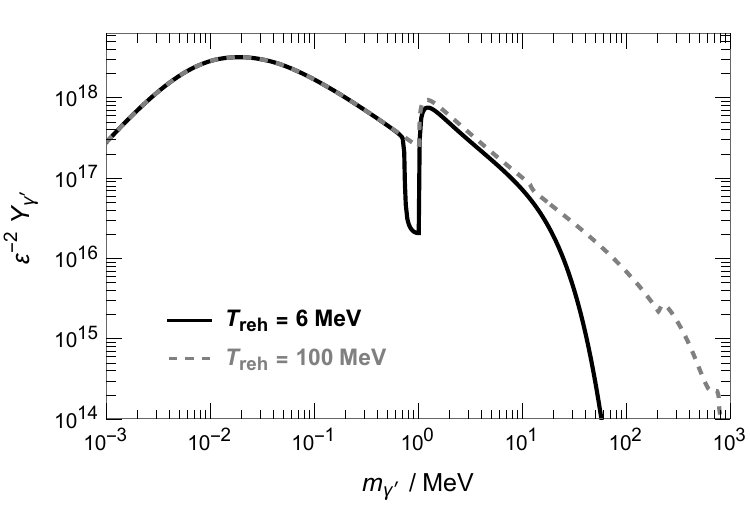}
 \caption{The comoving dark photon density $Y_{\gamma^\prime} = n_{\gamma^\prime}/s$ generated by thermal processes with the initial temperatures $T_{\rm res} = 6\,{\rm MeV}$ (black) and $T_{\rm res} = 100\,{\rm MeV}$ (gray dashed), 
 in the freeze-in scenario where thermal equilibrium is never achieved.} 
\label{fig:YDP}
\end{figure}

Nevertheless, once the temperature drops to the lower temperature regime of $\omega_{\rm pl} <  m_{\gamma^\prime}$, the plasma effect turns out to be insignificant.
At such low temperatures, the effective coupling of DPs becomes effectively independent of polarization and behaves equivalent to the vacuum case.
In light of this and considering that calculations are relatively insensitive to the initial temperature, the conventional Lorentz-invariant method remains a reliable approximation.
In Appendix.~\ref{app:crosssection}, we provide the cosmological DP production rate $\gamma_{\gamma^\prime}$ for the relevant scattering processes (semi-Compton scatterings and $e^-e^+$ annihilation) and the inverse-decay processes (the so-called `coalescence') in terms of the Lorentz invariant variables.


Fig.~\ref{fig:YDP} shows the freeze-in DP abundance $Y_{\rm \gamma^\prime} = n_{\gamma^\prime}/s$ for the initial temperatures $T_{\rm RH} = 6\,{\rm MeV}$ (black solid) and $T_{\rm RH} = 100\,{\rm MeV}$ (gray dashed).
To compute $Y_{\rm \gamma^\prime}$, we perform the Boltzmann equation in Eq.~\eqref{eq:Boltzmann} with the production rate given in Eq.~\eqref{eq:ThermalDPrate}, incorporating the plasma effect described in Sec.~\ref{sec:DPcoupling}.
We take the effective degrees of freedom for the entropy ($g_{*s}$) and energy ($g_*$) densities within the standard cosmology from Ref.~\cite{Saikawa:2018rcs}.

For $T_{\rm RH} = 6\,{\rm MeV}$ (black solid), the DP abundance at masses below $\sqrt{3/2}\,\omega_{\rm pl}(6\,{\rm MeV}) \simeq 0.74\,{\rm MeV}$ is dominated by resonant production, while at higher masses scattering processes without the plasma effect control the abundance.
A sizable contribution from the inverse electron-positron decay arises only when it becomes kinematically allowed, i.e., for $m_{\gamma^\prime} > 2m_e$.
For the higher temperature of $T_{\rm RH} = 100\,{\rm MeV}$, resonance production can occur even for $m_{\gamma^\prime} > 2m_e$, but its contribution is subdominant compared to inverse-decay process.

When the DP mass is sufficiently small compared to the initial temperature, the thermal relic abundance becomes insensitive to the initial temperature due to the IR-dominated nature of the production, as noted above.
On the other hand, for too heavy DP masses, its abundance is exponentially suppressed by the Boltzmann factor, as clearly shown in Fig.~\ref{fig:YDP}. These results do not account for the subsequent decay of DPs; thus, the actual density at a given time should include an additional exponential suppression factor from decay.

The scaling behavior of the freeze-in DP abundance, in regimes where it becomes independent of the initial temperature (i.e., for $m_{\gamma^\prime} \ll T_{\rm RH}$), can be understood analytically as follows. 
For masses above the electron-positron threshold ($2m_e \simeq 1.02{\rm MeV}$), the dominant production mechanism is inverse decay, the rate of which is $\gamma_{\gamma^\prime} \sim m_{\gamma^\prime}^2 \,T\, \Gamma_{e\bar{e}} K_1(m_{\gamma^\prime}/T) \propto \varepsilon^2 m_{\gamma^\prime}T^2$ at $T> m_{\gamma^\prime}$; see Eq.~\eqref{eq:gammaID}. With this expression, we find
\begin{eqnarray}
Y_{\gamma'} &\simeq&
\frac{135}{1.66\,8\pi^{3}}\,
\frac{g_{{\gamma'}}\, M_{\rm Pl}\, \Gamma_{\gamma'\to e^+e^-}}
     {g_{s}\, g_*^{1/2}\, m_{\gamma'}^{2}}\,
S\!\left(x\right)\nonumber\\ &\simeq& \frac{8.2 \times 10^{17}}{m_{\gamma'}/{\rm MeV}}\Big(\frac{10.75}{g_s}\Big)^{3/2}\\
&& \times S\!\left(x\right)
\left(1+\frac{2m_e^2}{m_{\gamma'}^2}\right)\sqrt{1-4\frac{m_e^2}{m_{\gamma^\prime}^2}}\nonumber
\end{eqnarray}
where $S(x) \simeq 
[\,0.266 + 0.734\,e^{-x}\,]\, e^{-x}\,(1 + x^{2})^{5/4}$, and $x \equiv m_{\gamma'}/T_{\rm RH}$. Notice that for $x \gg 1$ (equivalently, $m_{\gamma'} \gg T_{\rm RH}$), then the abundance gets suppressed as $S(x) \sim x^{5/2}e^{-x}$, while for $x \ll 1$, then $S(x) \sim 1$. Derivation of these results can be found in Appendix~\ref{app:CosDPres}.
For lower masses, production is governed by the resonant scattering, which relies on the resonant temperature $T_{\rm res}$.
From Eq.~\eqref{eq:YDPTres}, the resonant contribution scales as
$[Y_{\gamma^\prime}]^{\rm res} \propto \varepsilon^2 m_{\gamma^\prime}^2 T_{\rm res}^{-3}$.
When $m_{\gamma^\prime} > 10^{-2}\,{\rm MeV}$, electrons and positrons are still relativistic at $T_{\rm res}$ that the plasma frequency scales as $\omega_{\rm pl} \propto T$.
This results in $Y_{\gamma^\prime} \propto \varepsilon^2 m_{\gamma^\prime}^{-1}$, consistent with the inverse decay regime.
In the intermediate mass range of $10^{-8}\,{\rm MeV} < m_{\gamma^\prime} < 10^{-2}\,{\rm MeV}$, resonance occurs during the epoch when positrons are being annihilated.
As illustrated in Fig.~\ref{fig:plasmafrequency}, $T_{\rm res}$ depends weakly on $m_{\gamma^\prime}$ across this range, which leads to $Y_{\gamma^\prime} \propto \varepsilon^2 m_{\gamma^\prime}^{2}$.
At last, for $m_{\gamma^\prime} < 10^{-8}\,{\rm MeV}$, positrons are depleted and the remaining asymmetric electron population determines the plasma frequency that $\omega_{\rm pl} \propto n_e^{1/2} \propto T^{3/2}$.
This results in a mass-independent abundance $Y_{\gamma^\prime} \propto \varepsilon^2$.

In the main text, we focus on the irreducible DP relic abundance for the initial temperature of $6\,{\rm MeV}$, 
corresponding to the minimal onset temperature consistent with the standard cosmology ($5.96\,{\rm MeV}$ at the 95\% confidence level~\cite{Barbieri:2025moq,Hasegawa:2019jsa}), and explore the associated cosmological and astrophysical sensitivities to probe their signals.
We also examine a higher temperature case as a more general case, with a detailed disucssion provided in Appendix~\ref{app:highertemperature}.

The presence of a cosmological DP relic and its subsequent decay into the visible electromagnetic (EM) sector could leave an imprint on cosmological observations.
In particular, cosmological probes are appropriate to searching for DPs with lifetimes shorter than the age of the universe.
Such short lifetimes are naturally realized for DP masses above the electron-positron threshold, where tree-level decay into electron-positron pairs becomes kinematically allowed with the rate described in Eq.~\eqref{eq:DPdecayee}.
Hence, the phenomenological impacts of these decays during the early cosmological history become crucial for probing relatively heavy DPs.
We examine the effects of decaying DPs with an irreducible abundance on various cosmological observations, including the Cosmic Microwave Background (CMB) spectrum and the primordial abundance of light elements from Big Bang nucleosynthesis (BBN).

\subsection{Overproduction}
\label{sec:overprod}

Excessive production of DPs can alter the expansion history of the universe by increasing the total energy density beyond what is expected in the standard $\Lambda$CDM cosmology.
Once DPs are generated, they become non-relativistic around $T \lesssim m_{\gamma^\prime}$ and begin contributing to the matter density, which redshifts more slowly than radiation. An excess in matter abundance leads to a faster expansion rate in the early universe and, in particular, brings forward the matter-radiation equality epoch. This shift in the expansion history affects various cosmological observables, including the growth of perturbations and the shape of the CMB power spectrum (e.g., through modifications to the early integrated Sachs-Wolfe effect and the acoustic peak structure).

\begin{figure}[t!]
\centering 
\includegraphics[width=.9\linewidth]{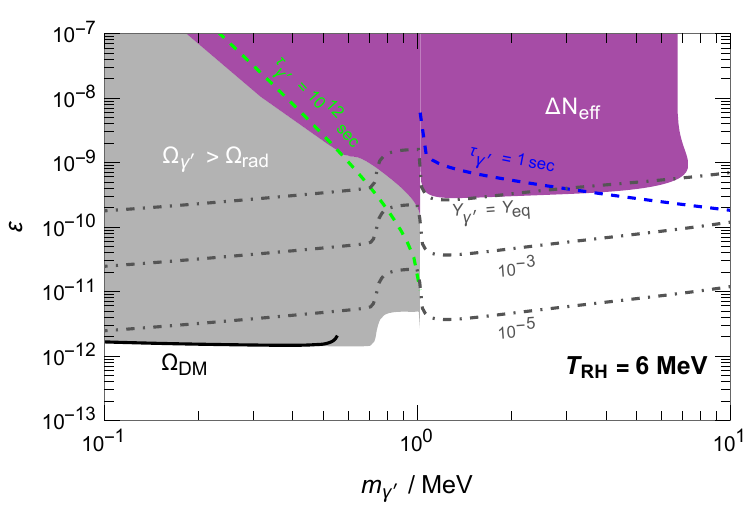}
 \caption{The gray-shaded region is excluded by the overproduction of dark photons.
 The black solid line indicates the parameter values that reproduce the present dark matter abundance. The purple-shaded region is excluded by the $\Delta N_{\rm eff}$ constraint ($2\sigma$ confidence level). The green and blue dashed lines correspond to the DP lifetime of $10^{12}\,{\rm sec}$ (around the CMB epoch) and $1\,{\rm sec}$ (around the neutrino decoupling), respectively. The gray dot-dashed lines indicate contours of the dark photon yield, $Y_{\gamma^\prime} = 10^{-5}$, $10^{-3}$, and the equilibrium value from bottom to top.}
\label{fig:DPDMNeff}
\end{figure}

In Fig. \ref{fig:DPDMNeff}, the gray-shaded region labeled `$\Omega_{\gamma^\prime} > \Omega_{\rm rad}$' (corresponding to the olive-shaded region in Fig.~\ref{fig:moneyplot}) is excluded based on the requirement that DPs must not dominate the energy density prior to matter-radiation equality.
This overproduction bound appears only for masses below the electron-positron threshold, since at higher masses the DP lifetime is too short compared to the matter-radiation equality time $t_{\rm eq} \simeq 1.2 \times 10^{12}\,{\rm sec}$.
Even for lower masses, DPs can decay into three photons rapidly when the coupling $\varepsilon$ is large, which prevents them from ever dominating the total energy density.
However, as we discuss in the next subsection, there is an additional constraint arising from changes to the effective number of neutrino species on top of the SM contribution due to photon entropy injection by DP decay (red-shaded region in Fig.~\ref{fig:moneyplot} and purple-shaded region in Fig.~\ref{fig:DPDMNeff}).

\subsection{$\Delta N_{\rm eff}$}

The decay of DP relics before the recombination era can leave observable imprints on the early universe, such as the power spectrum of CMB anisotropies. If DPs decay after the neutrino decoupling epoch ($T\sim 2 {\rm MeV}$), their energy is injected exclusively into the electron-photon plasma, while the neutrino sector remains unaffected.
This selective heating increases the photon temperature relative to that of neutrinos, effectively reducing the fractional contribution of neutrinos to the total energy density.
This effect is conventionally quantified by the effective number of neutrino species, $N_{\rm eff}= (8/7)(11/4)^{4/3}(\rho_{\nu_e}+\rho_{\nu_\mu}+\rho_{\nu_\tau})/\rho_\gamma$, where $\rho_{\nu_i}$ and $\rho_\gamma$ are the energy densities of $\nu_i$ flavor and photons, respectively.
The SM prediction is $N_{\rm eff} = 3.0440$~\cite{Froustey:2020mcq,Bennett:2020zkv}, while the most stringent observational constraint from the Planck collaboration yields $N_{\rm eff}^{\rm SM} = 2.99 \pm 0.17$ at the time of recombination ($t_{\rm rec} \simeq 1.2 \times 10^{13}\,{\rm sec}$).
In the DP mass range of interest, where $m_{\gamma^\prime} \gg T_{\rm CMB}$, DPs cannot contribute to additional (invisible) radiation degrees of freedom.
Instead, their decay reduces $N_{\rm eff}$ by injecting entropy solely into the photon bath.

The Planck constraint on $N_{\rm eff}$ implies limits on DPs with large kinetic mixing $\varepsilon$, where DPs thermalize with the SM plasma and subsequently decay, resulting in a suppressed $N_{\rm eff}$. In this regime, the freeze-in approximation is no longer valid, and we compute the relic abundance and entropy injection by solving the full Boltzmann equations.
We follow the procedure outlined in Ref.~\cite{Ibe:2019gpv}, while properly incorporating the plasma effect as discussed earlier.
The purple-shaded region in Fig.~\ref{fig:DPDMNeff} shows the constraint from the Planck collaboration on $|\Delta N_{\rm eff} = N_{\rm eff} -  N_{\rm eff}^{\rm SM}| < 0.394$ at $2\sigma$.

For DP masses above $2m_e$, the $\Delta N_{\rm eff}$ constraint remains relevant up to about $7\,{\rm MeV}$.
Above this mass scale, the DP lifetime easily becomes much shorter than the neutrino decoupling time ($\sim 1\,{\rm sec}$), so that the entropy injected by the decay of earlier-produced DP relics is efficiently distributed between the electron-photon plasma and the neutrino sector, which leaves negligible impact on $N_{\rm eff}$.
The small bump observed around $m_{\gamma'} \simeq 7\,{\rm MeV}$ arises from the interplay between the DP lifetime and abundance, both of which control the efficiency of entropy injection.

We also derive the $\Delta N_{\rm eff}$ constraint for DP masses below $2m_e$, where the relic abundance is substantially enhanced by resonant photon--DP conversion in the plasma, an effect that was not fully captured in earlier analyses~\cite{Ibe:2019gpv}. 
In this low-mass regime, however, the loop-induced decay $\gamma^\prime \to 3\gamma$ results in considerably long lifetimes. 
When the lifetime becomes comparable to or exceeds the recombination time, the corresponding entropy injection occurs too late to significantly modify the photon-to-neutrino temperature ratio. 
This interplay leads to the sharp, lifetime-dependent exclusion for $m_{\gamma^\prime} < 2m_e$ seen in Fig.~\ref{fig:DPDMNeff}. Nevertheless, in this parameter region, DPs are so efficiently produced and long-lived that they already conflict with standard cosmology and its associated observables.
In other words, the $\Delta N_{\rm eff}$ constraint for $m_{\gamma^\prime} < 2m_e$ lies entirely within the parameter space excluded by overproduction, as discussed in Sec.~\ref{sec:overprod}.
For this reason, we do not include it in the main summary plot, Fig.~\ref{fig:moneyplot}.

Additional constraints related to $\Delta N_{\rm eff}$ arise from Big Bang nucleosynthesis (BBN), which proceeds at temperatures of $T \sim 0.01$–$1\,{\rm MeV}$.
Due to higher temperatures, dark radiation from light DPs could, in principle, contribute to an increase of $N_{\rm eff}$ (i.e., $\Delta N_{\rm eff} > 0$) during this BBN epoch, which impacts the expansion rate of the Universe as the cosmological parameter to govern dynamics in BBN, such as the freeze-out of the neutron-to-proton ratio.
However, the production of lighter DPs is typically delayed relative to the freeze-out of weak interactions because plasma effects strongly suppress DP--photon mixing at high temperatures.
As a result, the direct contribution of DPs themselves to $N_{\rm eff}$ during BBN is negligible.

Furthermore, entropy injection from decaying DPs (particularly for $m_{\gamma^\prime} > 2m_e$) can imprint observable effects on BBN.
Such preferential entropy injection into the EM sector accounts for a reduction of the relative neutrino energy density (i.e., $\Delta N_{\rm eff} <0$), which alters not only the expansion rate but also neutrino-induced weak processes relevant for the proton-neutron interconversion~\cite{Li:2020roy,Ganguly:2025mdi,Escudero:2025avx}.
In addition, entropy injection dilutes comoving baryon densities, which subtly shifts the predicted primordial element abundances~\cite{Li:2020roy,Escudero:2025avx}.
We find that the resulting BBN constraints~\cite{Li:2020roy} are comparable to our Planck $\Delta N_{\rm eff}$ bounds, as both depend on similar requirements for the DP relic abundance and lifetime.
Therefore, while BBN provides complementary information in terms of $\Delta N_{\rm eff}$, in our scenario it does not impose additional constraints beyond those derived from the CMB.
Beyond the $\Delta N_{\rm eff}$ considerations, energetic cascade photons produced by DP decay after BBN can dissociate light nuclei; these photodissociation bounds are discussed in detail in Sec.~\ref{sec:BBN}.

\subsection{CMB anisotropy}

An EM energy deposition through DP decay into electron-positron pairs, occurring after recombination, would heat up and ionize the neutral gas in the universe.
Such modifications to the cosmological thermal history during the dark ages leave observable imprints on the CMB, particularly in the temperature and polarization anisotropy spectra.
As extensively discussed in literature~\cite{Slatyer:2016qyl, Langhoff:2022bij, Cang:2020exa, Bolliet:2020vqu, Balazs:2022tjl, Poulin:2016anj}, precise CMB measurements place stringent constraints on the properties of decaying (or annihilated) dark matters associated with energy injection.

\begin{figure}[t!]
\centering 
\includegraphics[width=.9\linewidth]{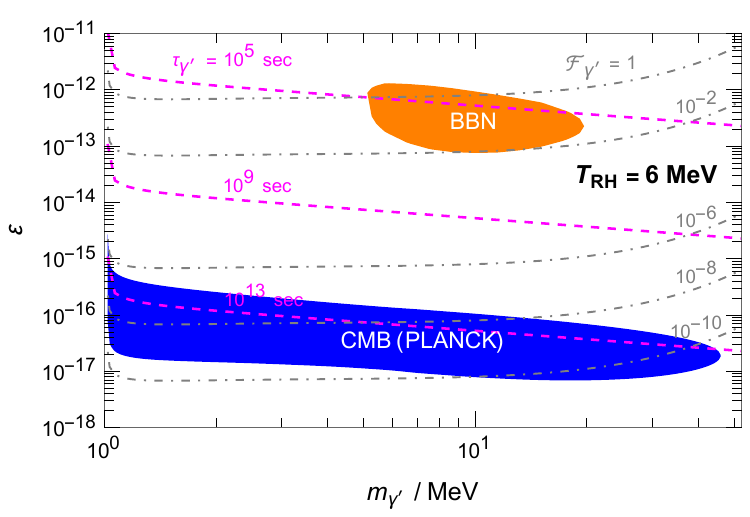}
 \caption{The blue-shaded region corresponds to the CMB constraint from the lastest Planck results.
 The orange-shaded region is excluded by BBN. 
 } 
\label{fig:DPDMcosmo}
\end{figure}

To derive the CMB constraints on DPs, we follow the prescription outlined in Ref.~\cite{Langhoff:2022bij}, which provides a consistent fit to the results in the literature, using PLANCK 2018 data~\cite{Planck:2018vyg}.
The CMB constraints on DPs reads
\bea
\frac{m_{\gamma^\prime}Y_{\gamma^\prime}}{\rho_{\rm DM}/s_0} \leq 
\frac{\tau_{\gamma'}}{\tau_{\rm min}} \exp\left[\left(\frac{t_{\text{CMB}}}{\tau_{\gamma'}}\right)^{2/3}\right] \,,
\label{eq:CMBcon}
\eea
where $\tau_{\gamma^\prime} = \Gamma_{e\bar{e}}^{-1}$ is the lifetime of non-relativistic DPs, with the decay rate $\Gamma_{e\bar{e}}$ given by 
\bea
\Gamma_{e\bar{e}}= \varepsilon^2\frac{\alpha}{3}\frac{m_{\gamma^\prime}^2}{\omega} \left(1+\frac{2m_e^2}{m_{\gamma'}^2}\right)\sqrt{1-4\frac{m_e^2}{m_{\gamma^\prime}^2}}\, .
\label{eq:DPdecayee},
\eea
where at rest ($\omega = m_{\gamma^\prime}$).
Here, $\tau_{\rm min}$ refers to the corresponding lifetime constraints for dark matters decaying into electron-positron pairs with a mass of $m_{\gamma^\prime}$, as derived in Ref.~\cite{Cang:2020exa}, and we take $t_{\rm CMB}\simeq 8.7 \times 10^{13}\,{\rm sec}$. The blue-shaded region in Figs.~\ref{fig:moneyplot} and \ref{fig:DPDMcosmo} describes the CMB constraint.

Notice that the lower part of the curve can be derived analytically. Ignoring the exponential in Eq.~\ref{eq:CMBcon}, and considering that the bound from Ref.~\cite{Cang:2020exa} is almost flat between $1\text{-}10$ MeV (around $\sim 3\times10^{24}\, \rm s$), one finds
\begin{equation}
    \varepsilon \gtrsim 1.5 \times 10^{-17}\left(\frac{\rm MeV}{m_{\gamma'}}\right)^{1/4},
\end{equation}
which is in good numerical agreement with the plot. The upper part of the curve instead coincides with having $\tau_{\gamma'}\simeq t_{\rm CMB}$.

\subsection{BBN}
\label{sec:BBN}

BBN offers a powerful probe of physics beyond the Standard Model, as it took place only a few seconds to minutes after the Big Bang, when light nuclei first formed. The observed primordial abundances of light elements are in remarkable agreement with the predictions of standard cosmology, and even small deviations can thus impose stringent constraints on new particles that could alter nuclear abundances.

The decay of DPs into the electromagnetic (EM) sector can modify the final outcome of BBN. Once the universe cools below $T \simeq 0.1\,{\rm MeV}$, photons energetic enough to dissociate nuclei (especially deuterium) become too rare to counteract nuclear synthesis. At this stage, roughly three minutes after the Big Bang, the formation of light elements ceases and their primordial abundances become fixed.

The decay of DPs into electron–positron pairs triggers an EM cascade, as the injected electrons, positrons, and photons interact with the background plasma. Assuming that the cascade develops instantaneously, the resulting zero-generation photon spectrum takes the form of a broken power law with a sharp cutoff at the minimum value of $\{m_e^2/22T\,, m_{\gamma^\prime}/2\}$, reflecting photon–photon pair creation processes~\cite{Kawasaki:1994sc}, That is to say, high-energy photons are efficiently dissipated before they can interact with nuclei. For details of this analytic spectrum, see Refs.~\cite{Berezinsky:1990qxi,Protheroe:1994dt}.

Because nuclear binding energies exceed $\sim 1 \, {\rm MeV}$, photo-dissociation by the EM cascade becomes relevant only after BBN has completed. We therefore use the standard primordial abundances as initial conditions, adopting the light-element yields computed with \texttt{PArthENoPE} 3.0~\cite{Pisanti:2007hk,Consiglio:2017pot,Gariazzo:2021iiu}.

We solve the Boltzmann equation, including nuclear destruction terms from energetic cascade photons produced by DP decays. The stationary distribution of these cascade photons depends on the DP number density and decay rate, as well as on the zero-generation photon spectrum discussed above. It results from the balance between photon injection and energy degradation through scatterings with the background medium~\cite{Pospelov:2010cw,Fradette:2014sza}. Analytic cross sections for all relevant photo-dissociation channels, validated against experimental data, are taken from Ref.~\cite{Cyburt:2002uv}.

For the observationally inferred light-element abundances, we adopt the recommended PDG values~\cite{ParticleDataGroup:2024cfk}:
\bea
{\rm D}/{\rm H} & = & \left(25.47 \pm 0.29\right)\times 10^{-6}  \\
Y_p & = & 0.245 \pm 0.003 \,.
\eea
Accounting for both observational and theoretical uncertainties - which for ${\rm D}/{\rm H}$ amount to $\sigma_D^{\rm th} = 0.6\times 10^{-6}$ (arising from nuclear reaction rates), and are negligible for $Y_p$ compared with the observational one — we derive the corresponding BBN constraints at the $2\sigma$ level. For DP masses around $m_{\gamma^\prime}\sim10 \,{\rm MeV}$—relevant given the low initial temperature $T_i \simeq 6 \,{\rm MeV}$—the dominant constraint arises from the photo-dissociation of primordial deuterium.

\subsection{X-Rays}
\label{sec:FI-astro}

The decay of cosmological DPs into the electromagnetic (EM) sector can contribute to various astrophysical observations, such as X-ray flux from the galactic center or diffuse X-ray background.
For the decay of DPs to leave a significant observational imprint, their lifetime is required to be comparable to or longer than the age of the Universe.
This long-lifetime condition can be naturally achieved for DPs with masses below $2m_e$, where the leading decay channel, $\gamma^\prime \rightarrow \gamma\gamma\gamma$, is induced at the one-loop level.
The corresponding rest-frame decay rate $\Gamma_{3\gamma}$ is given by~\cite{Pospelov:2008jk,McDermott:2017qcg}
\bea
\Gamma_{3\gamma} = \varepsilon^2 \frac{17 \alpha^4}{2^7 3^6 5^3 \pi^3}\frac{m_{\gamma^\prime}^9}{m_e^8} \mathcal{G}\left(m_{\gamma^\prime}\right)\, ,
\label{eq:DPdecay3gamma}
\eea
where $\mathcal{G}(m_{\gamma^\prime})$ accounts for deviations from the Euler-Heisenberg approximation at higher masses; see Ref.~\cite{McDermott:2017qcg}.
The inclusive photon spectrum per DP decay, defined as $dN_\gamma/d\hat{x} \equiv 3(d\Gamma_{3\gamma}/d\hat{x})/\Gamma_{3\gamma}$ with $\hat{x} = E_\gamma/(m_{\gamma^\prime}/2)$ in the range of $[0,1]$, are described by 
\bea
\frac{d N_\gamma}{d\hat{x}}  =  3\times \frac{\hat{x}^3}{51}\left(1715 - 3105 \hat{x} + \frac{2919}{2}\hat{x}^2\right) \,,
\label{eq:DPdecay3gammaSpec}
\eea
where the factor of 3 accounts for the three photons produced in each decay.

\begin{figure}[t!]
\centering 
\includegraphics[width=.9\linewidth]{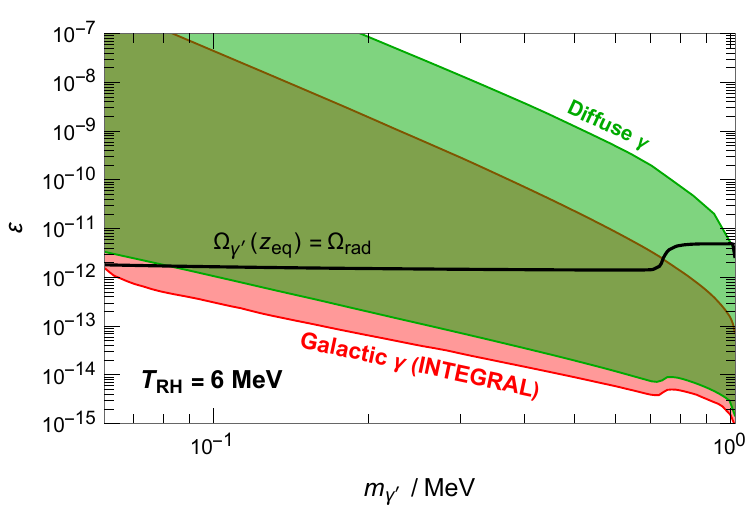}
 \caption{
 The red and green shaded regions are excluded by galactic and extragalactic X-ray observations, respectively, which constrain the X-ray flux from dark photon decays not to exceed the observed background within the corresponding instrumental sensitivities.
 The black solid line indicates the upper limit from the overproduction argument.
 } 
\label{fig:DPDMastro}
\end{figure}

Photons stemming from the DP radiative decay can result in an excess over the expected photon background flux.
Given that current X-ray observations are consistent with SM predictions within instrumental sensitivities, any such excess provides a sensitive probe of DP relics.
The expected differential galactic X-ray flux from DP decay reads
\bea
\frac{d \Phi_{\gamma}}{dE_\gamma d\Omega} = c\left[ \frac{2}{m_{\gamma^\prime}}\frac{d N_\gamma}{d\hat{x}} \right] \times \left[\frac{\Gamma_{3 \gamma}}{4\pi m_{\gamma^\prime}}\int ds \, \rho_{\gamma^\prime 0}\right]\,,
\eea
where $c$ is the speed of light, and $\rho_{\gamma^\prime 0} = m_{\gamma^\prime} Y_{\gamma^\prime} s_0 e^{-\Gamma_{3\gamma} t_0}$ is the current dark photon relic density, accounting for exponential depletion via decay.
Here, $Y_{\gamma^\prime}$ is the yield in the absence of subsequent decays, $s_0$ is the current entropy density, and $s$ denotes the line of sight.
Integrating over the area of interest introduces the parameter $D_{\gamma^\prime} = \int ds \, d\Omega \, \rho_{\gamma^\prime 0}$, which indicates the dark photon column density for the corresponding observed region.
In our analysis, we assume that the local dark photon density profile follows the morphology of the total dark matter density; i.e. that the local and global ratios of dark photons with respect to the whole DM abundance remain consistent as $D_{\gamma
^\prime}/D = \rho_{\gamma^\prime 0}/\hat{\rho}_{\rm DM}$ with the global dark matter density $\hat{\rho}_{\rm DM}$.

There are several X-ray observations of various galaxies, where the observed flux is consistent with the SM prediction within the instrumental sensitivity limits: XMM-Newton~\cite{Boyarsky:2006fg,Boyarsky:2006ag,Boyarsky:2007ay,Foster:2021ngm}, NuSTAR~\cite{Ng:2019gch,Roach:2019ctw,Roach:2023}, and INTEGRAL~\cite{Calore:2022pks}.
These studies provide constraints on the lifetime of single-component dark matter decaying into two photons, which imposes a monochromatic photon energy spectrum at the energy of half the dark matter mass for each decay.

For DPs with sub-$\rm MeV$ masses, however, the trident decay channel $\gamma^\prime \to 3\gamma$ produces a continuum photon spectrum rather than a line, resulting in a broader X-ray feature.
We adopt the analysis of Ref.~\cite{Linden:2024fby}, which derived galactic X-ray constraints on DP dark matter using INTEGRAL data.
Since Ref.~\cite{Linden:2024fby} assumes that DPs constitute the entire dark matter abundance, we rescale their lifetime limit as
\bea
\frac{m_{\gamma^\prime}Y_{\gamma^\prime}}{\rho_{\rm DM}/s_0} \leq 
\frac{\tau_{\gamma'}}{\tau_{\rm DM}} \exp\left[\frac{t_{\rm U}}{\tau_{\gamma'}}\right] \,,
\eea
where $t_{\rm U}$ is the age of the Universe and $\tau_{\rm DM}$ denotes the corresponding lifetime constraint reported in Ref.~\cite{Linden:2024fby}.
The red-shaded region in Fig.~\ref{fig:DPDMastro} is excluded by the equation above.

Besides their contribution to the present galactic X-ray flux, the radiative decay of DPs during earlier cosmic epochs can also impact the diffuse extragalactic X-ray background.
For masses around the $\rm MeV$ scale of interest, the production of DPs occurs well before the recombination era.
Moreover, DPs become non-relativistic at the time of decay, which happens much later than the recombination era to contribute to the cosmic radiation background.
Under these conditions, the diffuse photon background flux from the decay of cosmic DPs reads
\bea
\frac{d\Phi_\gamma}{dE_\gamma d\Omega} & = & \frac{Y_{\gamma^\prime} s_0 \Gamma_{3\gamma}}{4\pi E_{\gamma}}\int_0^1 d\hat{x}\,\frac{c}{H(\hat{z})}\,e^{-\Gamma_{3\gamma}t\left(\hat{x}\right)}\frac{dN_\gamma}{d\hat{x}} \nonumber\\
&&\quad \times e^{-\kappa\left(\hat{z}\,,E_{\gamma}\right)}\Theta\left(\hat{x} -\frac{E_\gamma}{m_{\gamma^\prime}/2}\right) \,,
\eea
with
\bea
t\left(\hat{x}\right) = \frac{2}{3H_0\sqrt{\Omega_{\rm m}}}\sinh^{-1}\sqrt{\frac{\Omega_\Lambda}{\Omega_{\rm m}\left(\hat{x}\left(m_{\gamma^\prime}/2\right)/E_\gamma\right)^3}} \, .
\eea
Here, $\kappa(\hat{z}\,,E_\gamma)$ accounts for the optical depth, which factors in the attenuation of photons during propagation with the present energy $E_\gamma$ emitted at a redshift $\hat{z} = \hat{x}(m_{\gamma^\prime}/2)/E_\gamma$; we exploit the result from Ref.~\cite{Cirelli:2010xx} to describe this attenuation, in particular using their public Mathematica notebooks.
Based on the given upper limits on the cosmic background radiation in  Ref.~\cite{Hill:2018trh}, the green-shaded region in Fig.~\ref{fig:DPDMastro} illustrates the constraint from the diffuse X-ray background.

\section{Supernova Dark Photons}
\label{sec:astro}

During a core-collapse supernova (CCSN), the iron core of a massive red supergiant becomes gravitationally unstable and collapses under its own weight~\cite{BetheWilson1985_ShockRevival, Janka2012_ExplosionMechanisms, BoccioliRoberti2024_CCSNePhysics, BurrowsVartanyan2021_CCSNeTheory}. This collapse halts when nuclear densities are reached, leading to the formation of a stiff, neutron-rich object known as a protoneutron star (PNS). In the first few seconds after core bounce, the temperature of the PNS can reach approximately $\sim 30-50\,{\rm MeV}$.
Hence, massive dark photons (DPs) with masses in the $\rm MeV$ to several hundred $\rm MeV$ range can be copiously produced within the PNS.
Based on Eq.~\eqref{eq:DPdifferentialrate}, the DP luminosity per volume is given by
\bea
\frac{d \mathcal{P}_{\gamma^\prime}}{d V} & = & \int d^3\vec{k}\frac{d \mathcal{N}_{\gamma^\prime}}{dV \, dt \, d^3\vec{k}} \, \omega \nonumber\\
& = & \frac{\varepsilon^2 m_{\gamma^\prime}^4}{2\pi^2 } \int \frac{d\Omega_{\gamma^\prime}}{4\pi} d\omega \frac{\omega^2 v_{\gamma^\prime}}{e^{\omega/T}-1} \nonumber\\
&&\times \sum_{i={\rm T,L}}g_i   \frac{\left|{\rm Im}\,\pi_i\right|}{\left(m_{\gamma^\prime}^2 - {\rm Re}\,\pi_i\right)^2+\left|{\rm Im}\,\pi_i\right|^2}
\label{eq:SNDPvolumeEm}
\eea
with $v^\prime=\sqrt{1-m_{\gamma^\prime}^2/\omega^2}$.
Here, we retain the angular integral of $d\Omega_{\gamma^\prime}/4\pi$.
As we will see, although produced DPs radiate isotropically, their absorptive width along the propagation trajectory, i.e., optical depth, exhibits angular dependence.

In PNS, electrons are highly degenerate and relativistic, $(3\pi n_e)^{1/3} \gg T \,, m_e^{\rm eff}$, where $n_e$ is the electron number density and $m_e^{\rm eff}$ the electron effective mass in the medium.
The plasma frequency, $\omega_{\rm pl}^2\simeq (64\pi/3)^{1/3}\alpha n_e^{2/3}$, is of order $\mathcal{O}(10)\,{\rm MeV}$. Due to the form of the in-medium photon polarization tensors, which we report in the Appendix, Eqs.~\eqref{eq:RePiT} and \eqref{eq:RePiL} that remain valid in the dense PNS environment as the hard-dense-loop approximation~\cite{Stetina:2017ozh}, we notice that resonances for the transverse mode occur when the DP mass is $\mathcal{O}(10)\,{\rm MeV}$. This is because the value of ${\rm Re}\,\pi_{\rm T}$ is on the order of the plasma frequency across the entire energy range. On the other hand, for longitudinal modes,  ${\rm Re}\,\pi_{\rm L} \sim 3\omega_{\rm pl}^2 m_{\gamma^\prime}^2/\omega^2$, and a resonance energy $\omega\sim \omega_{\rm pl}$ always exists for the low mass regime, $m_{\gamma^\prime} < \omega_{\rm pl}$.

The dominant production channel for DPs inside PNS is bremsstrahlung in neutron-proton scatterings ($n+p \rightarrow n + p +\gamma^\prime$), by which vector bosons can be radiated at the leading dipole order~\cite{Rrapaj:2015wgs,Shin:2021bvz} when the center-of-mass is not consistent with the center-of-charge.
We employ two valid approximations that make the computations simpler, the soft-radiation approximation~\cite{Low:1958sn,Nyman:1973gw} and the non-degenerate and non-relativistic conditions of the nuclear medium (for details, see Refs.~\cite{Rrapaj:2015wgs,Chang:2016ntp}). 
Under these approximations, the scattering amplitude can be separated into the neutron-proton scattering part and the photon interaction with the dipole current.
We exploit the neutron-proton scattering cross-section derived from empirical data as shown in Ref.~\cite{Rrapaj:2015wgs}. 

Furthermore, we take into account the multiple coherent scattering effects~\cite{Landau:1953um,Migdal:1956tc}, which are significant in a high-density medium of PNS.
The absorption photon rate via the inverse neutron-proton bremsstrahlung is given by
\bea
\Gamma_i^{\rm ibr} & = & \frac{32\alpha}{3\pi}\frac{n_n n_p}{\omega^3} \left(\frac{\pi T}{m_N}\right)^{3/2} \, \nonumber\\
&&\times \left<\sigma_{np}^{(2)}\left(T\right)\right> \left(\frac{\omega^2}{\omega^2 + \Gamma_{np}^2/4} \right) \times \left[\frac{m_{\gamma^\prime}^2}{\omega^2}\right]_{\rm i = L}
\label{eq:npBrem}
\eea
where $n_{n(p)}$ denotes the number density of neutrons (protons), and $\big<\sigma_{np}^{(2)}(T)\big>$ is the thermally averaged neutron-proton scattering cross section taking from Ref.~\cite{Rrapaj:2015wgs}.
The last term appears only for the longitudinal component due to the current conservation.
The factor of $\omega^2/(\omega^2 + \Gamma_{np}^2/4)$ with $\Gamma_{np} \approx n_n \big<\sigma_{np}^{(2)}(T)\big> \left(3 T/m_N\right)^{1/2}$ represents the Lorentzian shape ansatz accounting for the high-density behavior in neutron-proton bremsstrahlung scatterings~\cite{Raffelt:1991pw}.

We remark that negatively charged particles heavier than electrons, most notably muons~\cite{Bollig:2017lki} and charged pions~\cite{Fore:2019wib}, may also contribute to DP production.
Their thermal abundance in the PNS can reach the $\sim 1\%$ level relative to nucleons due to sizable chemical potentials comparable to their masses $(\sim 100\,{\rm MeV})$.
While their overall contribution to the total emissivity is subdominant, these species can populate the high-energy tail of the DP spectrum, potentially leaving observable signatures that depend sensitively on the spectral shape rather than the integrated luminosity.
However, since the abundances of these particles carry significant model and equation-of-state uncertainties, we do not include them in our numerical analysis; for a discussion of their possible impact, see Ref.~\cite{Shin:2022ulh}, and Ref.~\cite{Caputo:2021rux} for the case of axions.

In the region far outside the PNS, DP production (and also absorption) is governed by the semi-Compton process, $e^- +\gamma \rightarrow e^- + \gamma^\prime$, which is less sensitive to the medium density than bremsstrahlung scatterings.
While semi-Compton scatterings in these outer layers contribute negligibly to the total DP luminosity, they can nonetheless have a nontrivial impact on supernova dynamics, particularly shock revival for a successful explosion~\cite{Caputo:2025aac}.
In Appendix.~\ref{app:semiCompton}, we present an approximate analytic derivation of the semi-Compton absorption rate, which is related to the production rate via detailed balance, under various relevant conditions.

The integral over the momentum range of the DP production rate in Eq.~\eqref{eq:SNDPvolumeEm} can be divided into two regions, around the resonance energy and outside the resonance energy,
\bea
\frac{d \mathcal{P}_{\gamma^\prime}}{d V} = \left. \frac{d \mathcal{P}_{\gamma^\prime}}{d V}\right|_{\rm res} +\left. \frac{d \mathcal{P}_{\gamma^\prime}}{d V}\right|_{\rm out} \, .
\eea
Due to ${\rm Im}\,\pi_i \ll {\rm Re}\,\pi_i$ at the resonance energy, denoted by $\omega_{\rm res}$, we can approximate 
\beq
\begin{split}
& \frac{1}{(m_{\gamma^\prime}^2-{\rm Re}\,\pi_i)^2+\left|{\rm Im}\,\pi_i\right|^2} \\
\simeq & ~\frac{1}{\left(\frac{\partial \, {\rm Re}\,\pi_i}{\partial \omega }\right)^2\left(\omega - \omega_{\rm res}\right)^2 + \left|{\rm Im}\,\pi_i\right|^2}\\
\simeq & ~ \delta (\omega-\omega_{\rm res}) \times \pi\left(\left|\frac{\partial \, {\rm Re}\,\pi_i}{\partial \omega} \right||{\rm Im}\,\pi_i|\right)^{-1}\,.
\end{split}
\label{eq:resonanceAPP}
\eeq

Therefore, the production rate from the resonance reads
\bea
\left. \frac{d \mathcal{P}_{\gamma^\prime}}{d V}\right|_{\rm res} & \simeq &\frac{\varepsilon^2 m_{\gamma^\prime}^4}{2\pi}  \int \frac{d\Omega_{\gamma^\prime}}{4\pi}  \nonumber\\
&&\times \sum_{i={\rm T,L}}g_i \left|\frac{\partial {\rm Re}\,\pi_i}{\partial \omega}\right|^{-1} \frac{\omega_{\rm res}^2 v_{\rm res}}{e^{\omega_{\rm res}/T}-1} \,,
\eea
where $v_{\rm res} = \sqrt{1-m_{\gamma^\prime}^2/\omega_{\rm res}^2}$ and
\bea
\left|\frac{\partial {\rm Re}\,\pi_i}{\partial \omega}\right| = \frac{m_{\gamma^\prime}^2}{\omega_{\rm res}v_{\rm res}^2}\left(2+\frac{m_{\gamma^\prime}^2}{\omega_{\rm res}^2}-\frac{3\omega_{\rm pl}^2}{m_{\gamma^\prime}^2}\left[\frac{m_{\gamma^\prime}^2}{\omega_{\rm res}^2}\right]_{i={\rm L}}\right) \, .
\eea
The contribution from outside the resonance energy is explicitly calculated from Eq.~\eqref{eq:SNDPvolumeEm} with the neutron-proton bremsstrahlung scattering rate given in Eq~\eqref{eq:npBrem}.

\subsection{Implications}

\begin{figure}[t!]
\centering 
\includegraphics[width=.8\linewidth]{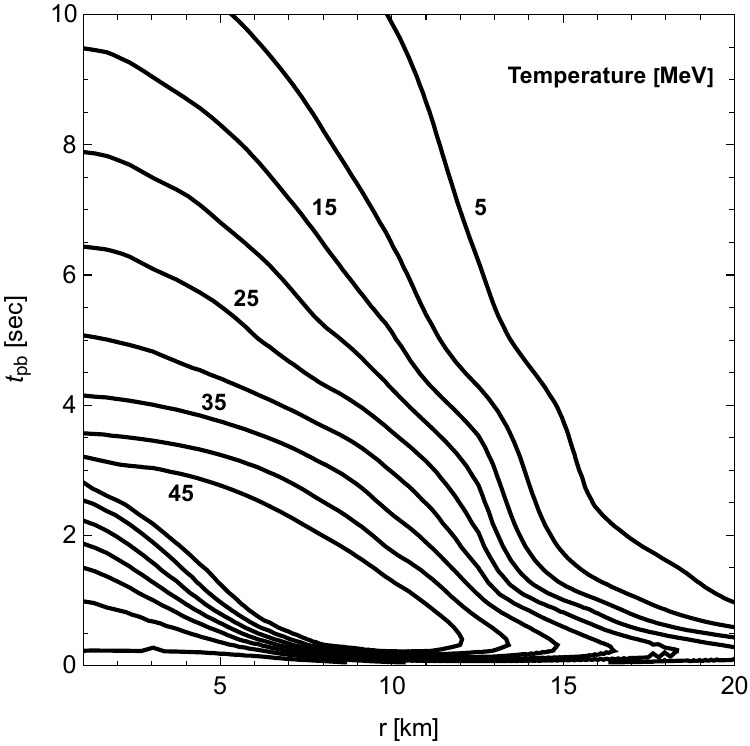}\\
\includegraphics[width=.8\linewidth]{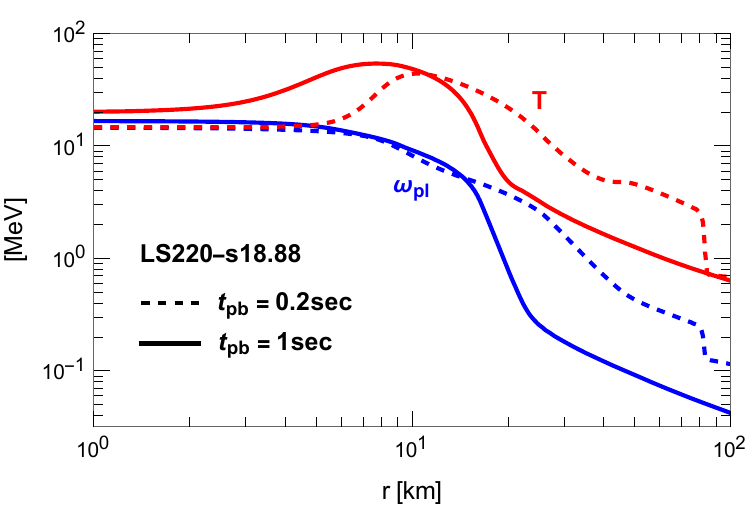}
 \caption{Upper: time evolution of the temperature in the LS220-s18.88 supernova model; contour lines correspond to the indicated value in MeV. Lower: radial profiles of temperature (red) and plasma frequency (blue) at $0.2\,{\rm sec}$ (dashed) and $1\,{\rm sec}$ (solid) post-bounce, respectively.
 } 
\label{fig:snprofile}
\end{figure}

Dark photons (DPs) produced in CCSNe can affect several observable aspects of the explosion and its aftermath. 
Building on the DP production rate described above, we consider multiple supernova observables to derive constraints on the DP parameter space. 
For our fiducial analysis, we adopt the LS220-s18.88 supernova simulation profile, which represents the one-dimensional counterpart of a self-consistent three-dimensional explosion model with a protoneutron star (PNS) mass of $1.81\,{\rm M}_\odot$~\cite{Bollig:2020phc}. 
Fig.~\ref{fig:snprofile} illustrates the time evolution of the temperature (upper panel) and the radial profiles of the temperature (red) and plasma frequency (blue) at $0.2\,{\rm s}$ (dashed) and $1\,{\rm s}$ (solid) post-bounce (lower panel). The profiles at $t_{\rm pb} \simeq 0.2\,{\rm s}$ are particularly relevant for the DP case, since for some DP masses the cooling rate can be larger at earlier times due to the nature of resonant production (this is somewhat peculiar; another similar case is the Majoron model studied in Ref.~\cite{Fiorillo:2022cdq}). We also compare our results with four additional models from the Garching supernova archive, thus bracketing the uncertainties associated with the SN modelling.

We also want to stress that no clear evidence of a compact remnant has yet been observed at the location of SN~1987A, and in particular no pulsar-like signal. In principle, this could call into question a neutrino-driven explosion mechanism for SN~1987A, although the remnant could simply be a non-pulsar neutron star, and the hot blob recently reported by the ALMA radio telescope and the James Webb Space Telescope can be interpreted as a hint of such a source (see section 8.6 in Ref.~\cite{Caputo:2024oqc} for a more detailed discussion about this). In any case, among the arguments that we examine here, only the standard cooling argument and the SN failure argument rely on a specific, neutrino-driven explosion mechanism. By contrast, low-energy supernovae, fireball formation, supernova $\gamma$-ray signals, and Galactic positron injection are independent on the explosion mechanism and, taken together, supersede the standard cooling argument, making SNe a robust probe. Nevertheless, we recall that much above the free-streaming boundaries, self-consistent simulations would be desirable.

\subsubsection{Cooling argument}
\label{sec:SNcooling}

The gravitational binding energy of the nascent PNS ($\approx 3\times10^{53}\,{\rm erg}$) is predominantly released in the form of neutrinos within the Standard Model. 
For the first few seconds after core bounce, neutrinos remain trapped in the dense stellar matter surrounding the PNS, forming a neutrino sphere. Energy diffusion through this region leads to a cooling phase lasting \(\mathcal{O}(10)\) seconds. The observed duration of the neutrino burst from SN1987A, detected by Kamiokande-II, IMB, and Baksan~\cite{Bionta:1987qt,Kamiokande-II:1987idp,Alekseev:1987ej}, agrees remarkably well with this prediction. 
To remain consistent with these observations, any exotic energy-loss mechanism must be subdominant to the standard neutrino cooling. This requirement is encapsulated in the so-called \emph{Raffelt criterion}~\cite{Raffelt:1996wa, Caputo:2021rux}, which constrains the specific energy-loss rate of new particles, ensuring that the overall cooling timescale is not significantly shortened.

DPs produced in PNS can contribute to extra energy leakage if they escape the region cooled down by neutrino diffusion.
In the course of propagation, DPs may either be absorbed into the nuclear medium through scatterings or decay into electron-positron pairs if the mass exceeds the two-electron mass threshold.
This absorption process depends on the production location of a DP and its direction of travel.
The optical depth for a DP produced at radius $r$ with energy $\omega$, traveling at an angle $\theta_{\gamma^\prime}$ relative to the radial direction, is given by
\bea
\tau_{\rm cool} = \int dl  \, \hat{\Gamma}_{\gamma^\prime} \left(\sqrt{r^2+l^2+2rl\cos\theta_{\gamma^\prime}}\right)/v^{\prime}\,,
\label{eq:taucool}
\eea
where $l$ represents the path length from the production point ($l\geq 0$), and $\hat{\Gamma}_{\gamma^\prime}$ is the DP absorptive rate.
The effective cooling rate due to DPs is then expressed as
\bea
\left. \mathcal{P}_{\gamma^\prime}\right|_{\rm cool} & = & \int dV \frac{d \mathcal{P}_{\gamma^\prime}}{d V} e^{-\tau_{\rm cool}} \,
\eea
with the volume emissivity $d\mathcal{P}_{\gamma^\prime}/dV$ given by Eq.~\eqref{eq:SNDPvolumeEm}.

For DP masses below $2m_e$, the optical depth is primarily determined by scatterings near PNS due to relatively higher density, allowing integration over arbitrarily large distances.
On the other hand, for masses above $2m_e$, DPs can decay into electron-positron pairs outside the PNS which leads to a significant contribution to the optical depth without suppression at large radii.
In our analysis, we set an upper integration limit for radius $r_{\rm up} = 20\,{\rm km}$ (i.e., the corresponding upper limit of $l=\sqrt{r_{\rm up}^2-r^2\sin^2\theta_{\gamma^\prime}}-r_{\rm up}\cos\theta_{\gamma^\prime}$), which appropriately accounts for the geometric size of the PNS and the neutrino sphere.

The DP absorption rate is written by
\bea
\hat{\Gamma}_{\gamma^\prime} & \simeq & \dfrac{\varepsilon^2 m_{\gamma^\prime}^4}{\left(m_{\gamma^\prime}^2 - {\rm Re}\,\pi_i\right)^2+\left|{\rm Im}\,\pi_i\right|^2} \nonumber\\
&& \times \left(1-e^{-\omega/T}\right) \left[\Gamma_i^{\rm ibr}+\Gamma_i^{\rm sC} +\Gamma_i^{e\bar{e}}\right] \, 
\label{eq:SNDPabs}
\eea
for each polarization mode $i={\rm T,L}$.
Inside the PNS, the absorption of DPs is controlled by the inverse neutron-proton bremsstrahlung, the rate ($\Gamma_i^{\rm ibr}$) of which is described in Eq.~\eqref{eq:npBrem}.
On the other hand, the semi-Compton scattering process, $e^- \gamma^\prime \rightarrow e^- \gamma$, dominates the DP absorption outside of the PNS due to its weaker power dependence of the number density (i.e., $\propto n_e$).
For DP masses above $2m_e$, an additional absorptive channel opens up via decay into electron-positron pairs.
Ignoring the positron abundance, the decay rate $\Gamma_i^{e\bar{e}}$ is given by
\bea
\Gamma_i^{e\bar{e}} \simeq \frac{1}{16\pi\sqrt{\omega^2 - m_{\gamma^\prime}^2}} \int_{x_-}^{x_+} dx\frac{\left|\bar{\mathcal{M}}_i\right|^2}{e^{\frac{\mu_e-\omega x}{T}}+1}\,,
\eea
where $\mu_e$ is the chemical potential of electrons, and $x_\pm = \left(1 \pm \sqrt{(1-4m_e^2/m_{\gamma^\prime}^2)(1-m_{\gamma^\prime}^2 / \omega^2)}\right)/2$ defines the integration range.
The squared matrix elements read
\bea
\left|\bar{\mathcal{M}}_i\right|^2 & = & 8\pi \alpha \times\left\{ 
\begin{tabular}{ll}
     $\left(m_{\gamma^\prime}^2 +2 m_e^2\right) - \xi \left(x\right)$ & \, $\left[i={\rm T}\right]$ \\
     $ 2 \xi \left(x\right)$ & \, $\left[i={\rm L}\right]$
\end{tabular}
\right.
\eea
with $\xi (x)  = \left(2 x(1-x) \omega^2 -m_{\gamma^\prime}^2/2\right)m_{\gamma^\prime}^2/(\omega^2-m_{\gamma^\prime}^2)$; for the detailed derivation, see Ref.~\cite{Chang:2016ntp}.
\begin{figure}[t!]
\centering 
\includegraphics[width=.9\linewidth]{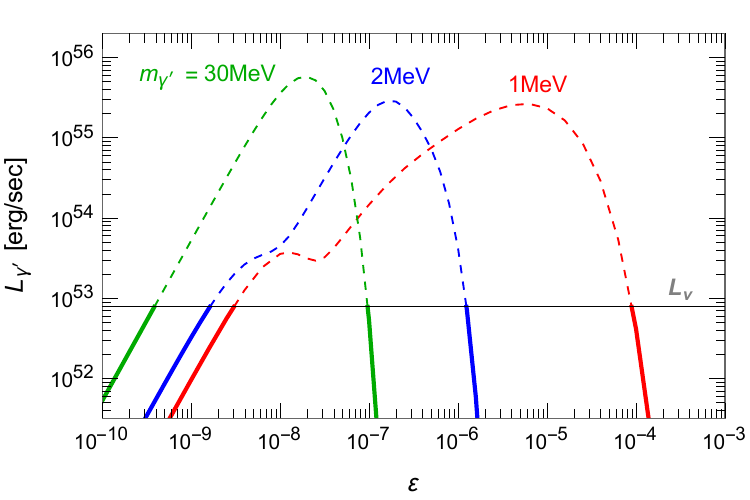}
 \caption{
 Dark photon luminosity as a function of the kinetic mixing parameter $\varepsilon$ for masses of $1\,{\rm MeV}$ (red), $2\,{\rm MeV}$ (blue), and $30\,{\rm MeV}$ (green), evaluated at 1s post-bounce. The gray line indicates the neutrino luminosity.} 
\label{fig:LumSNDP}
\end{figure}

Fig.~\ref{fig:LumSNDP} shows the DP luminosity as a function of the kinetic mixing parameter $\varepsilon$ for several representative masses.
In the low-coupling regime, where DPs freely escape from the PNS, the luminosity scales as $\varepsilon^2$.
At larger couplings, the typical optical depth in Eq.~\eqref{eq:taucool} exceeds unity, leading to a suppression of the effective DP luminosity due to increased trapping and reabsorption.
For lighter masses, a peak appears at intermediate couplings, where resonantly produced DPs are also resonantly reabsorbed near their production region, as shown by the red and blue lines in Fig.~\ref{fig:LumSNDP} for $m_{\gamma^\prime} = 1\,{\rm MeV}$ and $2\,{\rm MeV}$, respectively.
This implies that at higher couplings, the DP luminosity becomes dominated by non-resonant production processes.
In this context, heavier DPs (such as the case with $m_{\gamma^\prime} = 30\,{\rm MeV}$ shown by the green line) exhibit only a single peak due to the absence of resonance effects.
For DP masses below the $2m_e$ threshold, where decay into electron–positron pairs is kinematically forbidden, absorption is dominated by scattering processes; above this threshold, the additional decay channel into electron-positron pairs significantly enhances the optical depth.

\begin{figure}[t!]
\centering 
\includegraphics[width=.9\linewidth]{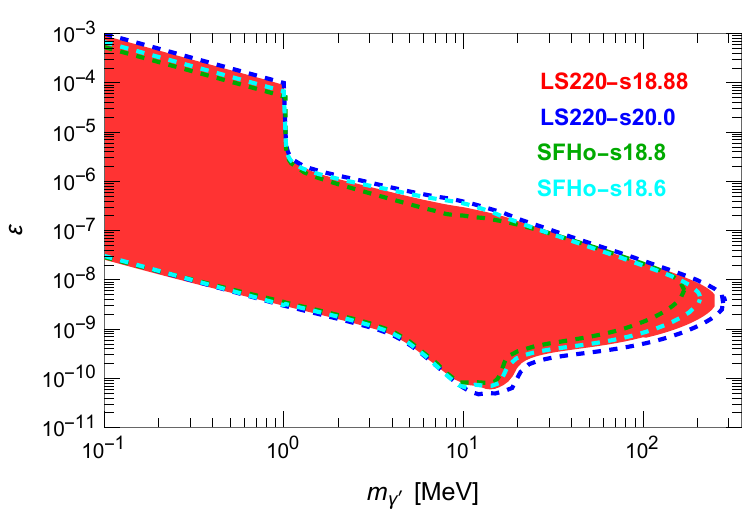}
 \caption{
 Constraints from the SN1987A cooling argument for different SN models: LS220-s18.88 (red-shaded), LS220-s20.0 (blue dashed), SHFo-s18.8 (green dashed), SFHo-s18.6 (cyan dashed).} 
\label{fig:DPcooling}
\end{figure}

Consequently, we derive cooling constraints from the SN1987A neutrino observations by comparing the effective DP cooling rate at 1\,s post-bounce with the corresponding neutrino luminosity (see, e.g., the gray line in Fig.~\ref{fig:LumSNDP}). 
The red region in Fig.~\ref{fig:DPcooling} illustrates the parameter space where the effective DP emission rate exceeds the diffusive neutrino luminosity, based on the LS220-s18.88 simulation profile. 
We additionally show cooling constraints obtained from other supernova models of the Garching group~\cite{Bollig:2020xdr,CCSNarchive} with different progenitor masses and/or nuclear equations of state (EOS): LS220-s20.0 (blue dashed), SFHo-s18.8 (green dashed), and SFHo-s18.6 (cyan dashed). 
The LS220 EOS employs a Lattimer–Swesty compressible liquid-drop model with an incompressibility of $K=220\,{\rm MeV}$, while the SFHo EOS~\cite{Steiner:2012rk} is based on relativistic mean-field theory and yields a more compact PNS with slightly higher central densities and temperatures. 
The comparison across these simulations, covering progenitor masses between $18.6$ and $20.0\,{\rm M}_\odot$, allows us to gauge the sensitivity of our results to uncertainties in the progenitor structure and the nuclear microphysics.
Overall, the resulting constraints are largely consistent across most of the parameter space, with some deviations only in the trapping regime for $m_{\gamma^\prime} \lesssim 1\,{\rm MeV}$—where the result is sensitive to the target density for scattering—and in the heavy-mass region, where the constraint depends more strongly on the maximum temperature of the profile.

\begin{figure}[t!]
\centering 
\includegraphics[width=.9\linewidth]{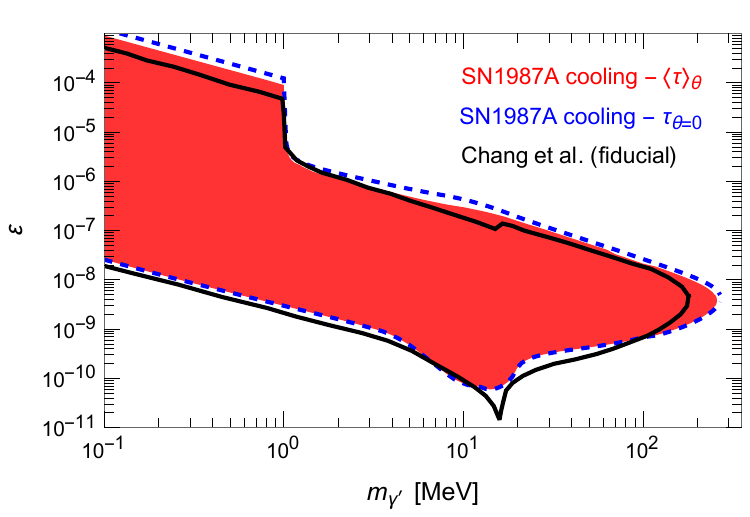}
 \caption{Our SN1987A cooling constraints (red) using the Garching LS220-s18.88 model, compared with the previous results from Ref.~\cite{Chang:2016ntp}. We also show results (dashed blue) obtained by evaluating the optical depth assuming purely radial propagation. As pointed out by Ref.~\cite{Caputo:2021rux}, such an incorrect procedure leads to an overestimation of the constraints.} 
\label{fig:DPcooling_comparison}
\end{figure}

We also compare our updated results with those previously reported in the literature~\cite{Chang:2016ntp}.
Fig.~\ref{fig:DPcooling_comparison} shows this comparison, where the black solid line represents the result from Ref.~\cite{Chang:2016ntp} obtained using their fiducial supernova profile.
As pointed out by Ref.~\cite{Caputo:2021rux}, their evaluation of the optical depth assumes a purely radial propagation (i.e., $\cos\theta_{\gamma^\prime} = 1$ in Eq.~\eqref{eq:taucool}) with an additional geometric correction.
This prescription tends to overestimate the optical depth, leading to a stronger constraint in the trapping regime.
Nevertheless, their bound for $m_{\gamma^\prime} \lesssim 1,{\rm MeV}$ is less stringent than ours because their fiducial profile contains a higher electron density outside the PNS, which enhances the optical depth through the semi-Compton scattering contribution.
For comparison, we confirm that adopting a purely radial optical depth treatment (blue dashed line) indeed reproduces this overestimation.

Finally, we note that the Stefan–Boltzmann (SB) approach provides a good approximation to the cooling bound in the trapping regime for low-mass DPs, where $m_{\gamma^\prime}$ is smaller than the typical temperature at the PNS boundary ($T \sim 10\,{\rm MeV}$).
The differential SB luminosity~\cite{Caputo:2021rux} reads
\bea
\frac{d\mathcal{P}_{\gamma^\prime}^{\rm SB}}{d\omega} = \sum_{i={\rm T,L}}4\pi R_{\omega}^2 \frac{g_i}{8\pi^2} \frac{\omega^3}{e^{\omega/T(R_\omega)}-1} \,,
\eea
where $R_\omega$ denotes the radius at which the radial optical depth becomes $2/3$ (i.e., [$\tau_{\rm cool}]_{\cos\theta_{\gamma^\prime} = 1} =2/3$).
Since the DP absorption rates from bremsstrahlung and semi-Compton processes depend on the DP energy $\omega$, each energy corresponds to a different diffusion radius $R_\omega$.
In particular, the absorption rate decreases with increasing $\omega$, implying that the dominant emitted DP spectrum in the trapping regime shifts toward higher energies as $\varepsilon$ increases~\cite{Chang:2016ntp}.
For very heavy DPs, however, the SB approximation ceases to be valid.
In this regime, the dominant production region lies deep inside the PNS, near the maximum temperature, where the transport of energy even in the trapping limit remains essentially ballistic rather than diffusive.

\subsubsection{Low-energy supernova explosion}
\label{sec:SNexplosion}

Even if DPs successfully escape the PNS, they can still interact with materials in the outer layer of the progenitor.
Such interactions may deposit energy into the mantle, subsequently increasing the total kinetic energy of SN explosive ejecta~\cite{Sung:2019xie,Caputo:2022mah, Falk:1978kf}.

The explosion energy of a CCSN is inferred from light-curve properties and spectroscopic analysis.
While the typical range of the SN explosion energy is around $1\,{\rm bethe}$ ($1{\rm bethe} = 10^{51}\,{\rm erg}$), some SN cases imply an order of magnitude lower explosion energy, on the order of $0.1\,{\rm bethe}$~\cite{Pastorello:2003tc, Spiro_2014}. We exploit observations of low-energy supernova explosions to constrain DPs, following Ref.~\cite{Caputo:2022mah}, which pioneered this method using the axion–photon coupling as an illustrative example.

DPs deposit energy in the mantle through scatterings with materials or via their decay into visible particles.
Scattering processes, which are proportional to the number density of charged particles, become significantly less efficient outside the PNS, where the density drastically decreases.
Therefore, the absorption of DPs via scatterings would be inefficient in transferring their energy to the mantle.
On the other hand, the decay of DPs is almost independent of the thermal properties of the outer stellar medium (e.g., electron degeneracy).

For DP masses below $2m_e$, decays occur only via the radiative three-photon channel, which is extremely suppressed and leads to lifetimes too long to contribute to SN explosion energy.
On the other hand, for $m_{\gamma^\prime} > 2m_e$, their decay into electron-positron pairs is kinematically allowed, and the lifetime can be short enough to deposit their energy into the mantle.
Hence, low-energy supernova explosion observations can give a relevant constraint for $m_{\gamma^\prime} > 2m_e$.

We follow the procedure in Ref.~\cite{Fiorillo:2025yzf} to derive the constraint from low-energy SN explosion observations.
Based on the energy-deposition framework developed in Refs.~\cite{Sung:2019xie,Caputo:2022mah}, Ref.~\cite{Fiorillo:2025yzf} provides an improved treatment that properly accounts for the effective energy transfer from feebly interacting particles produced inside the PNS.
In particular, it offers a consistent description of the strong-coupling regime, where DPs are efficiently absorbed or decay within the dense PNS medium, leading to diffusive energy transport.
The energy deposited in the mantle by DP decays can be expressed as~\cite{Fiorillo:2025yzf}
\bea
\mathcal{E}_{\rm mantle} = \int dt \left[\mathcal{P}_{\rm PNS}^{(1)} - \mathcal{P}_{\rm PNS}^{(2)} - \mathcal{P}_{\rm prog}^{(1)}- \mathcal{P}_{\rm prog}^{(2)}\right] \,.
\eea
Here, $\mathcal{P}_{\rm PNS}^{(1)} = \mathcal{P}_{\gamma^\prime}|_{\rm cool}$ represents the total DP luminosity generated inside the PNS that is available for energy deposition, while $\mathcal{P}_{\rm PNS}^{(2)}$ correspond to the portion escaping the progenitor.
The terms $\mathcal{P}_{\rm prog}^{(1)}$ and $\mathcal{P}_{\rm prog}^{(2)}$ denote the DP production in the mantle, with decays occurring inside the PNS and outside the progenitor, respectively.

In the weak-coupling regime, where $\mathcal{P}_{\rm PNS}^{(i)} \gg \mathcal{P}_{\rm prog}^{(i)}$ and the mean free path of DPs exceeds the PNS radius (so that energy transport is ballistic), the deposited energy in the mantle reads
\bea
\mathcal{E}_{\rm mantle} = \int dt \int dV \frac{d \mathcal{P}_{\gamma^\prime}}{d V} \left(e^{-\tau_{\rm cool}} - e^{-\tau_{R_*}}\right) \,. 
\label{eq:LowEDP}
\eea
The DP volume emissivity ($d\mathcal{P}_{\gamma^\prime}/dV$), the optical depth at the boundary of the PNS ($=\tau_{\rm cool}$) are described above, and $\tau_{R_*}$ indicates the optical depth for the progenitor surface, which is dominated by the decay probability due to the large progenitor radius.
The volume integration range for DP production is limited to the PNS region ($R_{\rm PNS}=20\,{\rm km}$).
For $m_{\gamma^\prime} > 2m_e$, we find
\bea
\tau_{R_*} \simeq \Gamma_{e\bar{e}} \frac{R_* }{v^\prime}\,
\eea
with the vacuum decay rate of the electron-positron channel (neglecting the medium effect in the mantle)
\bea
\Gamma_{e\bar{e}}= \varepsilon^2\frac{\alpha}{3}\frac{m_{\gamma^\prime}^2}{\omega} \left(1+\frac{2m_e^2}{m_{\gamma'}^2}\right)\sqrt{1-4\frac{m_e^2}{m_{\gamma^\prime}^2}}\, .
\label{eq:DPdecayee}
\eea

In the small-$\varepsilon$ regime of our interest, DPs freely escape the PNS (i.e., $\tau_{\rm cool} \ll 1$) and decay predominantly within the progenitor (i.e., $\tau_{R_*} \gg 1$).
Eq.~\eqref{eq:LowEDP} then simplifies to
\bea
\mathcal{E}_{\rm mantle} \simeq \int dt \int dV \frac{d \mathcal{P}_{\gamma^\prime}}{d V}\,
\label{eq:SNDPlowEsmallcoupling}
\eea
corresponding to the total energy emitted in DPs from the PNS.
This allows a simple estimate of the bound by comparing the DP luminosity with the neutrino luminosity at $1$sec post-bounce (Sec.~\ref{sec:SNcooling}).

At large couplings, where the DP absorption or decay length becomes shorter than the PNS radius, the ballistic description in Eq.~\eqref{eq:LowEDP} may no longer apply, but the energy transport from the PNS would become diffusive~\cite{Fiorillo:2025yzf} (i.e., $\mathcal{P}_{\rm PNS}^{(1)} \sim \mathcal{P}_{\rm prog}^{(1)} \gg \mathcal{P}_{\rm PNS}^{(2)}\,, \mathcal{P}_{\rm prog}^{(2)}$).
Such a diffusive heating relies on the DP thermal conductivity, which is set by the mean free path $\lambda_{\gamma^\prime} \simeq v^\prime/\Gamma_{e\bar{e}} \propto \varepsilon^{-2}$.
The resulting diffusive energy deposition rate can be approximated analytically (see Ref.~\cite{Fiorillo:2025yzf} for detailed derivation)
\bea
\frac{d\mathcal{E}_{\rm mantle}}{dt} =\frac{2R_{\rm PNS}^2}{3\pi T^2} \left(-\frac{\partial T}{\partial r}\right)\int dE_{\gamma^\prime} \frac{\lambda_{\gamma^\prime} E_{\gamma^\prime}^3 p_{\gamma^\prime}e^{E_{\gamma^\prime}/T}}{\left(e^{E_{\gamma^\prime}/T}-1\right)^2}\,
\eea
evaluated at the PNS surface, $r=R_{\rm PNS}$.

However, this approximation breaks down for DP masses above $50\,{\rm MeV}$, since it assumes diffusive DP emission at the PNS radius where $T < 10\,{\rm MeV}$.
For heavier DPs, their emission are dominated deeper inside the PNS, where the temperature is higher, and thus even in the strong-coupling regime the transport remains effectively ballistic.
In this case, Eq.~\eqref{eq:SNDPlowEsmallcoupling} remains valid.

\begin{figure}[t!]
\centering
\includegraphics[width=.8\linewidth]{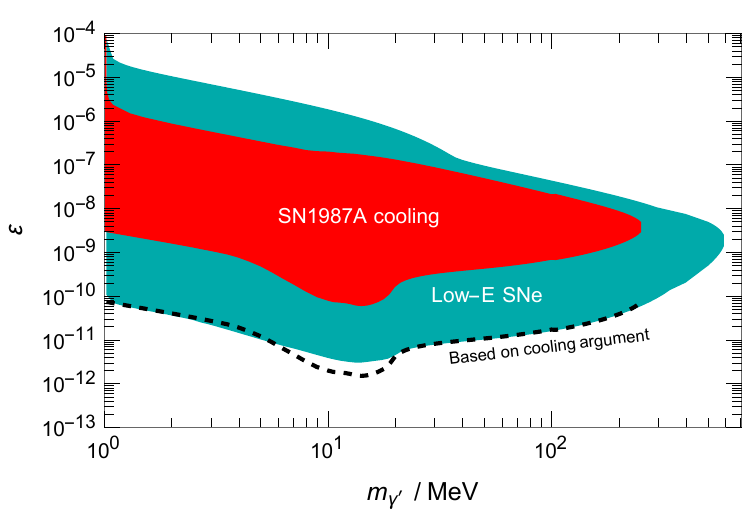}
 \caption{
 The cyan-shaded region shows the constraint from low-energy supernova explosions, derived from energy deposition in the mantle due to dark photon decays. The black dashed line indicates the result obtained using the approach based on the cooling argument (red), with an effective emission duration of 2 seconds.}
\label{fig:DPSNlowESNexp}
\end{figure}

To evaluate $\mathcal{E}_{\rm mantle}$, we use the SN simulation profile for the LS220-s18.88 model, incorporating all post-bounce time snapshots, and adopt a progenitor radius of $R_* \simeq 10^{14}\,{\rm cm}$.
The cyan region in Fig.~\ref{fig:DPSNlowESNexp} (and the corresponding `LESNe' region in Fig.~\ref{fig:moneyplot}) is excluded by the condition of $\mathcal{E}_{\rm mantle} < 0.1\,{\rm bethe}$.
As expected, the low-energy SN explosion provides a more stringent constraint on DPs than the cooling argument.

The black curve in Fig.~\ref{fig:DPSNlowESNexp} indicates the result from the simplified approach described in Eq.~\eqref{eq:SNDPlowEsmallcoupling} with the effective emission duration of 2 seconds.
This approximate bound agrees well with the full numerical result, except near $m_{\gamma^\prime} \sim 10\,{\rm MeV}$, where transverse resonance production becomes significant.
In the numerical calculation, this resonance is smeared out by accounting for the time-evolving profiles.

\subsubsection{Absence of prompt SN1987A $\gamma$-ray signals}
\label{sec:SNgamma}

In standard core-collapse SN scenarios, any energetic photons produced within the PNS are promptly trapped and thermalized due to the extremely high density of the stellar medium.
Consequently, no prompt $\gamma$-ray signal is expected to accompany the core collapse.
However, new particles such as DPs, which interact only feebly with SM particles, can escape the progenitor.
If they subsequently decay into photons, these events may yield detectable $\gamma$-ray signals, potentially observable with minimal delay relative to neutrino emission.

At the time when the diffusive neutrino burst from SN1987A was observed, the Gamma-Ray Spectrometer (GRS) aboard the Solar Maximum Mission satellite~\cite{1979ICRC....5..135R} was operational.
It monitored the $\gamma$-ray flux in three energy bands: $4.1\text{-}6.4\,{\rm MeV}$, $10\text{-}25\, {\rm MeV}$, and $25\text{-}100 \,{\rm MeV}$~\cite{Oberauer:1993yr}.
The observation period extended for $223.2\,\rm sec$ post-bounce before the satellite switched to calibration mode. No excess was detected in any energy range, leading to upper bounds on the $\gamma$-ray fluence:
\bea
\mathcal{F}_\gamma   & < & 6.11\,{\rm cm}^{-2}\quad  \left[ 4.1\text{-}6.4\,{\rm MeV} \right]\,, \\
\mathcal{F}_\gamma  & < & 1.48\,{\rm cm}^{-2} \quad \left[ 10\text{-}25\, {\rm MeV} \right]\,, \\
\mathcal{F}_\gamma  & < & 1.84\,{\rm cm}^{-2}\quad \left[ 25\text{-}100 \,{\rm MeV}\right]\,.
\eea

Since the region where DP-induced photons can arrive within the observation window is much closer to the progenitor than the SN1987A distance ($d_* = 51.4,\rm kpc$), the expected $\gamma$-ray fluence from DP decays can be approximated as
\bea
\mathcal{F}_\gamma & \simeq & \int dr \int d\cos\theta_\gamma \int dE_{\gamma} \int d\omega \frac{1}{4\pi d_{*}^2}\frac{d\mathcal{N}_{\gamma^\prime}}{d\omega} e^{-\Gamma_{\gamma^\prime}r/v^\prime}\nonumber\\
&&\times   \frac{1}{v^\prime}\frac{\partial^2 \Gamma_{\rm rad}}{dE_\gamma d\cos\theta_\gamma}
\Theta\left(t_{\rm delay} < 223.2\,{\rm sec}\right)\,,
\label{eq:SNgammaFluence}
 \eea
where $d\mathcal{N}_{\gamma^\prime}/d\omega$ is the spectrum of the total number of DPs emitted from the SN, integrated over the entire PNS and all post-bounce snapshots.
Dividing by $4\pi d_*^2$ gives the DP fluence at Earth in the absence of decay.
While $\Gamma_{\gamma^\prime}$ is the total decay rate of DPs with the energy $\omega$, $\Gamma_{\rm rad}$ denotes the partial decay rate into channels with a final-state photon.
The variables $E_\gamma$ and $\theta_\gamma$ are defined as the photon energy and angle relative to the DP boost in its rest frame, respectively.
The observed photon energy at Earth is then given by $E_\gamma \gamma^\prime (1+ v^\prime \cos\theta_\gamma)$ with $v^\prime = \sqrt{1-m_{\gamma^\prime}^2/\omega^2}$ and $\gamma^\prime = \omega/m_{\gamma^\prime}$.

The non-vanishing DP mass and a non-straight line of sight of radially emitted DPs (corresponding relative angle to the straight line is correlated with $\theta_\gamma$) impose a delay in the arrival of photon signals from the DP decay, which reads
\bea
t_{\rm delay} \simeq \frac{r}{v^\prime \gamma^{\prime 2}\left(1+v_{\gamma^\prime}\cos\theta_\gamma\right)} \, .
\eea
The observational constraint, $t_{\rm delay} < 223.2\,{\rm sec}$, in the last term of Eq.~\eqref{eq:SNgammaFluence} leads to the integration limits.
Furthermore, for the signals to be detectable, DPs must decay outside the photosphere, i.e., $r>R_*$.

We thus derive constraints on dark photons (DPs) from the non-observation of \emph{prompt} SN~1987A $\gamma$-ray emission.  For masses above the di-electron threshold, $m_{\gamma'} > 2 m_e$, the decay $\gamma' \to e^+ e^-$ is kinematically allowed, yielding a short DP lifetime with a total DP decay rate given by Eq.~\eqref{eq:DPdecayee}. 
The dominant radiative mode is final-state radiation (FSR), $\gamma' \to e^- e^+ \gamma$, for which we adopt the differential rate from Appendix~B of Ref.~\cite{DeRocco:2019njg}. 
The resulting constraint on the DP parameter space, derived from the absence of a SN~1987A $\gamma$-ray excess, is shown in orange in Fig.~\ref{fig:moneyplot} and labeled as ``SN~1987A $\gamma$.''

Apart from SN~1987A, one may also consider the $\gamma$-ray signals from Type~Ic SNe, which have much more compact progenitors and therefore allow one to access different regions of parameter space by stacking many such events detected by Fermi-LAT~\cite{Candon:2025ypl} (stacking multiple events is necessary, because if a clear timing of the onset of the explosion is not available, a single Type Ic SN cannot individually provide constraints). Furthermore, it would be interesting to study the impact of DPs on early-time observations of SN explosions, such as those of SN~2023ixf~\cite{Kozyreva:2024ksv}. We leave these studies for the future.

\subsubsection{Galactic positron injection}
\label{sec:SNpositron}

Injection of a significant number of positrons into the Milky Way can substantially enhance the diffuse $\gamma$-ray emission from the Galactic Center region.  As these positrons propagate through the interstellar medium, they lose energy via Coulomb interactions with ambient electrons, emit bremsstrahlung radiation, and eventually annihilate.  The annihilation of non-relativistic positrons, predominantly through the formation of positronium (a bound state of an electron and a positron), gives rise to the observed $511\,{\rm keV}$ $\gamma$-ray line flux~\cite{Prantzos:2010wi}. 
Precise measurements of this $511\,{\rm keV}$ line by the SPI spectrometer onboard the \textit{INTEGRAL} satellite~\cite{Siegert:2015knp,Siegert:2019tus, Strong:2005fr, Bouchet:2010dj} have placed stringent limits on the Galactic non-relativistic positron injection rate, constraining it to be no larger than $\mathcal{O}(10^{43})\,e^+\,{\rm s}^{-1}$.

The average number of positrons injected into the Galaxy per CCSN, which eventually thermalize and contribute to the $511\,{\rm keV}$ emission through the decay of SN-produced dark photons outside the progenitor, can be expressed as
\bea \mathcal{N}_{e^+}^{\rm SN} & = & \sum_{i=\text{I}\,,\text{II}} r_i \int dt \,d\omega \frac{d\mathcal{N}_{\gamma^\prime}}{dt d\omega} e^{-\tau_{R_*}} \nonumber \\ \nonumber & & \times \int dE_{e^+} \frac{dN_{e^+}}{dE_{e^+}} P_{E_{e^+}\rightarrow m_e} \,, \label{eq:Nep}\eea
where $P_{E_{e^+}\rightarrow m_e}$ denotes the survival probability for a positron with initial energy $E_{e^+}$ to become non-relativistic through interactions with the ambient Galactic medium~\cite{Beacom:2005qv}. 
Here, $dN_{e^+}/dE_{e^+}$ represents the positron spectrum from the decay of a boosted dark photon with energy $\omega$; the general transformation between the boosted and rest-frame spectra is presented in Appendix~\ref{app:spectrumBoost}. 

There are two main CCSN subclasses that contribute to this rate: Type~II and Type~Ib/c. 
They are distinguished by the presence (Type~II) or absence (Type~Ib/c) of hydrogen features in their spectra, which reflects the difference in progenitor envelope size and composition. 
Accordingly, the relative weight $r_i$ in the sum in Eq.~\ref{eq:Nep} accounts for the occurrence rate of each SN type in the Galaxy.

\begin{figure}[t!]
\centering
\includegraphics[width=1.\linewidth]{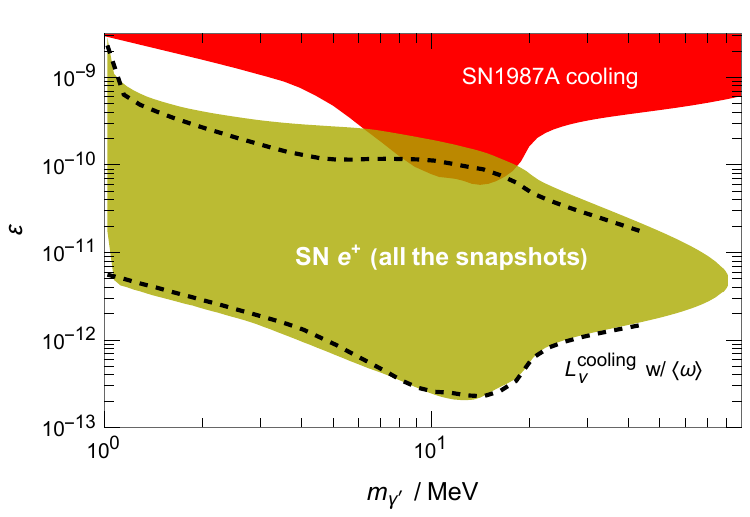}
 \caption{The yellow-shaded region is excluded by galactic positron injection leading to the observed $511\,{\rm keV}$ line. The black dashed lines represent the approximate estimate obtained by rescaling the SN1987A cooling constraint (red-shaded).}
\label{fig:DPSNpositron}
\end{figure}

Following Ref.~\cite{Calore:2021lih}, we adopt the upper limit on the number of positrons injected per CCSN,
\bea
N_{e^+}^{\rm SN} < 1.4 \times 10^{52}\,,
\label{eq:CONSTgalpos}
\eea
assuming the CCSN rate in the Milky Way galaxy of $2.3$ per century~\cite{Li:2010kd}.
For analysis, we take the progenitor radii of $R_* = 5\times 10^{12}\,{\rm cm}$ for Type Ib/c and $10^{14}\,{\rm cm}$ for Type II, with the relative fraction $r_\text{I} = 33\%$ and $r_\text{II} = 67\%$~\cite{Li:2010kd}, respectively.
Since the core-collapse and the subsequent PNS formation are essentially identical for both classes, we assume an equivalent DP production mechanism.
By integrating over all the time snapshots, the yellow-shaded region in Fig.~\ref{fig:DPSNpositron} (and the corresponding `511 keV line' region in Fig.~\ref{fig:moneyplot}) is excluded by the condition in Eq.~\eqref{eq:CONSTgalpos}.

Analogous to the low-energy supernova explosion constraint discussed in Sec.~\ref{sec:SNexplosion}, one can estimate the galactic positron injection constraint, based on a simple rescaling of the SN cooling limits.
In the regime where DPs freely escape the PNS, the constraint can be rescaled as
\bea
N_{e^+}^{\rm SN} \sim \sum_{i=\text{I}\,,\text{II}} r_i t_{\rm SN} \left(\frac{\varepsilon}{\varepsilon_{\rm cool}}\right)^2 \frac{L_\nu}{\left<\omega\right>} e^{-\tau_{R_*}(\left<\omega\right>)}\,,
\eea
where $\varepsilon_{\rm cool}$ denotes the kinetic mixing corresponding to the SN1987A cooling constraint (the boundary of the red-shaded region in Fig.~\ref{fig:DPSNpositron}), and $L_\nu$ is the total neutrino luminosity.
We take an effective duration of $t_{\rm SN} = 2\,{\rm sec}$ and assume a monochromatic DP spectrum with the average energy $\langle \omega \rangle$; one can evaluate this latter numerically, but a decent approximation (especially for DP masses above $\sim 20 \, \rm MeV$ is $\langle \omega \rangle\sim m_{\gamma^\prime} + T$.
The black dashed lines in Fig.~\ref{fig:DPSNpositron} show this approximation, which agrees well with the full numerical result except around $m_{\gamma^\prime} \sim 10\,{\rm MeV}$, where the broad transverse-resonance spectrum becomes relevant.
For masses below $3\,{\rm MeV}$, the production is dominated by longitudinal resonance, which imposes a sharply peaked DP spectrum centered near $\langle \omega \rangle$.

While here we have considered only the annihilation of non-relativistic positrons, predominantly through the formation of positronium, one can also consider relativistic positrons losing energy by scattering on free and bound electrons in the ISM, and then leading to a continuum emission above 511 keV~\cite{DelaTorreLuque:2024zsr}. This type of radiation leads to constraints (using COMPTEL and EGRET data) which are slightly better or slightly worse than the ones considered here, depending on the treatment of the background (see Fig.~9 of Ref.~\cite{Balaji:2025alr}).

\subsubsection{Fireball}
\label{sec:SNfireball}

The fireball model was originally proposed to explain the characteristics of $\gamma$-ray burst (GRB), intense flashes of energetic photons peaking in the sub-${\rm MeV}$ range~\cite{Piran:1999kx}. In this model, a compact source drives a relativistic outflow that carries energy either through the kinetic energy of ultra-relativistic particles or through electromagnetic Poynting flux. The initially optically thick plasma surrounding the source is referred to as a \emph{fireball}. As the outflow expands and becomes optically thin, its internal energy is released as $\gamma$-rays. (See Ref.~\cite{Piran:1999kx} and references therein for a comprehensive review.)

Novel particles produced in core-collapse supernovae (SNe) and decaying electromagnetically—such as the dark photons (DPs) considered here—can act as an ``engine'' for forming a fireball around the progenitor star~\cite{Kazanas:2014mca}; see Refs.~\cite{Diamond:2023scc,Diamond:2023cto} for an analogous study in the axion-like particle scenario. Following the approach of Ref.~\cite{Diamond:2023scc}, we assess whether DPs could trigger fireball formation.

When DPs decay into electron--positron pairs, $\gamma' \rightarrow e^+ e^-$, they inject a dense population of charged particles just outside the stellar surface. 
The characteristic radius at which most decays occur is $R_{\rm FB} \simeq \frac{v_{\gamma'}}{\Gamma_{e^+ e^-}}$, where $\Gamma_{e^+ e^-}$ is the DP decay rate to an $e^+e^-$ pair, 
and the spread in the decay radii defines the shell thickness $\Delta$, which typically satisfies $\Delta \ll R_{\rm FB}$. The number density of electrons (and positrons) then reads
\bea
n_{e^\pm} \approx \frac{\mathcal{N}_{\gamma^\prime}{ e^{-R_*/R_{\rm FB}}}}{4\pi R_{\text{FB}}^2 \Delta} \,,
\label{eq:FBne}
\eea
where $\mathcal{N}_{\gamma^\prime}$ denotes the total number of DPs emitted from PNS.

\begin{figure}[t!]
\centering
\includegraphics[width=1.\linewidth]{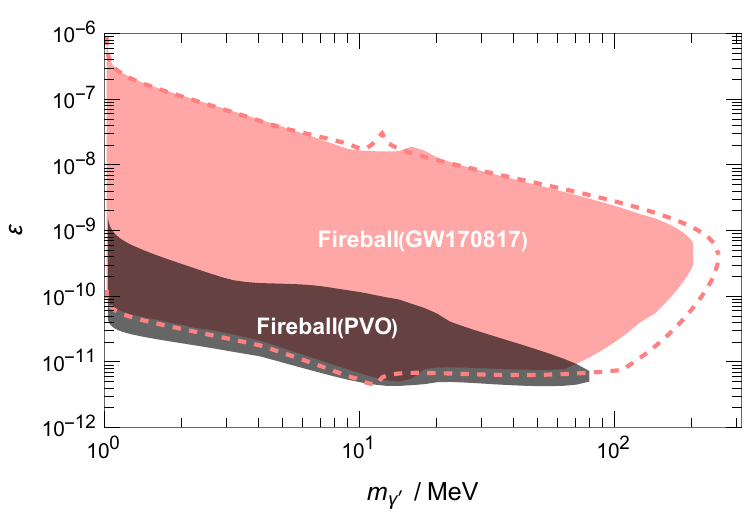}
 \caption{Fireball constraints on the dark photon scenario from both the \emph{Pioneer Venus Orbiter} at the time of SN~1987A (dark gray shaded region) and X-ray observations of GW170817 by CALET/CGBM, Konus-Wind, and Insight-HXMT/HE (pink region for the LS220-s18.88
SN profile and pink dashed curve for the one-zone model).}
\label{fig:DPSNfireball}
\end{figure}

A fireball can form only if the electron--positron plasma is sufficiently dense to maintain thermal equilibrium and chemical equilibrium. 
The first requirement is that the optical depth to pair annihilation, $e^- e^+ \to \gamma\gamma$, be large:
\bea
n_{e^\pm}\,\sigma_{e^+e^-\to\gamma\gamma}\,\Delta \gg 1\,,
\label{eq:FBoptdepth}
\eea
where $\sigma_{e^+e^-\to\gamma\gamma}$ is the annihilation cross section. 
In the comoving frame of the expanding plasma, the center-of-mass energy of the $e^\pm$ pairs is set by the DP mass, yielding the approximation
$\sigma_{e^+e^-\to\gamma\gamma} \simeq 4\pi \alpha^2/m_{\gamma'}^2$~\cite{Diamond:2021ekg}. A second condition ensures that number-changing processes such as bremsstrahlung, $e^- e^+ \to e^- e^+ \gamma$, are efficient:
\bea
2n_{e^\pm}\,\sigma_{ee\to ee\gamma}\,\Delta \gg 1\,,
\label{eq:FBnumchange}
\eea
with $\sigma_{ee\to ee\gamma} = 8\alpha^2[\log(m_{\gamma'}/3m_e)+\gamma_{\rm E}+5/4]/m_e^2$~\cite{Diamond:2021ekg}, assuming an initial plasma temperature of order $m_{\gamma'}$.
Substituting Eq.~\eqref{eq:FBne} into Eqs.~\eqref{eq:FBoptdepth} and \eqref{eq:FBnumchange} shows that both conditions are independent of $\Delta$. The black-shaded region labeled ``Fireball (PVO)'' in Figs.~\ref{fig:moneyplot} and~\ref{fig:DPSNfireball} indicates the portion of the DP parameter space in which either of the above conditions is satisfied and a fireball can form around the SN progenitor.

Fireball formation has important implications for interpreting several observational constraints on DPs~\cite{DeRocco:2019njg}. In deriving bounds from the SN~1987A $\gamma$-ray search by the GRS, we assumed that photons from final-state radiation escape freely and retain their original spectrum.  If a fireball forms, however, this radiation is trapped and thermalized, and is instead released as a GRB-like signal in the sub-MeV range—outside the GRS sensitivity window. 
Similarly, fireball formation affects the Galactic positron-injection constraints from SN DPs: if bremsstrahlung efficiently increases the particle number [Eq.~\eqref{eq:FBnumchange}], the plasma cools while maintaining energy conservation, and most positrons annihilate once the fireball becomes optically thin at $T\ll{\rm MeV}$~\cite{Piran:1999kx,DeRocco:2019njg,Diamond:2023scc,Diamond:2023cto}. As a result, the SN~1987A $\gamma$-ray and Galactic positron bounds do not apply within the black-shaded ``fireball'' region in Figs.~\ref{fig:moneyplot} and~\ref{fig:DPSNfireball}.

Beyond evading other constraints, the thermalized emission from a fireball can itself provide a probe of new particles. At the time of SN~1987A, the \emph{Pioneer Venus Orbiter} (PVO) was sensitive to photons in the $0.2$--$2\,{\rm MeV}$ range—consistent with the expected thermalized fireball spectrum. 
The non-observation of a $\gamma$-ray excess by PVO constrains the total fireball energy to 
$E_{\rm FB}^{\rm SN1987A} \lesssim 3\times10^{43}\,{\rm erg}$~\cite{Diamond:2023scc}. 
In the dark gray region of Figs.~\ref{fig:moneyplot} and~\ref{fig:DPSNfireball}, DPs would deposit more energy than this limit, leading to an exclusion.

A similar mechanism can occur in binary neutron-star mergers if a long-lived remnant forms before collapsing into a black hole.
The event GW170817 likely produced a hypermassive neutron star surviving for $\sim1\,{\rm s}$~\cite{LIGOScientific:2017vwq, Murguia-Berthier:2020tfs}. To estimate DP emission in this context, we consider two representative thermal profiles:
(i) the LS220-s18.88 SN profile at $t_{\rm pb}=8\,{\rm s}$, and 
(ii) a simplified one-zone model with $T=18\,{\rm MeV}$, $R=16\,{\rm km}$, and duration $\delta t=1\,{\rm s}$ following Ref.~\cite{Diamond:2023cto}. 
For both, we evaluate the thermalization criteria in Eqs.~\eqref{eq:FBoptdepth} and~\eqref{eq:FBnumchange}, adopting Eq.~\eqref{eq:FBne} for $n_{e^\pm}$ and requiring decays occur beyond $R_{\rm min}=1000\,{\rm km}$. 
X-ray observations of GW170817 by CALET/CGBM, Konus-Wind, and Insight-HXMT/HE place an upper limit of $E_{\rm FB}^{\rm GW170817} \lesssim 3\times10^{46}\,{\rm erg}$ on any GRB-like fireball emission. 
The corresponding exclusion based on the LS220-s18.88 profile is shown as the pink-shaded region in Figs.~\ref{fig:moneyplot} and~\ref{fig:DPSNfireball}, while the pink dashed curve denotes the result for the one-zone model.

\subsubsection{Diffuse $\gamma$-ray background}
\label{sec:SNdiffGamma}

CCSNe have occurred isotropically throughout cosmic history, and the emission of DPs from each event cumulatively contributes to a cosmological DP population.
The contribution to the comoving DP density per redshift interval is given by 
\bea
\frac{dn_{\gamma^\prime}}{dz} = \frac{d n_{\rm cc}}{dz} \int d \omega \frac{d\mathcal{N}_{\gamma^\prime}}{d\omega}\,,
\eea
where $dn_{\rm cc}/dz$ represents the CCSNe event rate per comoving volume. Here we ignore any attenuation from the intergalactic medium, since the optical depth at these energies and low redshift is negligible. The subsequent radiative decay of DPs tones up the diffuse cosmic $\gamma$-ray background, which should not exceed the measured value~\cite{Hill:2018trh}. In the following we consider only DPs with masses below $2\,m_e$ and their direct radiative decay into photons. For larger masses, the DPs can decay into electron-positron pairs, with associated final state radiation. The associated bounds in this mass region, however, are already excluded by other probes and we will not include them.

For the masses of interest here (smaller than $2\,m_e$) DPs remain relativistic at the time of decay, and the resulting diffuse photon flux can be written as 
\bea
\frac{d \Phi_{\gamma}}{dE_\gamma} & \simeq & \frac{c}{4\pi} \int_0^\infty dz \frac{d n_{\rm cc}}{dz}  \int_{E_\gamma(1+z)}^\infty d \omega \frac{d\mathcal{N}_{\gamma^\prime}}{d\omega} \left(\frac{1+z}{\omega}\right)\nonumber\\
&&\times \left(1 - \exp\left[-\int_0^z dz^\prime \Gamma_{\gamma^\prime}\left(z^\prime\right)\left|\frac{dt}{dz^\prime}\right|\right]  \right)\left(\frac{\Gamma_{\rm rad}}{\Gamma_{\gamma^\prime}}\right)    \nonumber\\
&&\times \left. \frac{dN_\gamma}{dx}\right|_{x=E_\gamma(1+z)/\omega} \,.
\eea
where $\Gamma_{\gamma^\prime}(z^\prime)$ is the total DP decay rate evaluated at the redshifted energy $\omega^\prime = \omega (1+z^\prime)/(1+z)$, and $\Gamma_{\rm rad}$ is the partial decay rate for the radiative channel.
The Hubble time is expressed as $|dt/dz| = 1/(1+z)H$ with the Hubble rate $H$
\bea
H\left(z\right) = H_0 \sqrt{\Omega_{\rm m} \left(1+z\right)^3 +\Omega_\Lambda}\,,
\label{eq:HubbleRate}
\eea
where $H_0$ is the present Hubble constant, and $\Omega_{\rm m}$ and $\Omega_\lambda$ are the matter and dark energy densities, respectively.
We neglect photon absorption during propagation, as the core-collapse SN rate $dn_{\rm cc}/dz$ is concentrated at relatively low redshifts ($z \lesssim 3$).
The last term $dN_\gamma/dx$ represents the photon energy spectrum per radiative decay of a boosted DP with the initially produced energy $\omega$ in terms of the covariant ratio of $x=E_\gamma(1+z)/\omega \in [0,1]$.
The explicit derivation of $dN_\gamma/dx$ from the given energy spectrum in the rest frame of the DP, denoted by $dN_{\gamma}/d\hat{x}$ with $\hat{x} = E/(m_{\gamma^\prime}/2)$ for the photon energy $E$ in the rest frame of the DP, is provided in Appendix.~\ref{app:spectrumBoost}.
In the highly boosted limit (i.e., $\omega \gg m_{\gamma^\prime}$), the boosted spectrum simplifies to
\bea
\frac{dN_\gamma}{dx} \simeq \int^1_{x} \frac{d\hat{x}}{\hat{x}}\frac{dN_\gamma}{d\hat{x}}\,.
\eea

For masses below ${\rm MeV}$, DPs are dominantly produced by resonant conversion of longitudinal plasmons at the energy around the plasma frequency, approximately $20\,{\rm MeV}$, which results in DPs being relativistic during decay.
Therefore, the description above is relevant with $d\mathcal{N}_{\gamma^\prime}/d\omega \approx \mathcal{N}_{\gamma^\prime}\delta(\omega -\omega_R)$.

In the highly boosted limit, the spectrum in the lab frame becomes
\bea
\frac{d N_\gamma}{dx} =  \frac{5239 - 34300 x^3 + 46575 x^4-17514 x^5}{1020} \, .
\eea
The factor of $3$ accounts for the total number of photons per the decay.
Substituting into the general expression yields the diffuse photon flux from SN-originated DPs
\bea
\frac{d \Phi_{\gamma}}{dE_\gamma} & \simeq & \frac{c}{4\pi} \int_0^\infty dz \frac{d n_{\rm cc}}{dz}  \mathcal{N}_{\gamma^\prime}\Theta\left(\frac{\omega_R}{1+z} - E_\gamma\right) \left(\frac{1+z}{\omega_R}\right)\nonumber\\
&&\times \left(1 - \exp\left[-\int_0^z dz^\prime \frac{\hat{\Gamma}_{3\gamma}}{\omega^\prime/m_{\gamma^\prime}}\left|\frac{dt}{dz^\prime}\right|\right]  \right)    \nonumber\\
&&\times \left. \frac{dN_\gamma}{d\hat{x}}\right|_{\hat{x}=E_\gamma(1+z)/\omega_R} \,,
\label{eq:SNDPdiffGamma}
\eea
where $\omega_R$ denotes the approximately monochromatic DP energy from resonant production.

\begin{figure}[t!]
\centering
\includegraphics[width=1.\linewidth]{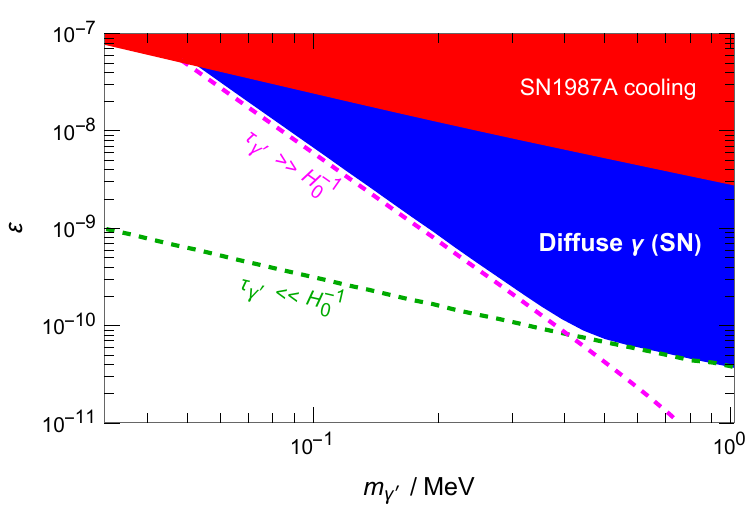}
 \caption{The blue-shaded region is excluded by the extragalactic $\gamma$-ray background, contributed by radiative decay of dark photons from cosmic core-collapse supernovae. The magenta- and green-dashed lines represent approximate analytical limits corresponding to lifetimes much longer and much shorter than the age of the universe, respectively.}
\label{fig:DPSNdiff}
\end{figure}

The blue-shaded region in Fig.~\ref{fig:DPSNdiff} (and the corresponding `Diffuse $\gamma$ (SN)' region in Fig.~\ref{fig:moneyplot}) illustrates the constraint derived from the diffuse $\gamma$-ray background.
The cosmic core-collapse rate is related to the comoving star formation rate $\dot{\rho}_* (z)$, through $dn_{\rm cc} /dz = \dot{\rho}_* (z) k_{\rm cc}$ with the coefficient $k_{\rm cc}$ for core-collapse ratio per mass.
For analysis, we adopt the star formation rate from Ref.~\cite{Madau:2014bja} and take $k_{\rm cc} = (143 M_\odot)^{-1}$~\cite{Beacom:2010kk}, which yields the most conservative (i.e., lowest) estimate of the cosmic core-collapse rate among those discussed in Ref.~\cite{Caputo:2021rux}

The expression in Eq.~\eqref{eq:SNDPdiffGamma} can be further simplified in terms of the DP lifetime, following the approach for axions in Ref.~\cite{Caputo:2021rux}.
For short-lived DPs ($\Gamma_{\gamma^\prime} \gg H_0$), most decay shortly after production, and the exponential suppression can be neglected.
Since the cosmic SN rate peaks near $z \approx 1$, we approximate
\bea
\left[\frac{d \Phi_{\gamma}}{dE_\gamma}\right]_{\Gamma_{\gamma^\prime} \gg H_0} & \approx & \frac{c}{4\pi} \left[\frac{d n_{\rm cc}}{dz} \right]_{z=1}\mathcal{N}_{\gamma^\prime}\Theta\left(\frac{\omega_R}{2} - E_\gamma\right) \left(\frac{2}{\omega_R}\right)\nonumber\\
&&\times \left. \frac{dN_\gamma}{d\hat{x}}\right|_{\hat{x}=2 E_\gamma/\omega_R} \,.
\label{eq:SNDPdiffshort}
\eea
Considering the approximately measured diffuse photon flux in the $\gamma$-ray spectrum range as $E_\gamma^2(d\Phi_\gamma/dE_\gamma) \simeq 2\times 10^{-3}\,{\rm MeV}\,{\rm cm}^{-2}\,{\rm s}^{-1}\,{\rm ster}^{-1}$, the dashed green line in Fig.~\ref{fig:DPSNdiff} shows the resulting limit in the short-lifetime regime, consistent with the full numerical calculation for $m_{\gamma^\prime} \gtrsim 0.5,\mathrm{MeV}$, scaling as $\varepsilon \propto m_{\gamma^\prime}^{-1}$.

In the long-lived regime ($\Gamma_{\gamma^\prime} \ll H_0$), most DPs survive until late times, and the photon flux is instead approximated as
\bea
\left[\frac{d \Phi_{\gamma}}{dE_\gamma}\right]_{\Gamma_{\gamma^\prime} \ll H_0}  & \approx & \frac{c}{4\pi} \int_0^\infty dz \frac{d n_{\rm cc}}{dz}  \mathcal{N}_{\gamma^\prime}\Theta\left(\frac{\omega_R}{1+z} - E_\gamma\right) \nonumber\\
&&\times \left(\frac{1+z}{\omega_R}\right)\int_0^z dz^\prime \frac{\hat{\Gamma}_{3\gamma}}{\omega^\prime/m_{\gamma^\prime}}\left|\frac{dt}{dz^\prime}\right|  \nonumber\\
&&\times \left. \frac{dN_\gamma}{d\hat{x}}\right|_{\hat{x}=E_\gamma(1+z)/\omega_R} \,.
\label{eq:SNDPdifflong}
\eea
The corresponding estimate is shown by the dashed magenta curve in Fig.~\ref{fig:DPSNdiff}, which exhibits a scaling $\varepsilon \propto m_{\gamma^\prime}^{-3}$.

In principle, for $m_{\gamma^\prime} > 2m_e$ one could also consider constraints from photons produced via FSR in the decay $\gamma' \to e^- e^+ \gamma$, similarly to the SN1987A gamma-ray analyses. In the diffuse case, however, the corresponding limits are already ruled out by either SN1987A observations or by the 511\,keV line constraints, so we do not show them.

\subsubsection{Can dark photons prevent SN explosions?}
\label{sec:preventSNex}

Traditional SN1987A DP bounds come from cooling arguments: too much energy loss from the PNS via DP emission would shorten the observed neutrino burst. Moreover, for most cases the emission of new particles is computed around $\sim 1 \, \rm s$ after bounce and it comes from the central core, where the PNS is hot and dense enough, and the typical emission of particles such as the QCD axion is at its maximum. However, DP energy loss is rather peculiar due to the DP resonant production; thus cooling can be largest in the gain layer below the stalled shock wave and counter-act the usual shock rejuvenation by neutrino energy deposition and thus prevent the explosion~\cite{Caputo:2025aac}. In fact, in the Bethe--Wilson framework~\cite{Bethe:1985sox} of CCSN, a \emph{gain region} exists between the gain radius and the stalled shock front, $r_{\rm gain} < r < r_{\rm shock}$, where neutrino absorption dominates over emission and deposits energy that helps re-energize the shock. DPs, in contrast, provide an additional energy-loss channel that drains energy from the same region, potentially suppressing shock revival if the cooling is sufficiently strong. In this regime, arguments based on subsequent PNS cooling become irrelevant: the mere requirement that the explosion occurs already leads to more restrictive and robust limits, independent of the SN~1987A neutrino signal.

To investigate the impact of this cooling channel, one can compute the differential DP luminosity  ($d\mathcal{P}_{\gamma^\prime}/dr$)  as a function of radius and compared it with neutrino heating and cooling at various times before shock revival~\cite{Janka:2000bt}. 

The net neutrino heating rate per unit volume in the gain region can be approximated as~\cite{Janka:2000bt}
\begin{eqnarray}
Q_{\nu}^{+} &\approx& 
\frac{160~{\rm MeV}}{\rm s}\,
\frac{\rho}{m_u}\,
\frac{L_{\nu_e,52}}{r_7^2 \langle \mu_{\nu} \rangle}
\left( \frac{T_{\nu_e}}{4~{\rm MeV}} \right)^2 ,
\label{eq:heat}
\end{eqnarray}
where $\rho$ is the matter density, $m_u$ the atomic mass unit, $L_{\nu_e,52}$ the $\nu_e$ luminosity in units of $10^{52}\,{\rm erg\,s^{-1}}$, $r_7$ the radius in $10^7\,{\rm cm}$, and $\langle \mu_{\nu} \rangle$ the neutrino flux factor approaching unity in the free-streaming limit. 
This estimate assumes equal luminosities and spectral shapes for $\nu_e$ and $\bar{\nu}_e$ ($L_{\nu_e}\!\simeq\!L_{\bar{\nu}_e}$), Fermi–Dirac spectra without degeneracy ($\langle \epsilon^2_{\nu_e}\rangle\!\approx\!21T_{\nu_e}^2$), and typical composition $Y_n\!+\!2Y_p\simeq1$. The temperature $T_{\nu_e}$ is defined at the neutrinosphere, the energy-sphere—where neutrinos decouple energetically from the background. 

Along the same lines, one can derive the corresponding neutrino cooling rate per unit volume~\cite{Janka:2000bt}
\begin{eqnarray}
Q_{\nu}^{-} &\approx& 
\frac{145~{\rm MeV}}{\rm s}\,
\frac{\rho}{m_u}
\left( \frac{T}{2~{\rm MeV}} \right)^6 ,
\label{eq:cool}
\end{eqnarray}
where $T$ is the local matter temperature and $\rho$ the density in the gain region. 
This expression follows from
\begin{eqnarray}
Q_{\nu}^{-} &=&
\frac{\sigma_0\,T^6 \rho (3\alpha^2+1)}{8\pi^2 m_e^2m_u}
\left[Y_p\mathcal{F}_5(\eta_e)+Y_n\mathcal{F}_5(-\eta_e)\right]
\label{Eq:CoolingUsed}
\end{eqnarray}
with $\mathcal{F}_5$ the relativistic Fermi integral (see Eq.~(32) in Ref.~\cite{Janka:2000bt}). 
In the second line of Eq.~\eqref{eq:cool} we assumed $Y_n + Y_p \simeq 1$ and low electron degeneracy, appropriate for shock-heated layers rich in $e^\pm$ pairs. 
Unlike the heating rate, which scales as $T_{\nu_e}^2$ with the neutrino temperature at the neutrinosphere, the cooling rate depends on the sixth power of the \emph{local} temperature. 
The gain radius, $r_{\rm gain}$, is defined by the balance condition $Q_{\nu}^{+}=Q_{\nu}^{-}$, yielding~\cite{Janka:2000bt}
\begin{equation}
r_{{\rm gain},7}
\left( \frac{T_g}{2~{\rm MeV}} \right)^3
\simeq
1.05
\sqrt{\frac{L_{\nu_e,52}}{\langle \mu_{\nu} \rangle_{\rm g}}}
\left( \frac{T_{\nu_e}}{4~{\rm MeV}} \right),
\end{equation}
where $T_g \equiv T(r_{\rm gain})$ is the local temperature at the gain radius.

\begin{figure}[t!]
\centering
\includegraphics[width=1.\linewidth]{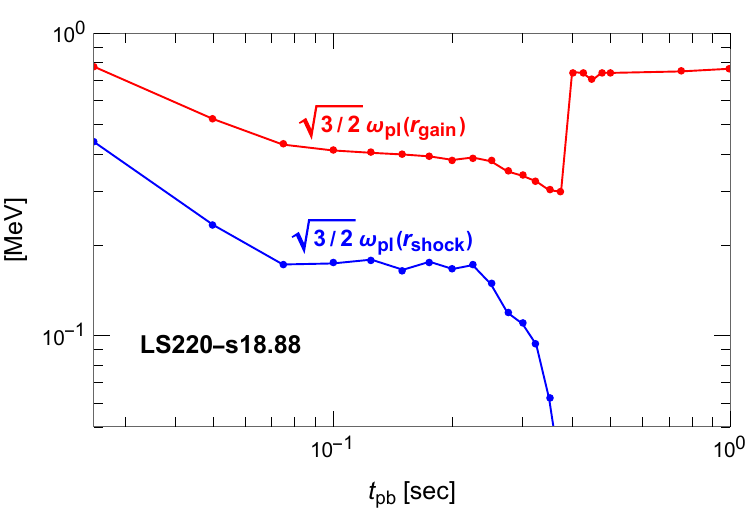}\\
\includegraphics[width=1.\linewidth]{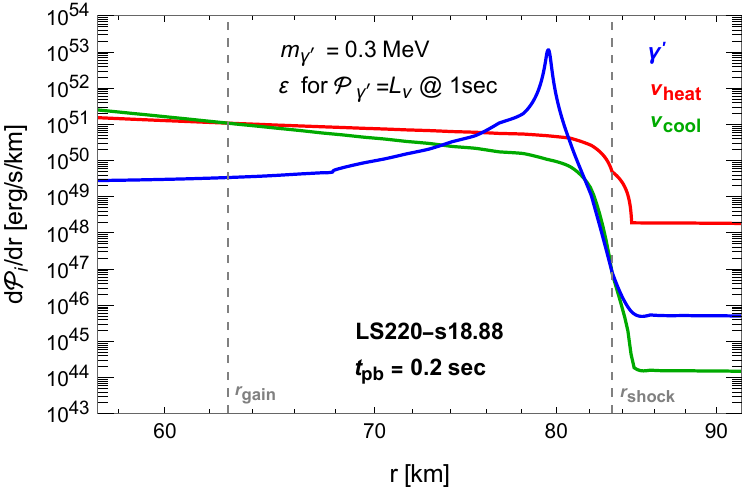}
 \caption{Upper panel: time evolution of the plasma frequency at the gain radius (red) and at the shock radius (blue), multiplied by a factor of $\sqrt{3/2}$. Lower panel: differential luminosity profiles as a function of radius at $0.2\,{\rm s}$ post-bounce for dark photons (blue), neutrino heating (red), and neutrino cooling (green).
For the dark photon emission, we adopt $m_{\gamma^\prime} = 0.3\,{\rm MeV}$ and the corresponding kinetic mixing at the supernova cooling limit.}
\label{fig:SNexpPrev}
\end{figure}

Fig.~\ref{fig:SNexpPrev} illustrates the time evolution of the plasma frequency at the gain and the shock radii (upper panel) and the corresponding radial profiles of differential luminosities at $0.2\,\mathrm{sec}$ post-bounce for the supernova explosion model LS220-s18.88; the DP parameters are fixed at $m_{\gamma^\prime} =0.3\,{\rm MeV}$ and $\varepsilon  \sim 10^{-8}$ corresponding to the supernova cooling limit.
We compute the DP emissivity and compare it with the neutrino cooling and heating rates. A resonance occurs within the gain region at $r \sim 80\,\mathrm{km}$, where DP cooling largely exceeds the effective neutrino heating. In such a case, a successful supernova explosion within the Bethe–Wilson framework can be prevented.
The resulting excluded parameter space for $0.1 \, \text{MeV} \lesssim m_{\gamma'} \lesssim 0.5 \, \rm MeV$ is shown in Fig.~\ref{fig:moneyplot} (``SN Failure"), although most of it is already ruled out by cosmological probes. More interesting regions of the parameter space could be probed studying the impact on the explosion for larger DP masses, $m_{\gamma'} \gtrsim 0.5 \, \rm MeV$, for which resonant production of T modes occurs in deeper regions below $r_{\rm gain}$, where it could alter the diffusive neutrino flux and affect the explosion mechanism. Such a case, however, deserves self-consistent numerical simulations, which we plan to perform in the future. Further details can be found in Ref.~\cite{Caputo:2025aac}.



\section{Constraints from colliders}

Although this is not the focus of our work, we find it useful to include for comparison also bounds from collider probes, given that DPs can affect both precision observables and direct searches at accelerators. 

One of the earliest probes arises from the anomalous magnetic moments of the electron and muon~\cite{Pospelov:2008zw}, 
since a kinetically mixed dark photon induces loop corrections proportional to \(\alpha \varepsilon^2 / 2\pi\). For $m_{\gamma'} > 2m_e$, dark photons decay to charged fermions and depending on parameters, the decay can be prompt or displaced, requiring complementary search strategies. \textit{Electron fixed-target and beam-dump experiments} are sensitive to dark photons in the MeV--GeV range~\cite{Bjorken:2009mm,Essig:2010xa,HPS:2018xkw,NA64:2018lsq}. Fixed-target setups search for narrow resonances in the invariant-mass distribution of 
\(e^+e^-\) pairs produced in \(eZ\to eZ A'\) reactions, while thick-target beam-dump experiments (e.g., E137, E141, E774~\cite{Bjorken:1988as,Riordan:1987aw}) probe smaller mixings by exploiting long-lived dark photons decaying downstream of the dump. \textit{Proton-beam dumps} also produce dark photons via meson decays 
such as \(\pi^0,\eta \to \gamma A'\)~\cite{Batell:2009di,Essig:2010gu},providing sensitivity to similar parameter regions. \textit{Electron--positron colliders} 
(e.g., BaBar, Belle, Belle-II, KLOE) search for \(e^+e^- \to \gamma A'\) followed by 
\(A' \to f^+f^-\), using bump-hunt techniques in the invariant-mass spectrum~\cite{BaBar:2014zli}. 
The production cross section scales as \(\sigma_{\gamma A'} \propto \varepsilon^2 / E_{\rm beam}^2\), 
making low-energy colliders particularly competitive for sub-GeV dark photon masses. At higher energies, \textit{proton--proton colliders} such as the LHC can probe heavier 
dark photons via Drell--Yan production, prompt low-mass dimuon resonances~\cite{CMS:2023hwl, CMS:2021sch} or (more model dependent) exotic Higgs decays~\cite{Curtin:2014cca, ATLAS:2023cjw}, while LHCb extends coverage to smaller masses through displaced-vertex searches 
in meson decays~\cite{Ilten:2015hya,Ilten:2016tkc}. \textit{Rare meson decays} such as 
\(\pi^0,\eta,K,\phi \to \gamma A'\)~\cite{Reece:2009un} 
provide additional, though currently subleading, constraints. Finally, dark sectors may contain a dark Higgs boson (\(h'\)) enabling 
associated production (\(e^+e^- \to A'h'\)), 
leading to distinctive multi-lepton or displaced signatures studied at 
BaBar~\cite{BaBar:2014zli}, Belle~\cite{Jaegle:2015sxl}, KLOE-2~\cite{KLOE-2:2015nli}, and Belle-II~\cite{Belle-II:2022jyy}. 

In Fig.~\ref{fig:moneyplot}, the grey-shaded region in  shows the compilation of constraints obtained using the \texttt{DarkCast} tool~\cite{DarkCast, Baruch:2022esd, Ilten:2018crw}; for the masses and couplings of interest for this work, the relevant bounds come from beam-dump experiments.



\section{Discussions and Conclusion}
\label{sec:conclusion}

In this work, we have re-examined the \textit{irreducible} constraints on DPs. 
On the cosmological side, we fixed the reheating temperature of the Universe to $T_{\rm RH} = 6~{\rm MeV}$, which corresponds to the lowest value still compatible with Big Bang Nucleosynthesis according to the most recent evaluation of light-element abundances and $N_{\rm eff}$~\cite{Barbieri:2025moq}. 
Within this cosmological setup, we consistently derived all related observables, including diffuse and galactic gamma-ray emission, CMB spectral distortions, variations of $N_{\rm eff}$, and the impact on primordial element abundances. 
After addressing these cosmological implications, we turned to the production of DPs in core-collapse supernova (CCSN) environments and derived the associated constraints, such as those from neutrino cooling, prompt and diffuse gamma-ray emission, fireball formation, galactic positron injection, low-energy supernova energetics, and the possible stalling of the SN shock wave. 
We also surveyed current collider limits, which provide a complementary probe to the cosmological and astrophysical bounds, particularly in the regime of large DP masses and sizable kinetic mixing. Taken together, these results delineate the most robust and model-independent exclusion region for dark photons over many orders of magnitude in mass, highlighting the complementarity between laboratory, astrophysical, and cosmological probes.

Finally, we recall that our constraints apply to the minimal DP model, and that any coupling to a dark sector~\cite{Chang:2018rso} may alter some of the bounds discussed. For example, if the DP can decay invisibly into dark sector particles, all limits from fireball formation, $\gamma$-rays, 511 keV line, LESNe must be rescaled. Similarly, cosmological bounds assume that production is driven by freeze-in via kinetic mixing with the Standard Model; however, the relic abundance can be significantly modified in the presence of a richer dark sector. We leave the study of these different scenarios for future work. 

\textit{Acknowledgement.} We thank Francesco D'Eramo for collaboration at different stages of this project and many useful discussions. We also thank Miguel Escudero Abenza, Damiano Fiorillo, Georg Raffelt, Yotam Soreq, Edoardo Vitagliano, and Meng-Ru Wu for useful comments on the draft. AC acknowledges the hospitality of the Weizmann Institute of Science and the support from the Benoziyo Endowment Fund for the Advancement of
Science. This project has
received funding from the European Research Council
(ERC) under the European Union’s Horizon Europe re-
search and innovation programme (grant agreement No.
101117510). Views and opinions expressed are however
those of the authors only and do not necessarily reflect those of the European Union. The European Union cannot be held responsible for them.
SY is supported by IBS under the project code, IBS-R018-D1.
SY acknowledges the hospitality of CERN during the visit. AC and SY acknowledge the support of the INFN Sezione di Roma for the Pollica Summer Workshop `The Strong CP Problem and Its Possible Solutions', where part of this work was carried out.

\appendix

\section{Properties of the SM plasma}
\label{app:plasma}

In this appendix, we summarize well-known results about the SM plasma. We focus on the scenario in which a maximal temperature is as low as $\mathcal{O} (1)\,{\rm MeV}$, as required to ensure the consistency of the standard cosmology; in this paper, we take $6\,{\rm MeV}$. Thus, the light degrees in the SM thermal bath contain photons, electrons, and neutrinos.
Although there exists a relatively small baryon asymmetry in the form of nucleons, its impact on our discussion is negligible due to its massiveness. Accordingly, the real parts of $\pi_{\rm T,L}$ are mainly contributed by the coherent scatterings with surrounding electrons/positrons and read~\cite{Altherr:1992jg,Braaten:1993jw}
\bea
{\rm Re}\,\pi_{\rm T} & = & \omega_{\rm pl}^2 \left[1+\frac{1}{2}G\left(v_*^2k^2/\omega^2\right)\right] \label{eq:RePiT}\,,\\
{\rm Re}\,\pi_{\rm L} & = & \omega_{\rm pl}^2 \frac{m_{\gamma^\prime}^2}{\omega^2}\frac{1-G\left(v_*^2k^2/\omega^2\right)}{1-v_*^2k^2/\omega^2} \,,
\label{eq:RePiL}
\eea
where $\omega$ and $k$ denote the energy and momentum of the external DP state, respectively.
The plasma frequency $\omega_{\rm pl}$ is given by
\bea
\omega_{\rm pl}^2 = \sum_{i=e^{\mp}} \frac{4\alpha}{\pi}\int_0^\infty dp \, f_i \, \frac{p^2}{E} \, \left(1- \frac{1}{3}v^2\right) \,,  
\eea
where $f_i$ denotes the phase space distribution of  $i$ with their energy $E$ and momentum $p$, $v = p/E$, and $v_*$ is the typical electron velocity given by
\bea
v_*^2 = \frac{\int_0^\infty dp \, f_p \, p \left(5v^3/3- v^5\right)}{\int_0^\infty dp \, f_p \, p \left(v- v^3/3\right)} \, .
\eea
The function G is defined as
\bea
G\left(x\right) = \frac{3}{x}\left[1-\frac{2x}{3} - \frac{1-x}{2\sqrt{x}}\log\frac{1+\sqrt{x}}{1-\sqrt{x}}\right] \, .
\eea

\section{Thermal production in the early universe}
\label{app:CosDPres}

In this Appendix, we provide details about resonant and non resonant production of DPs. 

\subsection{Resonant production}
\label{app:cosmicDPres}

We start providing the analytic derivation of the thermal production of DPs at the resonance point.
The following derivation is valid for narrow resonance width cases, which are typically satisfied.

\subsubsection*{Transverse modes}

For the transverse polarization, ${\rm Re}\,\pi_{\rm T}$ is of the order of $\omega_{\rm pl}^2$ across the entire momentum range.
Since the typical energy range of produced DPs is an order of temperature, which is larger than $\omega_{\rm pl}$, we can employ the relativistic approximation for the energy-momentum of DPs (i.e., $\omega \simeq k$) leading to
\bea
{\rm Re}\,\pi_{\rm T} \simeq \omega_{\rm pl}^ 2\left[ 1 + \frac{1}{2}G\left(v_*^2\right) \right] \, .
\label{eq:RePiTRes}
\eea
Although each momentum of transverse DPs may experience resonance at different epochs in principle, the time difference within the dominant energy range of temperature is shorter than the Hubble scale; taking momentum-independent ${\rm Re}\,\pi_{\rm T}$ in Eq.~\eqref{eq:RePiTRes} is a valid simplification.
Therefore, the resonance of the transverse polarization occurs approximately at the specific temperature $T_{\rm res}$.
Plugging Eq.~\eqref{eq:DPdifferentialrate} with ${\rm Re}\,\pi_{\rm T}$ in Eq.~\eqref{eq:RePiTRes} into the Boltzmann equation in Eq.~\eqref{eq:Boltzmann}, we can derive
\bea
\left[Y_{\gamma^\prime}\right]_{\rm T}^{\rm res} 
& \simeq & \frac{\varepsilon^2 m_{\gamma^\prime}^4}{\pi^2} \int dT \frac{1}{s T H}\left(1-\frac{1}{3}\frac{d\log g_{*s}}{d\log T^{-1}}\right) \nonumber\\
&& \times\frac{|{\rm Im}\,\pi_{\rm T}|}{\left(\partial \, {\rm Re}\,\pi_{\rm T}/\partial T \right)^2\left(T - T_{\rm res}\right)^2 + \left|{\rm Im}\,\pi_{\rm T}\right|^2} \nonumber\\
&& \times  \int d\omega \frac{\sqrt{\omega^2 - m_{\gamma^\prime}^2}}{e^{\omega/T} -1} \nonumber \\
& \simeq & \varepsilon^2 m_{\gamma^\prime}^2 \frac{\pi^3}{12\zeta(3)} \Big[\frac{n_{\gamma}}{s}\frac{\left(\partial \, {\rm Re}\,\pi_{\rm T}/\partial T \right)^{-1}}{ T/m_{\gamma^\prime}^2}\frac{1}{TH} \nonumber\\
&&\left(1-\frac{1}{3}\frac{d\log g_{*s}}{d\log T^{-1}}\right) \Big]_{T=T_{\rm res}} \, .
\label{eq:YDPTres}
\eea
Here, we approximate $(\left(\partial \, {\rm Re}\,\pi_{\rm T}/\partial T \right)^2\left(T - T_{\rm res}\right)^2 + \left|{\rm Im}\,\pi_{\rm T}\right|^2)^{-1}\simeq \delta (T-T_{\rm res}) \times \pi\left(\partial \, {\rm Re}\,\pi_{\rm T}/\partial T \right)^{-1}/|{\rm Im}\,\pi_{\rm T}|$, which is valid within the narrow resonance width condition ${\rm Im}\,\pi_{\rm T} \ll {\rm Re}\,\pi_{\rm T}$.
Moreover, we neglect $m_{\gamma^\prime}$ for integral over $\omega$ due to $\omega_{\rm pl} \ll T$, enabling it to be expressed by the photon equilibrium number density $n_{\gamma}$.



\subsubsection*{Longitudinal modes}

The resonance feature for the longitudinal polarization differs from the case of the transverse polarization.
According to Eq.~\eqref{eq:RePiL}, the real part of the longitudinal polarization can be approximated as
\bea
{\rm Re}\,\pi_{\rm L} \simeq m_{\gamma^\prime}^2 \frac{\tilde{\omega}^2}{\omega^2} \,,
\eea
where $\tilde{\omega}^2 = \mathcal{O}(\omega_{\rm pl}^2)$~\footnote{When electrons/positrons are relativistic (i.e., $T\gg m_e$), $\tilde{\omega}^2$ is an order of $\omega_{\rm pl}^2$ increased by a factor $\log[\omega^2/m_{\gamma^\prime}^2]$, which is never a large number for the cases of interest.}.
This indicates that the resonance for the longitudinal polarization, ${\rm Re}\,\pi_{\rm L} = m_{\gamma^\prime}^2$, occurs at the specific energy $\omega = \tilde{\omega}$ at any temperature, as long as $m_{\gamma^\prime} > \omega_{\rm pl}$.
Following the transverse case above, the contribution to the DP abundance from the resonance of the longitudinal polarization reads
\bea
\left[Y_{\gamma^\prime}\right]_{\rm L}^{\rm res} 
& \simeq & \frac{\varepsilon^2 m_{\gamma^\prime}^2}{2\pi^2} \int dT \frac{1}{s T H}\left(1-\frac{1}{3}\frac{d\log g_{*s}}{d\log T^{-1}}\right)  \nonumber\\
&& \times  \int d\omega \frac{\sqrt{\omega^2 - m_{\gamma^\prime}^2}}{e^{\omega/T} -1} \nonumber\\
&&\frac{|{\rm Im}\,\pi_{\rm L}/m_{\gamma^\prime}^2|}{\left(2/\tilde{\omega}\right)^2\left(\omega - \tilde{\omega}\right)^2 + \left|{\rm Im}\,\pi_{\rm L}/m_{\gamma^\prime}^2\right|^2} \nonumber\\
&\simeq & \frac{\varepsilon^2 m_{\gamma^\prime}^2}{2\pi} \int dT \frac{1}{s T H}\left(1-\frac{1}{3}\frac{d\log g_{*s}}{d\log T^{-1}}\right) \nonumber\\
&&\times \frac{\tilde{\omega}\sqrt{\tilde{\omega}^2-m_{\gamma^\prime}^2}}{2 \left(e^{\tilde{\omega}/T}-1\right)} \, ,
\label{eq:YDPLres}
\eea
where we approximate $\left|{\rm Im}\,\pi_{\rm L}/m_{\gamma^\prime}^2\right|/(\left(2/\tilde{\omega}\right)^2\left(\omega - \tilde{\omega}\right)^2 + \left|{\rm Im}\,\pi_{\rm L}/m_{\gamma^\prime}^2\right|^2) \simeq \delta (\omega-\tilde{\omega}) \times \pi\left(\tilde{\omega}/2 \right)$.

\begin{figure}[t!]
\centering
\includegraphics[width=.8\linewidth]{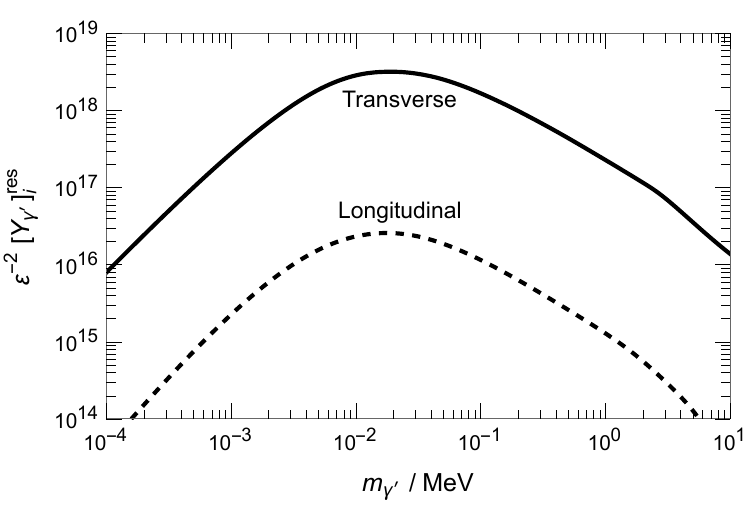}
 \caption{The freeze-in dark photon abundance partially contributed from the resonance, $[Y_{\gamma^\prime}]_i^{\rm res}$, for $i$ polarizaton.}
 \label{fig:Yres}
\end{figure}

Fig.~\ref{fig:Yres} illustrates $[Y_{\gamma^\prime}]_{\rm T,L}^{\rm res}$ for the transverse (solid) and longitudinal (dashed) polarizations.

\subsection{Non-resonant production}

\label{app:crosssection}

We provide the DP production rate for the following processes: $e^{\mp} + \gamma \rightarrow e^{\mp} + \gamma^\prime$ (Compton-like scatterings), $e^- + e^+ \rightarrow \gamma + \gamma^\prime$ (annihilation), and $e^- + e^+ \rightarrow \gamma^\prime$ (coalescence).
The following expressions and derivations can also be straightforwardly generalized to other charged particles as scattering targets.

In the following derivations, we neglect plasma effects, as they are irrelevant under the conditions where these processes become important.
Although this assumption is strictly valid only for temperatures satisfying $\omega_{\rm pl} < m_{\gamma^\prime}$, the IR-dominated nature of DP production renders this approximation reliable for our purposes (see the discussion in the main text)
In this limit, plasma effects can be ignored, and the polarization sum simplifies to
\bea
\sum_{\rm T,L} \epsilon^\mu_{\gamma^\prime}\epsilon^{* \nu}_{\gamma^\prime}  \rightarrow  -g^{\mu\nu} + \frac{k^\mu k^\nu}{m_{\gamma^\prime}^2}\, ,
\eea
where the contribution of the last term vanishes due to electromagnetic current conservation.

For a generic $2\to2$ process, $1 + 2 \rightarrow 3 + \gamma^\prime$, the DP production rate is given by
\beq
\begin{split}
\gamma_{\gamma^\prime} = &  \int 
\prod_{i=1,2,3} \frac{g_i d^3 \vec{p}_i}{\left(2\pi\right)^3 2 E_i}
\frac{g_{\gamma^\prime}d^3 \vec{k}}{\left(2\pi\right)^3 2 \omega} f_1f_2 \left(1\pm f_3\right) 
\\
& \times  \left| \mathcal{M}\right|^2 \, \left(2\pi\right)^4 \delta^{4}\left(p_1 + p_2 - p_3 - p_{\gamma^\prime}\right) \,,
\end{split}
\label{eq:22scattering}
\eeq
where $p_i = (E_i,\vec{p}_i)$, $f_i$ is the distribution function, and $g_i$ denotes the internal degrees of freedom of species $i$.
Here, $p_{\gamma^\prime} = (\omega, \vec{k})$ is the four-momentum of the DP, and $\mathcal{M}$ is the corresponding matrix element
The $\pm$ sign accounts for Bose enhancement or Pauli blocking in the final state, though these effects are negligible in the analysis below. 

Since plasma corrections are neglected, the phase-space integration can be evaluated in the center-of-mass (CM) frame following the standard method in Ref.~\cite{Gondolo:1990dk}.
Assuming the Boltzmann statistics, $f_i = e^{-E_i/T}$, the DP production rate can be written by
\bea
\gamma_{\gamma^\prime} & = & \frac{g_1g_2}{32\pi^4}T \int_{s_{\rm min}}^\infty ds \frac{\lambda\left(s\,,m_1\,,m_2\right)}{\sqrt{s}} \, K_1\left(\frac{\sqrt{s}}{T}\right) \nonumber \\
&&\qquad \qquad \times \sigma_{12\rightarrow 3\gamma^\prime}\left(s\right) 
\eea
where $\sigma_{12\rightarrow 3\gamma^\prime}(s)$ is the corresponding cross-section averaged over the angular dependence in terms of the squared center-of-mass energy $s$ with the integral range of $s_{\rm min} = {\rm Max}[(m_1+m_2)^2,(m_3+m_{\gamma^\prime})^2]$, and the K$\ddot{a}$ll\'{e}n function $\lambda$ is defined as
\bea
\lambda\left(x\,,y\,,z\right) = \left(x-(y+z)^2\right)\left(x-(y-z)^2\right) \;.
\eea
The corresponding differential cross section reads
\bea
\frac{d\sigma_{12\rightarrow 3\gamma^\prime}}{d\Omega} = \frac{1}{64g_ig_j\pi^2 s} \sqrt{\frac{\lambda(s,m_3,m_{\gamma^\prime})}{\lambda(s,m_1,m_2)}} \left| \mathcal{M} \right|^2 \, .
\eea



For the Compton-like scattering process $e^\mp + \gamma \rightarrow e^\mp + \gamma^\prime$, the cross section is given by
\bea
\sigma_{\rm sC} & = & \frac{\varepsilon^2 e^4}{16\pi s} \frac{\sqrt{\lambda(s,m_e,m_{\gamma^\prime})}}{s\left(s-m_e^2\right)^3} \nonumber\\
&& \times \Bigg(s^3 + \left(15 m_e^2 +7 m_{\gamma^\prime}^2\right) s^2 + m_e^2\left( 2m_{\gamma^\prime}^2- m_e^2 \right) s \nonumber\\&& + m_e^6 -m_e^4 m_{\gamma^\prime}^2 \nonumber\\
&& - \left(s^2 -\left(6m_e^2+2m_{\gamma^\prime}^2\right)s -3m_e^4 + 2m_e^2 m_{\gamma^\prime}^2+2m_{\gamma^\prime}^4\right) \nonumber\\
&& \times \frac{4s^2}{\sqrt{\lambda(s,m_e,m_{\gamma^\prime})}}{\rm Coth}^{-1} \left[\frac{m_{\gamma^\prime}^2-m_e^2-s}{\sqrt{\lambda(s,m_e,m_{\gamma^\prime})}}\right]\Bigg) \, .
\eea




The cross section for the electron-positron annihilation process reads
\bea
\sigma_{\rm ann} & = & \frac{\varepsilon^2 e^4}{4 \pi  s^{3/2}} \frac{1}{\sqrt{s-4m_e^2}\left(s-m_{\gamma^\prime}^2\right)} \nonumber \\
&&\times \Bigg(-s^2 -4 m_e^2 s - m_{\gamma^\prime}^4 \nonumber\\
&& + \frac{2\sqrt{s}\left(s^2 + 4 m_e^2 s -8 m_e^4 - 4 m_e^2 m_{\gamma^\prime}^2 + m_{\gamma^\prime}^4\right)}{\sqrt{s-4m_e^2}} \nonumber \\
&&\times {\rm Tanh}^{-1} \left[\sqrt{\frac{s-4m_e^2}{s}}\right]\Bigg) \, . \quad
\eea
Note that the factor $(s-m_{\gamma'}^2)^{-1}$ of the denominator in the production rate may cause IR divergence if $m_{\gamma'}>2m_e$.
Following the prescription in Ref.~\cite{Redondo:2008ec}, this can be regulated by replacing
\begin{equation}
    (s-m_{\gamma'}^2)^{-1} \quad \rightarrow \quad \left(s-m_{\gamma'}^2+2m_{\gamma'} m_{\gamma}(T)\right)^{-1}\, ,
\end{equation}
where $ m_{\gamma}(T)$ is the thermal photon mass from electron-photon plasma and is approximately given by $m_{\gamma}^2 \approx \omega_{\rm pl}^2$.

When kinematically allowed, inverse-decay (i.e., coalescence) processes, $e^- + e^+ \rightarrow \gamma^\prime$, can also contribute significantly to DP production.
Using detailed balance and following Ref.~\cite{Gondolo:1990dk}, the corresponding reaction density can be expressed in terms of the \emph{equilibrium} DP phase space and its vacuum decay width as
\begin{equation}
\gamma_{\gamma^\prime}(T) = \frac{g_{\gamma^\prime}\,m_{\gamma^\prime}^{2}T}{2\pi^{2}}\,
K_1\!\left(\frac{m_{\gamma^\prime}}{T}\right)\,
\Gamma_{e^+e^-}\,,
\label{eq:gammaID}
\end{equation}
where $g_{\gamma^\prime}=3$ for a massive vector, $K_1$ is the modified Bessel function, and the decay width at the rest frame reads
\begin{equation}
\Gamma_{e^+e^-} =
\frac{\alpha\,\varepsilon^{2}}{3}\,
m_{\gamma^\prime}\!
\left(1+\frac{2m_e^{2}}{m_{\gamma^\prime}^{2}}\right)
\sqrt{1-\frac{4m_e^{2}}{m_{\gamma^\prime}^{2}}} \, .
\label{eq:width}
\end{equation}

Here, we provide a useful analytical result for the DP abundance from electron-positron coalescence; for details, see Ref.~\cite{Hall:2009bx}.
Performing the Boltzmann equation in Eq.~\eqref{eq:Boltzmann} with defining \(x\equiv m_{\gamma^\prime}/T\), we obtain
\beq
\begin{split}
Y_{\gamma^\prime} &\simeq  \int_{T_0}^{T_{\rm RH}} \!\! \frac{\gamma_{\gamma^\prime}}{sHT}\,dT \\
&=
\frac{45}{(1.660) 4\pi^{4}}
\frac{g_{\gamma^\prime}M_{\rm Pl}\,\Gamma_{e^+e^-}}{g_{*s}\,g_*^{1/2}\,m_{\gamma^\prime}^{2}}\,
\int_{x_{\rm RH}}^{\infty} x^{3}K_{1}(x)\,dx
\end{split}
\label{eq:Ymaster}
\eeq
with \(x_{\rm RH}\equiv m_{\gamma^\prime}/T_{\rm RH}\). Since contributions at \(T\ll m_{\gamma^\prime}\) are exponentially suppressed, the lower integration limit \(T_0\) is irrelevant.
It is convenient to factor out the full-range integral \(\int_{0}^{\infty}x^{3}K_{1}(x)\,dx=3\pi/2\) and define
\begin{equation}
S(x)\;\equiv\;\frac{2}{3\pi}\int_{x}^{\infty} u^{3}K_{1}(u)\,du,
\qquad S(0)=1.
\label{eq:Sdef}
\end{equation}
Then, the yield takes the compact form
\begin{equation}
Y_{\gamma^\prime} =
\frac{135}{1.660\,8\pi^{3}}\,
\frac{g_{\gamma^\prime}\,M_{\rm Pl}\,\Gamma_{ e^+e^-}}
     {g_{*s}\,g_*^{1/2}\,m_{\gamma^\prime}^{2}}\,
S\!\left(\frac{m_{\gamma^\prime}}{T_{\rm RH}}\right).
\label{eq:Yfinal}
\end{equation}
For \(x\gg 1\), where \(K_{1}(x)\simeq\sqrt{\pi/(2x)}\,e^{-x}\), one finds
\begin{equation}
S(x)\;\simeq\;\frac{2}{3\sqrt{2\pi}}\,x^{5/2}e^{-x}
\quad (x\gg 1),
\label{eq:Sasy}
\end{equation}
which makes the expected Boltzmann suppression explicit when \(m_{\gamma^\prime}\gg T_{\rm RH}\).
A smooth analytic fit valid for all \(x\) that reproduces \(S(0)=1\) and Eq.~(\ref{eq:Sasy}) is
\begin{equation}
S(x)\;\simeq\;\bigl[\,0.266+0.734\,e^{-x}\,\bigr]\;e^{-x}\,(1+x^{2})^{5/4}.
\label{eq:Sfit}
\end{equation}



\section{Higher initial temperature limit}
\label{app:highertemperature}


We extend the analysis presented in the main text, where we considered the minimal cosmological initial temperature of $T_i = 6\,{\rm MeV}$, to explore higher-temperature scenarios as a more general case.
As an illustrative example, we take $T_i = 100\,{\rm MeV}$ and examine how the resulting cosmological bounds on DPs are modified.

Since the DP production rate governed by renormalizable interactions is infrared-dominated, the contribution of non-resonant scattering processes (see Appendix.~\ref{app:crosssection}) to the thermal freeze-in abundance is largely insensitive to $T_i$ when the DP mass is below this temperature.
Consequently, as shown in Fig.~\ref{fig:YDP}, for $m_{\gamma^\prime} \lesssim 0.74\,{\rm MeV}$ corresponding to $\sqrt{3/2}\,\omega_{\rm pl}(6\,{\rm MeV})$, the thermal DP abundance is essentially identical for both $T_i = 6\,{\rm MeV}$ (black solid) and $T_i = 100\,{\rm MeV}$ (gray dashed).
At higher masses, however, the abundance increases due to additional production channels.
Resonant conversion from photons occurs up to $\sqrt{3/2}\,\omega_{\rm pl}(10\,{\rm MeV}) \simeq 12\,{\rm MeV}$; while this mechanism dominates for sub-threshold masses as $m_{\gamma^\prime} < 2m_e$, its contribution becomes subdominant (about 10\% of the total) for heavier DPs, where kinematically allowed coalescence processes begin to dominate.
At even higher masses, additional coalescence channels involving both leptons and hadrons become kinematically accessible.
Their contribution to the total DP production rate can be expressed using Eq.~\eqref{eq:gammaID}, with the relevant decay widths for each channel.
For leptonic coalescence processes such as $\mu^-\mu^+ \to \gamma'$, the decay width follows Eq.~\eqref{eq:width} with the appropriate lepton mass.
For hadronic processes like $\pi^-\pi^+ \to \gamma'$, the decay width reads~\cite{Fradette:2014sza}
\bea
    \Gamma_{\pi\pi}=\frac{1}{12}\alpha_{\text{eff}}^{\pi\pi}\varepsilon^2 m_{\gamma'}\left(1-4\frac{m_\pi^2}{m_{\gamma'}^2}\right)^{3/2}\,,
\eea
where $\alpha_{\text{eff}}^{\pi\pi}$ denotes the effective scalar QED coupling~\cite{Fradette:2014sza, BaBar:2012bdw}.
We adopt the empirical fitting functions for $\alpha_{\rm eff}^{\pi\pi}$ in the cases of the charged pions~\cite{BaBar:2012bdw} and the charged kaons~\cite{BaBar:2013jqz}, based on measurements of the $e^-e^+ \to \pi^-\pi^+$ and $e^-e^+ \to K^-K^+$ from BaBar collaboration.

\begin{figure}[t!]
\centering
\includegraphics[width=.9\linewidth]{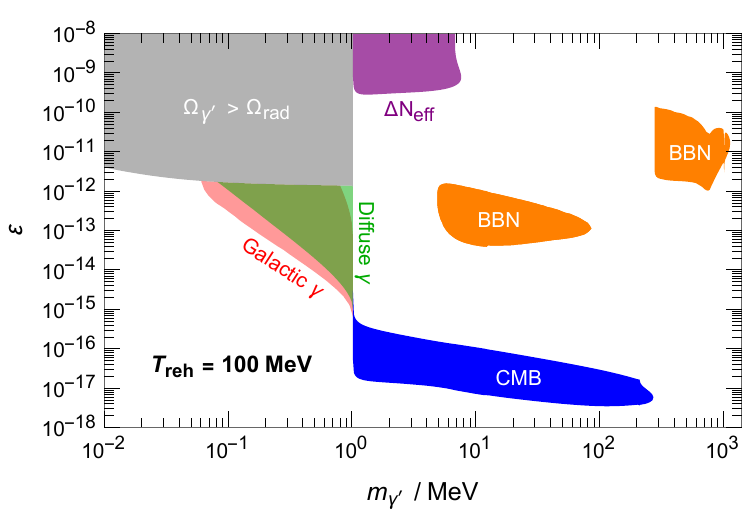}
 \caption{Constraints on the dark photon model with $T_{\mathrm{reh}}=100\;\mathrm{MeV}$. The blue-shaded region is excluded by CMB. The orange-shaded regions correspond to BBN. The purple-shaded region shows $\Delta N_{\mathrm{eff}}$ constraints. The gray-shaded region is excluded by the overproduction of dark photons. The red/green-shaded region is ruled out by galactic/extra-galactic X-ray flux.}
 \label{fig:DPDMcosmo100MeV}
\end{figure}


Building upon the derivation of the thermal DP abundance, we perform the same cosmological analyses as outlined in Sec.~\ref{sec:cosmo}.
The results for $T_i = 100\,{\rm MeV}$ are illustrated in Fig.~\ref{fig:DPDMcosmo100MeV}. For $m_{\gamma^\prime} < 2m_e$, the constraints from overproduction (gray-shaded) and galactic/extragalactic photon observations (red- and green-shaded regions, respectively) remain nearly identical to those in the $T_i = 6\,{\rm MeV}$ case, except for an enhancement in the range $m_{\gamma^\prime} \gtrsim 0.74\,{\rm MeV}$ due to additional resonant contributions to the DP abundance.
For intermediate masses, $2m_e \lesssim m_{\gamma^\prime} \lesssim 100\,{\rm MeV}$, the bounds from $\Delta N_{\rm eff}$, CMB energy injection, and BBN (mainly due to deuterium dissociation) are also comparable to those obtained for $T_i = 6\,{\rm MeV}$.

At higher masses, $m_{\gamma^\prime} \gtrsim 100\,{\rm MeV}$, the CMB constraint extends up to $m_{\gamma^\prime} \sim 300\,{\rm MeV}$, while new bounds emerge from hadronic effects during BBN once $m_{\gamma^\prime}$ exceeds the pion threshold, $2m_\pi$.
These additional BBN constraints originate from the injection of mesons (e.g., $\pi^\pm$) produced in DP decays.
Such mesons can alter the neutron–proton interconversion rates via reactions such as $p + \pi^- \leftrightarrow n + \pi^0$ and $n + \pi^+ \leftrightarrow p + \pi^0$, thus modifying the neutron freeze-out fraction prior to nucleosynthesis~\cite{Fradette:2017sdd}.
Since charged pions have short lifetimes ($\tau_\pi \sim 10^{-8}\,{\rm s}$) and are rapidly thermalized in the primordial plasma, their number density can be approximated by a quasi-static equilibrium form of $n_{\pi^\pm} \approx \tau_\pi\Gamma_{\gamma^\prime}n_{\gamma^\prime}$.
The resulting modification of the neutron–proton ratio can be evaluated by incorporating this meson population into the Boltzmann equations governing weak interactions; for details, see Refs.~\cite{Mukhanov:2003xs,Fradette:2017sdd}.
This alters the freeze-out neutron fraction and consequently the predicted ${}^4{\rm He}$ abundance, leading to the BBN constraint shown as the orange-shaded region in the high-mass regime of Fig.~\ref{fig:DPDMcosmo100MeV}.

\begin{figure}[t!]
\centering
\includegraphics[width=.9\linewidth]{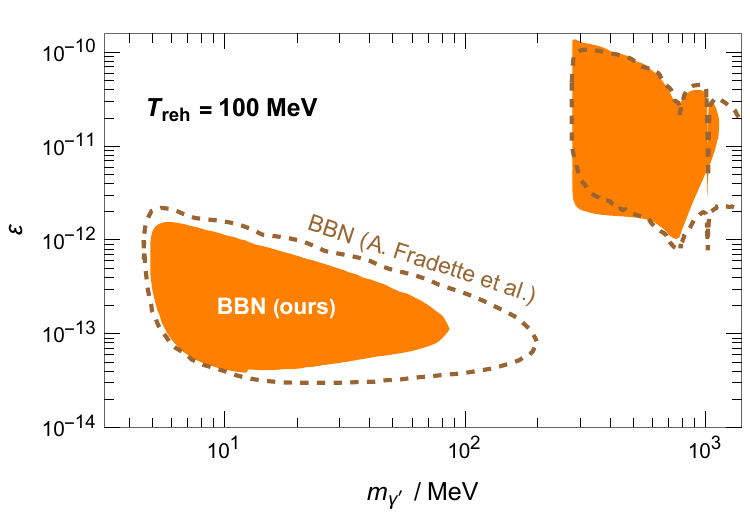}
 \caption{BBN constraints for an initial temperature of $T_{\mathrm{reh}}=100\,\mathrm{MeV}$ (orange-shaded) compared with the previous BBN result from Ref.~\cite{Fradette:2014sza} (brown dashed). We have verified that, when adopting the same values for the measured $D/H$ ratio and nuclear reaction rates as in Ref.~\cite{Fradette:2014sza}, our results below the GeV scale agree very well, except for a 10\% discrepancy of unknown origin around $m_{\gamma'} \sim 100~\mathrm{MeV}$.
}
 \label{fig:DPDMBBN100MeV}
\end{figure}

In Fig.~\ref{fig:DPDMBBN100MeV}, we compare our updated BBN constraint (orange-shaded) with the earlier result from Ref.~\cite{Fradette:2014sza} (brown dashed).
For lower masses, where reduction of the relic deuterium abundance is induced by the radiative DP decay, our bound is slightly weaker due to the inclusion of sizable theoretical uncertainties in the nuclear reaction rates affecting deuterium synthesis.
At higher masses, where hadronic energy injection occurs, our result is in good agreement with Ref.~\cite{Fradette:2014sza} up to $0.8\,{\rm GeV}$, above which the thermal DP abundance is exponentially suppressed. We have verified numerically that, when we adopt the same values for the measured $D/H$ ratio and nuclear reaction rates as in Ref.~\cite{Fradette:2014sza}, our results below the GeV scale agree, except for a 10\% discrepancy of unknown origin for $m_{\gamma'} \sim 100~\mathrm{MeV}$


\section{Compton-like scattering rate in supernova environment}
\label{app:semiCompton}

We compute the Compton-like scattering rate for DPs, $e^- + \gamma^\prime \rightarrow e^- + \gamma$, which contributes to both DP absorption and, by detailed balance, their inverse production process, under conditions relevant to core-collapse supernovae.
As discussed in Sec.~\ref{sec:DPcoupling}, DP interactions follow the same structure as electromagnetic interactions, up to an overall factor of the kinetic mixing parameter~$\varepsilon$.
The detailed balance relation connects the absorption and emission rates (see Eq.~\eqref{eq:ImPi}), allowing the Compton rate for photons to be reinterpreted.
Hence, we derive the Compton scattering rate for photons involving the dispersion of DPs and one can easily apply these results to evaluate the DP interaction rate in CCSNe.
In what follows, we focus on the transverse polarization modes; the longitudinal mode contribution can be obtained by multiplying the transverse result by a factor of $m_{\gamma^\prime}^2/\omega^2$, which follows from current conservation.

The general expression for the absorption via Compton-like scatterings reads
\beq
\begin{split}
 \Gamma_{\rm sC}&(\omega )  =  \frac{2}{2\omega}\int \frac{d^3  \vec{p}_i}{(2\pi)^3 2 E_i}f_i\int \frac{d^3 \vec{p}_f}{(2\pi)^3 2 E_f} \left(1-f_f\right)\\
    &  \int \frac{d^3 \vec{p}_\gamma}{(2\pi)^3 2E_\gamma}|\mathcal{M}|^2 \left(2\pi\right)^4 \delta^4 (k+p_i-p_f-p_\gamma) 
\end{split}
\eeq
where $p_{i(f)} = (E_{i(f)},\vec{p}_{i(f)})$ denotes the four-momentum of the incident (scattered) electron, $k$ is the four-momentum of the incoming photon with the DP dispersion, and $p_{\gamma} = (E_\gamma,\vec{p}_\gamma)$ is that of the outgoing photon state.
For the final photon state, we neglect the statistical factor and the plasma correction (i.e., $E_\gamma \simeq |\vec{p}_\gamma|$).
In the relativistic regime, the leading-order squared matrix element, averaged over initial spin degrees of freedom, reads
\begin{equation}
|\mathcal{M}|^2 = 32 \pi^2  \alpha^2
\left( \frac{p_i \cdot k}{p_i \cdot p_\gamma} \right)\,.
\end{equation}
Although a full analytic expression valid across all regimes is cumbersome, useful approximate results can be obtained for conditions characteristic of CCSNe.
Below we discuss two representative limits relevant for DP absorption and production.

DPs produced in the PNS may be reabsorbed in the outer stellar layers through Compton-like scattering, $\gamma^\prime + e^- \to \gamma+e^-$, since the bremsstrahlung process, being proportional to the square of the target density, becomes inefficient in the dilute outer regions.
In this case, the relevant DP energies are comparable to or higher than the PNS temperature of $30$-$50 \, \mathrm{MeV}$.
At the outer PNS boundary ($r \gtrsim 20\,\mathrm{km}$), the temperature decreases significantly, while the matter remains degenerate ($\mu_e \gg T$).
For the parameter space of interest, we typically have $\omega \gtrsim \mu_e$ in this case of absorption outside the PNS.

In this degenerate limit, the Fermi-Dirac distribution may be approximated by a Heaviside step function, implying that electrons are filled up to $\mu_e$, and scattering is only allowed for final electrons with $E_f > \mu_e$.
Performing the phase-space integration under these conditions yields
\bea
\Gamma_{\rm sC} \approx \frac{\alpha^2 }{2 \pi} \frac{\mu^2}{\omega} \log \left(\frac{\mu \omega}{m_e^2}\right) \, .
\eea
This expression provides a reliable estimate of the Compton-like absorption rate for DPs escaping the PNS core and is the dominant absorption mechanism outside the neutrino sphere.

DP production in the outer supernova layers can also impact the explosion dynamics, potentially preventing successful shock revival (see Sec.~\ref{sec:preventSNex}).
Analogous to the absorption discussed above, the DP production outside the PNS is governed by Compton-like scatterings.
In these regions, such as the gain region, the electron density is much lower, and the temperature exceeds the chemical potential ($T \gg \mu_e$), corresponding to the classical limit.
Pauli blocking can then be neglected, and the rate can be expressed as
\bea
    \Gamma_{\gamma^\prime} = 2 \int \frac{d^3 p_i}{\left(2\pi\right)^3} f_i \sigma_{\rm sC} v_{\rm mol} \,,
    \label{eq:intforabs2}
\eea
where $v_{\rm mol} = (s-m_e^2)/2\omega E_i$ is the Moller velocity, and the cross section takes the well-known form
\bea
 \sigma_{\rm sC} = \frac{2\pi \alpha^2}{s}\log\left(\frac{s}{m_e^2}\right) \,.
 \eea
Assuming the Maxwell-Boltzmann for electrons as a good approximation in this case results in
\bea
  \Gamma_{\rm sC} \approx \frac{\alpha^2}{\pi} \frac{T^2}{\omega} \log\left(\frac{\omega T}{m_e^2}\right).
\label{eq:scT}
\eea

Note that Eq.~\eqref{eq:scT}, derived under the assumptions $T \gg \mu_e$ and $\omega \sim T$, remains approximately valid even in the degenerate regime ($\mu_e \gg T$) as long as $T \sim \omega$, up to a factor of about 2.
This is because only electrons near the Fermi surface, within a narrow energy width $\sim T$, participate effectively in the scattering.
The dominant contribution arises from backward scatterings in the center-of-mass frame, where the outgoing electron acquires a higher energy, making Pauli suppression less. In this regime, however, the Compton-like process typically becomes subdominant compared to bremsstrahlung, owing to the higher target density dependence of the latter. Finally, we stress that for all the results of this paper, we have used the complete, fully numerical production and absorption rates for DPs. However, the analytical formulas reported here agree extremely well with the numerics across all the parameter space of interest.

\section{Spectrum of the decay of dark photons in the boosted frame}
\label{app:spectrumBoost}

We provide the energy spectrum of daughter particles from the decay of boosted dark photons, in terms of the spectrum in the rest frame.

Let us consider the dark photon decay channel into degenerate particles; for instance, $\gamma^\prime \rightarrow 3\gamma$ or $\gamma^\prime \rightarrow e^-e^+$.
Denoting the decay rate at the rest frame, $\Gamma_{\gamma^\prime}$, the spectrum of a daughter particle $i$ in the boosted dark photon frame with the energy $\omega$ is given by
\bea
\left.\frac{dN_i}{dE}\right|_{E=\omega} & = & \frac{N_i}{2\sqrt{\gamma^{\prime 2}-1}}\int_{E_0^{\rm min}}^{E_0^{\rm max}}  dE_0 \,\frac{1}{\sqrt{E_0^2-m^2}}\nonumber\\
&& \times \frac{1}{\Gamma_{\gamma^\prime}}\frac{d\Gamma_{\gamma^\prime}}{dE_0} \Theta\left(\frac{m_{\gamma^\prime}}{2}-E_0\right) \,,
\eea
where $\gamma^\prime = \omega/m_{\gamma^\prime}$, $m$ is the mass of daughter particles, and $N_i$ indicates the number of degree of $i$.
The integration range of $E_0$ reads
\bea
E_0^{\rm max,min}=E \gamma^\prime \left(1\pm \sqrt{1-\gamma_{\gamma^\prime}^{-2}}\sqrt{1-m^2/E^2}\right)\,.
\eea

\bibliography{bib}

\providecommand{\href}[2]{#2}\begingroup\raggedright\begin{thebibliography}{100}

\bibitem{Holdom:1985ag}
B.~Holdom, ``{Two U(1)'s and Epsilon Charge Shifts},''
  \href{http://dx.doi.org/10.1016/0370-2693(86)91377-8}{{\em Phys. Lett. B}
  {\bfseries 166} (1986) 196--198}.

\bibitem{Okun:1982xi}
L.~B. Okun, ``{Limits of electrodynamics: paraphotons?},'' {\em Sov. Phys.
  JETP} {\bfseries 56} (1982) 502. [Zh. Eksp. Teor. Fiz. 83 (1982) 892].

\bibitem{Galison:1983pa}
P.~Galison and A.~Manohar, ``{Two Z's or not two Z's?},''
  \href{http://dx.doi.org/10.1016/0370-2693(84)91161-4}{{\em Phys. Lett. B}
  {\bfseries 136} (1984) 279--283}.

\bibitem{Arkani-Hamed:2008hhe}
N.~Arkani-Hamed, D.~P. Finkbeiner, T.~R. Slatyer, and N.~Weiner, ``{A Theory of
  Dark Matter},'' \href{http://dx.doi.org/10.1103/PhysRevD.79.015014}{{\em
  Phys. Rev. D} {\bfseries 79} (2009) 015014},
  \href{http://arxiv.org/abs/0810.0713}{{\ttfamily arXiv:0810.0713 [hep-ph]}}.

\bibitem{Pospelov:2007mp}
M.~Pospelov, A.~Ritz, and M.~B. Voloshin, ``{Secluded WIMP dark matter},''
  \href{http://dx.doi.org/10.1016/j.physletb.2008.02.052}{{\em Phys. Lett. B}
  {\bfseries 662} (2008) 53--61},
  \href{http://arxiv.org/abs/0711.4866}{{\ttfamily arXiv:0711.4866 [hep-ph]}}.

\bibitem{DarkCast}
P.~Trener, ``Darkcast: Recasting collider bounds on dark photons.''
  \url{https://gitlab.com/philten/darkcast}.
\newblock Accessed: 2025-05-10.

\bibitem{Ilten:2018crw}
P.~Ilten, Y.~Soreq, M.~Williams, and W.~Xue, ``{Serendipity in dark photon
  searches},'' \href{http://dx.doi.org/10.1007/JHEP06(2018)004}{{\em JHEP}
  {\bfseries 06} (2018) 004}, \href{http://arxiv.org/abs/1801.04847}{{\ttfamily
  arXiv:1801.04847 [hep-ph]}}.

\bibitem{Baruch:2022esd}
C.~Baruch, P.~Ilten, Y.~Soreq, and M.~Williams, ``{Axial vectors in
  DarkCast},'' \href{http://dx.doi.org/10.1007/JHEP11(2022)124}{{\em JHEP}
  {\bfseries 11} (2022) 124}, \href{http://arxiv.org/abs/2206.08563}{{\ttfamily
  arXiv:2206.08563 [hep-ph]}}.

\bibitem{Bjorken:2009mm}
J.~D. Bjorken, R.~Essig, P.~Schuster, and N.~Toro, ``New fixed-target
  experiments to search for dark gauge forces,''
  \href{http://dx.doi.org/10.1103/PhysRevD.80.075018}{{\em Phys. Rev. D}
  {\bfseries 80} (2009) 075018},
  \href{http://arxiv.org/abs/0906.0580}{{\ttfamily arXiv:0906.0580 [hep-ph]}}.

\bibitem{Reece:2009un}
M.~Reece and L.-T. Wang, ``Searching for the light dark gauge boson in
  gev-scale experiments,''
  \href{http://dx.doi.org/10.1088/1126-6708/2009/07/051}{{\em JHEP} no.~07,
  (2009) 051}, \href{http://arxiv.org/abs/0904.1743}{{\ttfamily arXiv:0904.1743
  [hep-ph]}}.

\bibitem{Freytsis:2009bh}
M.~Freytsis, G.~Ovanesyan, and J.~Thaler, ``Dark force detection in low energy
  e-p collisions,'' \href{http://dx.doi.org/10.1007/JHEP01(2010)111}{{\em JHEP}
  no.~01, (2010) 111}, \href{http://arxiv.org/abs/0909.2862}{{\ttfamily
  arXiv:0909.2862 [hep-ph]}}.

\bibitem{APEX:2011dww}
{\bfseries APEX} Collaboration, S.~Abrahamyan {\em et~al.}, ``{Search for a New
  Gauge Boson in Electron-Nucleus Fixed-Target Scattering by the APEX
  Experiment},'' \href{http://dx.doi.org/10.1103/PhysRevLett.107.191804}{{\em
  Phys. Rev. Lett.} {\bfseries 107} (2011) 191804},
  \href{http://arxiv.org/abs/1108.2750}{{\ttfamily arXiv:1108.2750 [hep-ex]}}.

\bibitem{Merkel:2014avp}
H.~Merkel {\em et~al.}, ``{Search at the Mainz Microtron for Light Massive
  Gauge Bosons Relevant for the Muon g-2 Anomaly},''
  \href{http://dx.doi.org/10.1103/PhysRevLett.112.221802}{{\em Phys. Rev.
  Lett.} {\bfseries 112} no.~22, (2014) 221802},
  \href{http://arxiv.org/abs/1404.5502}{{\ttfamily arXiv:1404.5502 [hep-ex]}}.

\bibitem{A1:2011yso}
{\bfseries A1} Collaboration, H.~Merkel {\em et~al.}, ``{Search for Light Gauge
  Bosons of the Dark Sector at the Mainz Microtron},''
  \href{http://dx.doi.org/10.1103/PhysRevLett.106.251802}{{\em Phys. Rev.
  Lett.} {\bfseries 106} (2011) 251802},
  \href{http://arxiv.org/abs/1101.4091}{{\ttfamily arXiv:1101.4091 [nucl-ex]}}.

\bibitem{Batell:2009di}
B.~Batell, M.~Pospelov, and A.~Ritz, ``Exploring portals to a hidden sector
  through fixed targets,''
  \href{http://dx.doi.org/10.1103/PhysRevD.80.095024}{{\em Phys. Rev. D}
  {\bfseries 80} (2009) 095024},
  \href{http://arxiv.org/abs/0906.5614}{{\ttfamily arXiv:0906.5614 [hep-ph]}}.

\bibitem{Essig:2010gu}
R.~Essig, R.~Harnik, J.~Kaplan, and N.~Toro, ``Discovering new light states at
  neutrino experiments,''
  \href{http://dx.doi.org/10.1103/PhysRevD.82.113008}{{\em Phys. Rev. D}
  {\bfseries 82} (2010) 113008},
  \href{http://arxiv.org/abs/1008.0636}{{\ttfamily arXiv:1008.0636 [hep-ph]}}.

\bibitem{Pospelov:2008zw}
M.~Pospelov, ``{Secluded U(1) below the weak scale},''
  \href{http://dx.doi.org/10.1103/PhysRevD.80.095002}{{\em Phys. Rev. D}
  {\bfseries 80} (2009) 095002},
  \href{http://arxiv.org/abs/0811.1030}{{\ttfamily arXiv:0811.1030 [hep-ph]}}.

\bibitem{KLOE-2:2011hhj}
{\bfseries KLOE-2} Collaboration, F.~Archilli {\em et~al.}, ``{Search for a
  vector gauge boson in $\phi$ meson decays with the KLOE detector},''
  \href{http://dx.doi.org/10.1016/j.physletb.2011.11.033}{{\em Phys. Lett. B}
  {\bfseries 706} (2012) 251--255},
  \href{http://arxiv.org/abs/1110.0411}{{\ttfamily arXiv:1110.0411 [hep-ex]}}.

\bibitem{WASA-at-COSY:2013zom}
{\bfseries WASA-at-COSY} Collaboration, P.~Adlarson {\em et~al.}, ``{Search for
  a dark photon in the $\pi^0 \to e^+e^-\gamma$ decay},''
  \href{http://dx.doi.org/10.1016/j.physletb.2013.08.055}{{\em Phys. Lett. B}
  {\bfseries 726} (2013) 187--193},
  \href{http://arxiv.org/abs/1304.0671}{{\ttfamily arXiv:1304.0671 [hep-ex]}}.

\bibitem{HADES:2013nab}
{\bfseries HADES} Collaboration, G.~Agakishiev {\em et~al.}, ``{Searching a
  Dark Photon with HADES},''
  \href{http://dx.doi.org/10.1016/j.physletb.2014.02.035}{{\em Phys. Lett. B}
  {\bfseries 731} (2014) 265--271},
  \href{http://arxiv.org/abs/1311.0216}{{\ttfamily arXiv:1311.0216 [hep-ex]}}.

\bibitem{NA482:2015wmo}
{\bfseries NA48/2} Collaboration, J.~R. Batley {\em et~al.}, ``{Search for the
  dark photon in $\pi^0$ decays},''
  \href{http://dx.doi.org/10.1016/j.physletb.2015.04.068}{{\em Phys. Lett. B}
  {\bfseries 746} (2015) 178--185},
  \href{http://arxiv.org/abs/1504.00607}{{\ttfamily arXiv:1504.00607
  [hep-ex]}}.

\bibitem{PHENIX:2014duq}
{\bfseries PHENIX} Collaboration, A.~Adare {\em et~al.}, ``{Search for dark
  photons from neutral meson decays in $p + p$ and $d$ + Au collisions at
  $\sqrt{s_{NN}} =$ 200 GeV},''
  \href{http://dx.doi.org/10.1103/PhysRevC.91.031901}{{\em Phys. Rev. C}
  {\bfseries 91} no.~3, (2015) 031901},
  \href{http://arxiv.org/abs/1409.0851}{{\ttfamily arXiv:1409.0851 [nucl-ex]}}.

\bibitem{Williams:2011qb}
M.~Williams, C.~P. Burgess, A.~Maharana, and F.~Quevedo, ``{New Constraints
  (and Motivations) for Abelian Gauge Bosons in the MeV-TeV Mass Range},''
  \href{http://dx.doi.org/10.1007/JHEP08(2011)106}{{\em JHEP} {\bfseries 08}
  (2011) 106}, \href{http://arxiv.org/abs/1103.4556}{{\ttfamily arXiv:1103.4556
  [hep-ph]}}.

\bibitem{Blumlein:2011mv}
J.~Blumlein and J.~Brunner, ``{New Exclusion Limits for Dark Gauge Forces from
  Beam-Dump Data},''
  \href{http://dx.doi.org/10.1016/j.physletb.2011.05.046}{{\em Phys. Lett. B}
  {\bfseries 701} (2011) 155--159},
  \href{http://arxiv.org/abs/1104.2747}{{\ttfamily arXiv:1104.2747 [hep-ex]}}.

\bibitem{Gninenko:2012eq}
S.~N. Gninenko, ``{Constraints on sub-GeV hidden sector gauge bosons from a
  search for heavy neutrino decays},''
  \href{http://dx.doi.org/10.1016/j.physletb.2012.06.002}{{\em Phys. Lett. B}
  {\bfseries 713} (2012) 244--248},
  \href{http://arxiv.org/abs/1204.3583}{{\ttfamily arXiv:1204.3583 [hep-ph]}}.

\bibitem{Blumlein:2013cua}
J.~Bl{\"u}mlein and J.~Brunner, ``{New Exclusion Limits on Dark Gauge Forces
  from Proton Bremsstrahlung in Beam-Dump Data},''
  \href{http://dx.doi.org/10.1016/j.physletb.2014.02.029}{{\em Phys. Lett. B}
  {\bfseries 731} (2014) 320--326},
  \href{http://arxiv.org/abs/1311.3870}{{\ttfamily arXiv:1311.3870 [hep-ph]}}.

\bibitem{NA64:2018lsq}
{\bfseries NA64} Collaboration, D.~Banerjee {\em et~al.}, ``{Search for a
  Hypothetical 16.7 MeV Gauge Boson and Dark Photons in the NA64 Experiment at
  CERN},'' \href{http://dx.doi.org/10.1103/PhysRevLett.120.231802}{{\em Phys.
  Rev. Lett.} {\bfseries 120} no.~23, (2018) 231802},
  \href{http://arxiv.org/abs/1803.07748}{{\ttfamily arXiv:1803.07748
  [hep-ex]}}.

\bibitem{Fayet:2007ua}
P.~Fayet, ``{U-boson production in e+ e- annihilations, psi and Upsilon decays,
  and Light Dark Matter},''
  \href{http://dx.doi.org/10.1103/PhysRevD.75.115017}{{\em Phys. Rev. D}
  {\bfseries 75} (2007) 115017},
  \href{http://arxiv.org/abs/hep-ph/0702176}{{\ttfamily arXiv:hep-ph/0702176}}.

\bibitem{Batell:2009psz}
B.~Batell, M.~Pospelov, and A.~Ritz, ``Probing a secluded u(1) at
  b-factories,'' \href{http://dx.doi.org/10.1103/PhysRevD.79.115008}{{\em Phys.
  Rev. D} {\bfseries 79} (2009) 115008},
  \href{http://arxiv.org/abs/0903.0363}{{\ttfamily arXiv:0903.0363 [hep-ph]}}.

\bibitem{Essig:2009nc}
R.~Essig, P.~Schuster, and N.~Toro, ``Probing dark forces and light hidden
  sectors at low-energy e+e− colliders,''
  \href{http://dx.doi.org/10.1103/PhysRevD.80.015003}{{\em Phys. Rev. D}
  {\bfseries 80} (2009) 015003},
  \href{http://arxiv.org/abs/0903.3941}{{\ttfamily arXiv:0903.3941 [hep-ph]}}.

\bibitem{Curtin:2014cca}
D.~Curtin, R.~Essig, S.~Gori, and J.~Shelton, ``Illuminating dark photons with
  high-energy colliders,''
  \href{http://dx.doi.org/10.1007/JHEP02(2015)157}{{\em JHEP} {\bfseries 02}
  (2015) 157}, \href{http://arxiv.org/abs/1412.0018}{{\ttfamily arXiv:1412.0018
  [hep-ph]}}.

\bibitem{BaBar:2014zli}
{\bfseries BaBar} Collaboration, J.~P. Lees {\em et~al.}, ``{Search for a Dark
  Photon in $e^+e^-$ Collisions at BaBar},''
  \href{http://dx.doi.org/10.1103/PhysRevLett.113.201801}{{\em Phys. Rev.
  Lett.} {\bfseries 113} no.~20, (2014) 201801},
  \href{http://arxiv.org/abs/1406.2980}{{\ttfamily arXiv:1406.2980 [hep-ex]}}.

\bibitem{LHCb:2017trq}
{\bfseries LHCb} Collaboration, R.~Aaij {\em et~al.}, ``{Search for Dark
  Photons Produced in 13 TeV $pp$ Collisions},''
  \href{http://dx.doi.org/10.1103/PhysRevLett.120.061801}{{\em Phys. Rev.
  Lett.} {\bfseries 120} no.~6, (2018) 061801},
  \href{http://arxiv.org/abs/1710.02867}{{\ttfamily arXiv:1710.02867
  [hep-ex]}}.

\bibitem{BESIII:2017fwv}
{\bfseries BESIII} Collaboration, M.~Ablikim {\em et~al.}, ``{Dark Photon
  Search in the Mass Range Between 1.5 and 3.4 GeV/$c^2$},''
  \href{http://dx.doi.org/10.1016/j.physletb.2017.09.067}{{\em Phys. Lett. B}
  {\bfseries 774} (2017) 252--257},
  \href{http://arxiv.org/abs/1705.04265}{{\ttfamily arXiv:1705.04265
  [hep-ex]}}.

\bibitem{Anastasi:2015qla}
A.~Anastasi {\em et~al.}, ``{Limit on the production of a low-mass vector boson
  in $\mathrm{e}^{+}\mathrm{e}^{-} \to \mathrm{U}\gamma$, $\mathrm{U} \to
  \mathrm{e}^{+}\mathrm{e}^{-}$ with the KLOE experiment},''
  \href{http://dx.doi.org/10.1016/j.physletb.2015.10.003}{{\em Phys. Lett. B}
  {\bfseries 750} (2015) 633--637},
  \href{http://arxiv.org/abs/1509.00740}{{\ttfamily arXiv:1509.00740
  [hep-ex]}}.

\bibitem{Caputo:2021eaa}
A.~Caputo, A.~J. Millar, C.~A.~J. O'Hare, and E.~Vitagliano, ``{Dark photon
  limits: A handbook},''
  \href{http://dx.doi.org/10.1103/PhysRevD.104.095029}{{\em Phys. Rev. D}
  {\bfseries 104} no.~9, (2021) 095029},
  \href{http://arxiv.org/abs/2105.04565}{{\ttfamily arXiv:2105.04565
  [hep-ph]}}.

\bibitem{Baumgart:2009tn}
M.~Baumgart, C.~Cheung, J.~T. Ruderman, L.-T. Wang, and I.~Yavin,
  ``{Non-Abelian Dark Sectors and Their Collider Signatures},''
  \href{http://dx.doi.org/10.1088/1126-6708/2009/04/014}{{\em JHEP} {\bfseries
  04} (2009) 014}, \href{http://arxiv.org/abs/0901.0283}{{\ttfamily
  arXiv:0901.0283 [hep-ph]}}.

\bibitem{Lee:2016ief}
H.-S. Lee and S.~Yun, ``{Mini force: The $(B-L)+xY$ gauge interaction with a
  light mediator},'' \href{http://dx.doi.org/10.1103/PhysRevD.93.115028}{{\em
  Phys. Rev. D} {\bfseries 93} no.~11, (2016) 115028},
  \href{http://arxiv.org/abs/1604.01213}{{\ttfamily arXiv:1604.01213
  [hep-ph]}}.

\bibitem{Bauer:2022nwt}
M.~Bauer and P.~Foldenauer, ``{Consistent Theory of Kinetic Mixing and the
  Higgs Low-Energy Theorem},''
  \href{http://dx.doi.org/10.1103/PhysRevLett.129.171801}{{\em Phys. Rev.
  Lett.} {\bfseries 129} no.~17, (2022) 171801},
  \href{http://arxiv.org/abs/2207.00023}{{\ttfamily arXiv:2207.00023
  [hep-ph]}}.

\bibitem{Fayet:1990wx}
P.~Fayet, ``{Extra U(1)'s and New Forces},''
  \href{http://dx.doi.org/10.1016/0550-3213(90)90381-M}{{\em Nucl. Phys. B}
  {\bfseries 347} (1990) 743--768}.

\bibitem{Stueckelberg:1938hvi}
E.~C.~G. Stueckelberg, ``{Interaction energy in electrodynamics and in the
  field theory of nuclear forces},''
  \href{http://dx.doi.org/10.5169/seals-110852}{{\em Helv. Phys. Acta}
  {\bfseries 11} (1938) 225--244}.

\bibitem{Babu:1997st}
K.~S. Babu, C.~F. Kolda, and J.~March-Russell, ``{Implications of generalized Z
  - Z-prime mixing},'' \href{http://dx.doi.org/10.1103/PhysRevD.57.6788}{{\em
  Phys. Rev. D} {\bfseries 57} (1998) 6788--6792},
  \href{http://arxiv.org/abs/hep-ph/9710441}{{\ttfamily arXiv:hep-ph/9710441}}.

\bibitem{Redondo:2008ec}
J.~Redondo and M.~Postma, ``{Massive hidden photons as lukewarm dark matter},''
  \href{http://dx.doi.org/10.1088/1475-7516/2009/02/005}{{\em JCAP} {\bfseries
  02} (2009) 005}, \href{http://arxiv.org/abs/0811.0326}{{\ttfamily
  arXiv:0811.0326 [hep-ph]}}.

\bibitem{An:2013yfc}
H.~An, M.~Pospelov, and J.~Pradler, ``{New stellar constraints on dark
  photons},'' \href{http://dx.doi.org/10.1016/j.physletb.2013.07.008}{{\em
  Phys. Lett. B} {\bfseries 725} (2013) 190--195},
  \href{http://arxiv.org/abs/1302.3884}{{\ttfamily arXiv:1302.3884 [hep-ph]}}.

\bibitem{Stodolsky:1986dx}
L.~Stodolsky, ``{On the Treatment of Neutrino Oscillations in a Thermal
  Environment},'' \href{http://dx.doi.org/10.1103/PhysRevD.36.2273}{{\em Phys.
  Rev. D} {\bfseries 36} (1987) 2273}.

\bibitem{Weldon:1983jn}
H.~A. Weldon, ``{Simple Rules for Discontinuities in Finite Temperature Field
  Theory},'' \href{http://dx.doi.org/10.1103/PhysRevD.28.2007}{{\em Phys. Rev.
  D} {\bfseries 28} (1983) 2007}.

\bibitem{Redondo:2013lna}
J.~Redondo and G.~Raffelt, ``{Solar constraints on hidden photons
  re-visited},'' \href{http://dx.doi.org/10.1088/1475-7516/2013/08/034}{{\em
  JCAP} {\bfseries 08} (2013) 034},
  \href{http://arxiv.org/abs/1305.2920}{{\ttfamily arXiv:1305.2920 [hep-ph]}}.

\bibitem{Gondolo:1990dk}
P.~Gondolo and G.~Gelmini, ``{Cosmic abundances of stable particles: Improved
  analysis},'' \href{http://dx.doi.org/10.1016/0550-3213(91)90438-4}{{\em Nucl.
  Phys. B} {\bfseries 360} (1991) 145--179}.

\bibitem{Saikawa:2018rcs}
K.~Saikawa and S.~Shirai, ``{Primordial gravitational waves, precisely: The
  role of thermodynamics in the Standard Model},''
  \href{http://dx.doi.org/10.1088/1475-7516/2018/05/035}{{\em JCAP} {\bfseries
  05} (2018) 035}, \href{http://arxiv.org/abs/1803.01038}{{\ttfamily
  arXiv:1803.01038 [hep-ph]}}.

\bibitem{Barbieri:2025moq}
N.~Barbieri, T.~Brinckmann, S.~Gariazzo, M.~Lattanzi, S.~Pastor, and
  O.~Pisanti, ``{Current constraints on cosmological scenarios with very low
  reheating temperatures},'' \href{http://arxiv.org/abs/2501.01369}{{\ttfamily
  arXiv:2501.01369 [astro-ph.CO]}}.

\bibitem{Hasegawa:2019jsa}
T.~Hasegawa, N.~Hiroshima, K.~Kohri, R.~S.~L. Hansen, T.~Tram, and
  S.~Hannestad, ``{MeV-scale reheating temperature and thermalization of
  oscillating neutrinos by radiative and hadronic decays of massive
  particles},'' \href{http://dx.doi.org/10.1088/1475-7516/2019/12/012}{{\em
  JCAP} {\bfseries 12} (2019) 012},
  \href{http://arxiv.org/abs/1908.10189}{{\ttfamily arXiv:1908.10189
  [hep-ph]}}.

\bibitem{Froustey:2020mcq}
J.~Froustey, C.~Pitrou, and M.~C. Volpe, ``{Neutrino decoupling including
  flavour oscillations and primordial nucleosynthesis},''
  \href{http://dx.doi.org/10.1088/1475-7516/2020/12/015}{{\em JCAP} {\bfseries
  12} (2020) 015}, \href{http://arxiv.org/abs/2008.01074}{{\ttfamily
  arXiv:2008.01074 [hep-ph]}}.

\bibitem{Bennett:2020zkv}
J.~J. Bennett, G.~Buldgen, P.~F. De~Salas, M.~Drewes, S.~Gariazzo, S.~Pastor,
  and Y.~Y.~Y. Wong, ``{Towards a precision calculation of $N_{\rm eff}$ in the
  Standard Model II: Neutrino decoupling in the presence of flavour
  oscillations and finite-temperature QED},''
  \href{http://dx.doi.org/10.1088/1475-7516/2021/04/073}{{\em JCAP} {\bfseries
  04} (2021) 073}, \href{http://arxiv.org/abs/2012.02726}{{\ttfamily
  arXiv:2012.02726 [hep-ph]}}.

\bibitem{Ibe:2019gpv}
M.~Ibe, S.~Kobayashi, Y.~Nakayama, and S.~Shirai, ``{Cosmological constraint on
  dark photon from N$_{eff}$},''
  \href{http://dx.doi.org/10.1007/JHEP04(2020)009}{{\em JHEP} {\bfseries 04}
  (2020) 009}, \href{http://arxiv.org/abs/1912.12152}{{\ttfamily
  arXiv:1912.12152 [hep-ph]}}.

\bibitem{Li:2020roy}
J.-T. Li, G.~M. Fuller, and E.~Grohs, ``{Probing dark photons in the early
  universe with big bang nucleosynthesis},''
  \href{http://dx.doi.org/10.1088/1475-7516/2020/12/049}{{\em JCAP} {\bfseries
  12} (2020) 049}, \href{http://arxiv.org/abs/2009.14325}{{\ttfamily
  arXiv:2009.14325 [astro-ph.CO]}}.

\bibitem{Ganguly:2025mdi}
S.~Ganguly, T.~H. Jung, and S.~Yun, ``{Consistent $N_{\rm eff}$ fitting in big
  bang nucleosynthesis analysis},''
  \href{http://arxiv.org/abs/2507.23354}{{\ttfamily arXiv:2507.23354
  [hep-ph]}}.

\bibitem{Escudero:2025avx}
M.~Escudero, C.~Garcia-Perez, and M.~Ovchynnikov, ``{Nucleosynthesis and CMB
  bounds on photophilic ALPs: a fresh look},''
  \href{http://arxiv.org/abs/2511.00157}{{\ttfamily arXiv:2511.00157
  [hep-ph]}}.

\bibitem{Slatyer:2016qyl}
T.~R. Slatyer and C.-L. Wu, ``{General Constraints on Dark Matter Decay from
  the Cosmic Microwave Background},''
  \href{http://dx.doi.org/10.1103/PhysRevD.95.023010}{{\em Phys. Rev. D}
  {\bfseries 95} no.~2, (2017) 023010},
  \href{http://arxiv.org/abs/1610.06933}{{\ttfamily arXiv:1610.06933
  [astro-ph.CO]}}.

\bibitem{Langhoff:2022bij}
K.~Langhoff, N.~J. Outmezguine, and N.~L. Rodd, ``{Irreducible Axion
  Background},'' \href{http://dx.doi.org/10.1103/PhysRevLett.129.241101}{{\em
  Phys. Rev. Lett.} {\bfseries 129} no.~24, (2022) 241101},
  \href{http://arxiv.org/abs/2209.06216}{{\ttfamily arXiv:2209.06216
  [hep-ph]}}.

\bibitem{Cang:2020exa}
J.~Cang, Y.~Gao, and Y.-Z. Ma, ``{Probing dark matter with future CMB
  measurements},'' \href{http://dx.doi.org/10.1103/PhysRevD.102.103005}{{\em
  Phys. Rev. D} {\bfseries 102} no.~10, (2020) 103005},
  \href{http://arxiv.org/abs/2002.03380}{{\ttfamily arXiv:2002.03380
  [astro-ph.CO]}}.

\bibitem{Bolliet:2020vqu}
B.~Bolliet, J.~Chluba, and R.~Battye, ``Spectral distortions of the cmb and the
  history of the universe,''
  \href{http://dx.doi.org/10.1093/mnras/stab2344}{{\em Mon. Not. Roy. Astron.
  Soc.} {\bfseries 507} no.~3, (2021) 3148--3162},
  \href{http://arxiv.org/abs/2012.07292}{{\ttfamily arXiv:2012.07292
  [astro-ph.CO]}}.

\bibitem{Balazs:2022tjl}
C.~Bal{\'a}zs {\em et~al.}, ``Cosmological constraints on decaying dark
  photons,'' \href{http://arxiv.org/abs/2205.13549}{{\ttfamily arXiv:2205.13549
  [astro-ph.CO]}}. Preprint, submitted to journal.

\bibitem{Poulin:2016anj}
V.~Poulin, J.~Lesgourgues, and P.~D. Serpico, ``Cosmological constraints on
  exotic injection of electromagnetic energy,''
  \href{http://dx.doi.org/10.1088/1475-7516/2017/03/043}{{\em JCAP} {\bfseries
  03} (2017) 043}, \href{http://arxiv.org/abs/1610.10051}{{\ttfamily
  arXiv:1610.10051 [astro-ph.CO]}}.

\bibitem{Planck:2018vyg}
{\bfseries Planck} Collaboration, N.~Aghanim {\em et~al.}, ``{Planck 2018
  results. VI. Cosmological parameters},''
  \href{http://dx.doi.org/10.1051/0004-6361/201833910}{{\em Astron. Astrophys.}
  {\bfseries 641} (2020) A6}, \href{http://arxiv.org/abs/1807.06209}{{\ttfamily
  arXiv:1807.06209 [astro-ph.CO]}}. [Erratum: Astron.Astrophys. 652, C4
  (2021)].

\bibitem{Kawasaki:1994sc}
M.~Kawasaki and T.~Moroi, ``{Electromagnetic cascade in the early universe and
  its application to the big bang nucleosynthesis},''
  \href{http://dx.doi.org/10.1086/176324}{{\em Astrophys. J.} {\bfseries 452}
  (1995) 506}, \href{http://arxiv.org/abs/astro-ph/9412055}{{\ttfamily
  arXiv:astro-ph/9412055}}.

\bibitem{Berezinsky:1990qxi}
V.~S. Berezinsky, S.~V. Bulanov, V.~A. Dogiel, and V.~S. Ptuskin, {\em
  {Astrophysics of cosmic rays}}.
\newblock 1990.

\bibitem{Protheroe:1994dt}
R.~J. Protheroe, T.~Stanev, and V.~S. Berezinsky, ``{Electromagnetic cascades
  and cascade nucleosynthesis in the early universe},''
  \href{http://dx.doi.org/10.1103/PhysRevD.51.4134}{{\em Phys. Rev. D}
  {\bfseries 51} (1995) 4134--4144},
  \href{http://arxiv.org/abs/astro-ph/9409004}{{\ttfamily
  arXiv:astro-ph/9409004}}.

\bibitem{Pisanti:2007hk}
O.~Pisanti, A.~Cirillo, S.~Esposito, F.~Iocco, G.~Mangano, G.~Miele, and P.~D.
  Serpico, ``{PArthENoPE: Public Algorithm Evaluating the Nucleosynthesis of
  Primordial Elements},''
  \href{http://dx.doi.org/10.1016/j.cpc.2008.02.015}{{\em Comput. Phys.
  Commun.} {\bfseries 178} (2008) 956--971},
  \href{http://arxiv.org/abs/0705.0290}{{\ttfamily arXiv:0705.0290
  [astro-ph]}}.

\bibitem{Consiglio:2017pot}
R.~Consiglio, P.~F. de~Salas, G.~Mangano, G.~Miele, S.~Pastor, and O.~Pisanti,
  ``{PArthENoPE reloaded},''
  \href{http://dx.doi.org/10.1016/j.cpc.2018.06.022}{{\em Comput. Phys.
  Commun.} {\bfseries 233} (2018) 237--242},
  \href{http://arxiv.org/abs/1712.04378}{{\ttfamily arXiv:1712.04378
  [astro-ph.CO]}}.

\bibitem{Gariazzo:2021iiu}
S.~Gariazzo, P.~F.~de Salas, O.~Pisanti, and R.~Consiglio, ``{PArthENoPE
  revolutions},'' \href{http://dx.doi.org/10.1016/j.cpc.2021.108205}{{\em
  Comput. Phys. Commun.} {\bfseries 271} (2022) 108205},
  \href{http://arxiv.org/abs/2103.05027}{{\ttfamily arXiv:2103.05027
  [astro-ph.IM]}}.

\bibitem{Pospelov:2010cw}
M.~Pospelov and J.~Pradler, ``{Metastable GeV-scale particles as a solution to
  the cosmological lithium problem},''
  \href{http://dx.doi.org/10.1103/PhysRevD.82.103514}{{\em Phys. Rev. D}
  {\bfseries 82} (2010) 103514},
  \href{http://arxiv.org/abs/1006.4172}{{\ttfamily arXiv:1006.4172 [hep-ph]}}.

\bibitem{Fradette:2014sza}
A.~Fradette, M.~Pospelov, J.~Pradler, and A.~Ritz, ``{Cosmological Constraints
  on Very Dark Photons},''
  \href{http://dx.doi.org/10.1103/PhysRevD.90.035022}{{\em Phys. Rev. D}
  {\bfseries 90} no.~3, (2014) 035022},
  \href{http://arxiv.org/abs/1407.0993}{{\ttfamily arXiv:1407.0993 [hep-ph]}}.

\bibitem{Cyburt:2002uv}
R.~H. Cyburt, J.~R. Ellis, B.~D. Fields, and K.~A. Olive, ``{Updated
  nucleosynthesis constraints on unstable relic particles},''
  \href{http://dx.doi.org/10.1103/PhysRevD.67.103521}{{\em Phys. Rev. D}
  {\bfseries 67} (2003) 103521},
  \href{http://arxiv.org/abs/astro-ph/0211258}{{\ttfamily
  arXiv:astro-ph/0211258}}.

\bibitem{ParticleDataGroup:2024cfk}
{\bfseries Particle Data Group} Collaboration, S.~Navas {\em et~al.}, ``{Review
  of particle physics},''
  \href{http://dx.doi.org/10.1103/PhysRevD.110.030001}{{\em Phys. Rev. D}
  {\bfseries 110} no.~3, (2024) 030001}.

\bibitem{Pospelov:2008jk}
M.~Pospelov, A.~Ritz, and M.~B. Voloshin, ``{Bosonic super-WIMPs as keV-scale
  dark matter},'' \href{http://dx.doi.org/10.1103/PhysRevD.78.115012}{{\em
  Phys. Rev. D} {\bfseries 78} (2008) 115012},
  \href{http://arxiv.org/abs/0807.3279}{{\ttfamily arXiv:0807.3279 [hep-ph]}}.

\bibitem{McDermott:2017qcg}
S.~D. McDermott, H.~H. Patel, and H.~Ramani, ``{Dark Photon Decay Beyond The
  Euler-Heisenberg Limit},''
  \href{http://dx.doi.org/10.1103/PhysRevD.97.073005}{{\em Phys. Rev. D}
  {\bfseries 97} no.~7, (2018) 073005},
  \href{http://arxiv.org/abs/1705.00619}{{\ttfamily arXiv:1705.00619
  [hep-ph]}}.

\bibitem{Boyarsky:2006fg}
A.~Boyarsky, A.~Neronov, O.~Ruchayskiy, M.~Shaposhnikov, and I.~Tkachev,
  ``{Where to find a dark matter sterile neutrino?},''
  \href{http://dx.doi.org/10.1103/PhysRevLett.97.261302}{{\em Phys. Rev. Lett.}
  {\bfseries 97} (2006) 261302},
  \href{http://arxiv.org/abs/astro-ph/0603660}{{\ttfamily
  arXiv:astro-ph/0603660}}.

\bibitem{Boyarsky:2006ag}
A.~Boyarsky, J.~Nevalainen, and O.~Ruchayskiy, ``{Constraints on the parameters
  of radiatively decaying dark matter from the dark matter halo of the Milky
  Way and Ursa Minor},''
  \href{http://dx.doi.org/10.1051/0004-6361:20066774}{{\em Astron. Astrophys.}
  {\bfseries 471} (2007) 51--57},
  \href{http://arxiv.org/abs/astro-ph/0610961}{{\ttfamily
  arXiv:astro-ph/0610961}}.

\bibitem{Boyarsky:2007ay}
A.~Boyarsky, D.~Iakubovskyi, O.~Ruchayskiy, and V.~Savchenko, ``{Constraints on
  decaying Dark Matter from XMM-Newton observations of M31},''
  \href{http://dx.doi.org/10.1111/j.1365-2966.2008.13266.x}{{\em Mon. Not. Roy.
  Astron. Soc.} {\bfseries 387} (2008) 1361},
  \href{http://arxiv.org/abs/0709.2301}{{\ttfamily arXiv:0709.2301
  [astro-ph]}}.

\bibitem{Foster:2021ngm}
J.~W. Foster, M.~Kongsore, C.~Dessert, Y.~Park, N.~L. Rodd, K.~Cranmer, and
  B.~R. Safdi, ``{Deep Search for Decaying Dark Matter with XMM-Newton
  Blank-Sky Observations},''
  \href{http://dx.doi.org/10.1103/PhysRevLett.127.051101}{{\em Phys. Rev.
  Lett.} {\bfseries 127} no.~5, (2021) 051101},
  \href{http://arxiv.org/abs/2102.02207}{{\ttfamily arXiv:2102.02207
  [astro-ph.CO]}}.

\bibitem{Ng:2019gch}
K.~C.~Y. Ng, B.~M. Roach, K.~Perez, J.~F. Beacom, S.~Horiuchi, R.~Krivonos, and
  D.~R. Wik, ``{New Constraints on Sterile Neutrino Dark Matter from $NuSTAR$
  M31 Observations},'' \href{http://dx.doi.org/10.1103/PhysRevD.99.083005}{{\em
  Phys. Rev. D} {\bfseries 99} (2019) 083005},
  \href{http://arxiv.org/abs/1901.01262}{{\ttfamily arXiv:1901.01262
  [astro-ph.HE]}}.

\bibitem{Roach:2019ctw}
B.~M. Roach, K.~C.~Y. Ng, K.~Perez, J.~F. Beacom, S.~Horiuchi, R.~Krivonos, and
  D.~R. Wik, ``{NuSTAR Tests of Sterile-Neutrino Dark Matter: New Galactic
  Bulge Observations and Combined Impact},''
  \href{http://dx.doi.org/10.1103/PhysRevD.101.103011}{{\em Phys. Rev. D}
  {\bfseries 101} no.~10, (2020) 103011},
  \href{http://arxiv.org/abs/1908.09037}{{\ttfamily arXiv:1908.09037
  [astro-ph.HE]}}.

\bibitem{Roach:2023}
B.~M. Roach, S.~Rossland, K.~C.~Y. Ng, K.~Perez, J.~F. Beacom, B.~W.
  Grefenstette, S.~Horiuchi, R.~Krivonos, and D.~R. Wik, ``Long-exposure nustar
  constraints on decaying dark matter in the galactic halo,''
  \href{http://dx.doi.org/10.1103/PhysRevD.107.023009}{{\em Phys. Rev. D}
  {\bfseries 107} (Jan, 2023) 023009}.
  \url{https://link.aps.org/doi/10.1103/PhysRevD.107.023009}.

\bibitem{Calore:2022pks}
F.~Calore, A.~Dekker, P.~D. Serpico, and T.~Siegert, ``{Constraints on light
  decaying dark matter candidates from 16~yr of INTEGRAL/SPI observations},''
  \href{http://dx.doi.org/10.1093/mnras/stad457}{{\em Mon. Not. Roy. Astron.
  Soc.} {\bfseries 520} no.~3, (2023) 4167--4172},
  \href{http://arxiv.org/abs/2209.06299}{{\ttfamily arXiv:2209.06299
  [hep-ph]}}.

\bibitem{Linden:2024fby}
T.~Linden, T.~T.~Q. Nguyen, and T.~M.~P. Tait, ``{X-ray constraints on dark
  photon tridents},'' \href{http://dx.doi.org/10.1103/37gn-x3y1}{{\em Phys.
  Rev. D} {\bfseries 112} no.~2, (2025) 023026},
  \href{http://arxiv.org/abs/2406.19445}{{\ttfamily arXiv:2406.19445
  [hep-ph]}}.

\bibitem{Cirelli:2010xx}
M.~Cirelli, G.~Corcella, A.~Hektor, G.~Hutsi, M.~Kadastik, P.~Panci, M.~Raidal,
  F.~Sala, and A.~Strumia, ``{PPPC 4 DM ID: A Poor Particle Physicist Cookbook
  for Dark Matter Indirect Detection},''
  \href{http://dx.doi.org/10.1088/1475-7516/2012/10/E01}{{\em JCAP} {\bfseries
  03} (2011) 051}, \href{http://arxiv.org/abs/1012.4515}{{\ttfamily
  arXiv:1012.4515 [hep-ph]}}. [Erratum: JCAP 10, E01 (2012)].

\bibitem{Hill:2018trh}
R.~Hill, K.~W. Masui, and D.~Scott, ``{The Spectrum of the Universe},''
  \href{http://dx.doi.org/10.1177/0003702818767133}{{\em Appl. Spectrosc.}
  {\bfseries 72} no.~5, (2018) 663--688},
  \href{http://arxiv.org/abs/1802.03694}{{\ttfamily arXiv:1802.03694
  [astro-ph.CO]}}.

\bibitem{BetheWilson1985_ShockRevival}
H.~A. Bethe and J.~R. Wilson, ``Revival of a stalled supernova shock by
  neutrino heating,'' \href{http://dx.doi.org/10.1086/163343}{{\em Astrophys.
  J.} {\bfseries 295} (1985) 14}.

\bibitem{Janka2012_ExplosionMechanisms}
H.-T. Janka, ``Explosion mechanisms of core-collapse supernovae,'' {\em Annu.
  Rev. Nucl. Part. Sci.} {\bfseries 62} (2012) 407,
  \href{http://arxiv.org/abs/1206.2503}{{\ttfamily arXiv:1206.2503
  [astro-ph.SR]}}.

\bibitem{BoccioliRoberti2024_CCSNePhysics}
L.~Boccioli and L.~Roberti, ``The physics of core-collapse supernovae:
  Explosion mechanism and explosive nucleosynthesis,'' {\em Universe}
  {\bfseries 10} (2024) 148, \href{http://arxiv.org/abs/2403.12942}{{\ttfamily
  arXiv:2403.12942 [astro-ph.SR]}}.

\bibitem{BurrowsVartanyan2021_CCSNeTheory}
A.~Burrows and D.~Vartanyan, ``Core-collapse supernova explosion theory,'' {\em
  Nature} {\bfseries 589} (2021) 29,
  \href{http://arxiv.org/abs/2009.14157}{{\ttfamily arXiv:2009.14157
  [astro-ph.HE]}}.

\bibitem{Stetina:2017ozh}
S.~Stetina, E.~Rrapaj, and S.~Reddy, ``{Photons in dense nuclear matter:
  Random-phase approximation},''
  \href{http://dx.doi.org/10.1103/PhysRevC.97.045801}{{\em Phys. Rev. C}
  {\bfseries 97} no.~4, (2018) 045801},
  \href{http://arxiv.org/abs/1712.05447}{{\ttfamily arXiv:1712.05447
  [astro-ph.HE]}}.

\bibitem{Rrapaj:2015wgs}
E.~Rrapaj and S.~Reddy, ``{Nucleon-nucleon bremsstrahlung of dark gauge bosons
  and revised supernova constraints},''
  \href{http://dx.doi.org/10.1103/PhysRevC.94.045805}{{\em Phys. Rev. C}
  {\bfseries 94} no.~4, (2016) 045805},
  \href{http://arxiv.org/abs/1511.09136}{{\ttfamily arXiv:1511.09136
  [nucl-th]}}.

\bibitem{Shin:2021bvz}
C.~S. Shin and S.~Yun, ``{Dark gauge boson production from neutron stars via
  nucleon-nucleon bremsstrahlung},''
  \href{http://dx.doi.org/10.1007/JHEP02(2022)133}{{\em JHEP} {\bfseries 02}
  (2022) 133}, \href{http://arxiv.org/abs/2110.03362}{{\ttfamily
  arXiv:2110.03362 [hep-ph]}}.

\bibitem{Low:1958sn}
F.~E. Low, ``{Bremsstrahlung of very low-energy quanta in elementary particle
  collisions},'' \href{http://dx.doi.org/10.1103/PhysRev.110.974}{{\em Phys.
  Rev.} {\bfseries 110} (1958) 974--977}.

\bibitem{Nyman:1973gw}
E.~M. Nyman, ``{BREMSSTRAHLUNG IN NUCLEON-NUCLEON COLLISIONS},''
  \href{http://dx.doi.org/10.1016/0370-1573(74)90024-6}{{\em Phys. Rept.}
  {\bfseries 9} (1974) 179}.

\bibitem{Chang:2016ntp}
J.~H. Chang, R.~Essig, and S.~D. McDermott, ``{Revisiting Supernova 1987A
  Constraints on Dark Photons},''
  \href{http://dx.doi.org/10.1007/JHEP01(2017)107}{{\em JHEP} {\bfseries 01}
  (2017) 107}, \href{http://arxiv.org/abs/1611.03864}{{\ttfamily
  arXiv:1611.03864 [hep-ph]}}.

\bibitem{Landau:1953um}
L.~D. Landau and I.~Pomeranchuk, ``{Limits of applicability of the theory of
  bremsstrahlung electrons and pair production at high-energies},'' {\em Dokl.
  Akad. Nauk Ser. Fiz.} {\bfseries 92} (1953) 535--536.

\bibitem{Migdal:1956tc}
A.~B. Migdal, ``{Bremsstrahlung and pair production in condensed media at
  high-energies},'' \href{http://dx.doi.org/10.1103/PhysRev.103.1811}{{\em
  Phys. Rev.} {\bfseries 103} (1956) 1811--1820}.

\bibitem{Raffelt:1991pw}
G.~Raffelt and D.~Seckel, ``{Multiple scattering suppression of the
  bremsstrahlung emission of neutrinos and axions in supernovae},''
  \href{http://dx.doi.org/10.1103/PhysRevLett.67.2605}{{\em Phys. Rev. Lett.}
  {\bfseries 67} (1991) 2605--2608}.

\bibitem{Bollig:2017lki}
R.~Bollig, H.~T. Janka, A.~Lohs, G.~Martinez-Pinedo, C.~J. Horowitz, and
  T.~Melson, ``{Muon Creation in Supernova Matter Facilitates Neutrino-driven
  Explosions},'' \href{http://dx.doi.org/10.1103/PhysRevLett.119.242702}{{\em
  Phys. Rev. Lett.} {\bfseries 119} no.~24, (2017) 242702},
  \href{http://arxiv.org/abs/1706.04630}{{\ttfamily arXiv:1706.04630
  [astro-ph.HE]}}.

\bibitem{Fore:2019wib}
B.~Fore and S.~Reddy, ``{Pions in hot dense matter and their astrophysical
  implications},'' \href{http://dx.doi.org/10.1103/PhysRevC.101.035809}{{\em
  Phys. Rev. C} {\bfseries 101} no.~3, (2020) 035809},
  \href{http://arxiv.org/abs/1911.02632}{{\ttfamily arXiv:1911.02632
  [astro-ph.HE]}}.

\bibitem{Shin:2022ulh}
C.~S. Shin and S.~Yun, ``{Dark gauge boson emission from supernova pions},''
  \href{http://dx.doi.org/10.1103/PhysRevD.108.055014}{{\em Phys. Rev. D}
  {\bfseries 108} no.~5, (2023) 055014},
  \href{http://arxiv.org/abs/2211.15677}{{\ttfamily arXiv:2211.15677
  [hep-ph]}}.

\bibitem{Caputo:2021rux}
A.~Caputo, G.~Raffelt, and E.~Vitagliano, ``{Muonic boson limits: Supernova
  redux},'' \href{http://dx.doi.org/10.1103/PhysRevD.105.035022}{{\em Phys.
  Rev. D} {\bfseries 105} no.~3, (2022) 035022},
  \href{http://arxiv.org/abs/2109.03244}{{\ttfamily arXiv:2109.03244
  [hep-ph]}}.

\bibitem{Caputo:2025aac}
A.~Caputo, H.-T. Janka, G.~Raffelt, and S.~Yun, ``{Dark Photons can Prevent
  Core-Collapse Supernova Explosions},''
  \href{http://arxiv.org/abs/2502.01731}{{\ttfamily arXiv:2502.01731
  [hep-ph]}}.

\bibitem{Bollig:2020phc}
R.~Bollig, N.~Yadav, D.~Kresse, H.~T. Janka, B.~M{\"u}ller, and A.~Heger,
  ``{Self-consistent 3D Supernova Models From {\ensuremath{-}}7 Minutes to +7
  s: A 1-bethe Explosion of a {\ensuremath{\sim}}19 $M_\odot$ Progenitor},''
  \href{http://dx.doi.org/10.3847/1538-4357/abf82e}{{\em Astrophys. J.}
  {\bfseries 915} no.~1, (2021) 28},
  \href{http://arxiv.org/abs/2010.10506}{{\ttfamily arXiv:2010.10506
  [astro-ph.HE]}}.

\bibitem{Fiorillo:2022cdq}
D.~F.~G. Fiorillo, G.~G. Raffelt, and E.~Vitagliano, ``{Strong Supernova 1987A
  Constraints on Bosons Decaying to Neutrinos},''
  \href{http://dx.doi.org/10.1103/PhysRevLett.131.021001}{{\em Phys. Rev.
  Lett.} {\bfseries 131} no.~2, (2023) 021001},
  \href{http://arxiv.org/abs/2209.11773}{{\ttfamily arXiv:2209.11773
  [hep-ph]}}.

\bibitem{Caputo:2024oqc}
A.~Caputo and G.~Raffelt, ``{Astrophysical Axion Bounds: The 2024 Edition},''
  \href{http://dx.doi.org/10.22323/1.454.0041}{{\em PoS} {\bfseries
  COSMICWISPers} (2024) 041}, \href{http://arxiv.org/abs/2401.13728}{{\ttfamily
  arXiv:2401.13728 [hep-ph]}}.

\bibitem{Bionta:1987qt}
R.~M. Bionta {\em et~al.}, ``{Observation of a Neutrino Burst in Coincidence
  with Supernova SN 1987a in the Large Magellanic Cloud},''
  \href{http://dx.doi.org/10.1103/PhysRevLett.58.1494}{{\em Phys. Rev. Lett.}
  {\bfseries 58} (1987) 1494}.

\bibitem{Kamiokande-II:1987idp}
{\bfseries Kamiokande-II} Collaboration, K.~Hirata {\em et~al.}, ``{Observation
  of a Neutrino Burst from the Supernova SN 1987a},''
  \href{http://dx.doi.org/10.1103/PhysRevLett.58.1490}{{\em Phys. Rev. Lett.}
  {\bfseries 58} (1987) 1490--1493}.

\bibitem{Alekseev:1987ej}
E.~N. Alekseev, L.~N. Alekseeva, V.~I. Volchenko, and I.~V. Krivosheina,
  ``{Possible Detection of a Neutrino Signal on 23 February 1987 at the Baksan
  Underground Scintillation Telescope of the Institute of Nuclear Research},''
  {\em JETP Lett.} {\bfseries 45} (1987) 589--592.

\bibitem{Raffelt:1996wa}
G.~G. Raffelt, {\em Stars as Laboratories for Fundamental Physics: The
  Astrophysics of Neutrinos, Axions, and Other Weakly Interacting Particles}.
\newblock University of Chicago Press, Chicago, USA, 1996.

\bibitem{Bollig:2020xdr}
R.~Bollig, W.~DeRocco, P.~W. Graham, and H.-T. Janka, ``{Muons in Supernovae:
  Implications for the Axion-Muon Coupling},''
  \href{http://dx.doi.org/10.1103/PhysRevLett.125.051104}{{\em Phys. Rev.
  Lett.} {\bfseries 125} no.~5, (2020) 051104},
  \href{http://arxiv.org/abs/2005.07141}{{\ttfamily arXiv:2005.07141
  [hep-ph]}}. [Erratum: Phys.Rev.Lett. 126, 189901 (2021)].

\bibitem{CCSNarchive}
``Garching core-collapse supernova research archive.''
  \url{https://wwwmpa.mpa-garching.mpg.de/ccsnarchive/}.

\bibitem{Steiner:2012rk}
A.~W. Steiner, M.~Hempel, and T.~Fischer, ``Core-collapse supernova equations
  of state based on neutron star observations,''
  \href{http://dx.doi.org/10.1088/0004-637X/774/1/17}{{\em Astrophys. J.}
  {\bfseries 774} (2013) 17}, \href{http://arxiv.org/abs/1207.2184}{{\ttfamily
  arXiv:1207.2184 [astro-ph.SR]}}.

\bibitem{Sung:2019xie}
A.~Sung, H.~Tu, and M.-R. Wu, ``{New constraint from supernova explosions on
  light particles beyond the Standard Model},''
  \href{http://dx.doi.org/10.1103/PhysRevD.99.121305}{{\em Phys. Rev. D}
  {\bfseries 99} no.~12, (2019) 121305},
  \href{http://arxiv.org/abs/1903.07923}{{\ttfamily arXiv:1903.07923
  [hep-ph]}}.

\bibitem{Caputo:2022mah}
A.~Caputo, H.-T. Janka, G.~Raffelt, and E.~Vitagliano, ``{Low-Energy Supernovae
  Severely Constrain Radiative Particle Decays},''
  \href{http://dx.doi.org/10.1103/PhysRevLett.128.221103}{{\em Phys. Rev.
  Lett.} {\bfseries 128} no.~22, (2022) 221103},
  \href{http://arxiv.org/abs/2201.09890}{{\ttfamily arXiv:2201.09890
  [astro-ph.HE]}}.

\bibitem{Falk:1978kf}
S.~W. Falk and D.~N. Schramm, ``{Limits From Supernovae on Neutrino Radiative
  Lifetimes},'' \href{http://dx.doi.org/10.1016/0370-2693(78)90417-3}{{\em
  Phys. Lett. B} {\bfseries 79} (1978) 511}.

\bibitem{Pastorello:2003tc}
A.~Pastorello {\em et~al.}, ``{Low luminosity type II supernovae: spectroscopic
  and photometric evolution},''
  \href{http://dx.doi.org/10.1111/j.1365-2966.2004.07173.x}{{\em Mon. Not. Roy.
  Astron. Soc.} {\bfseries 347} (2004) 74},
  \href{http://arxiv.org/abs/astro-ph/0309264}{{\ttfamily
  arXiv:astro-ph/0309264}}.

\bibitem{Spiro_2014}
S.~Spiro, A.~Pastorello, M.~L. Pumo, L.~Zampieri, M.~Turatto, S.~J. Smartt,
  S.~Benetti, E.~Cappellaro, S.~Valenti, I.~Agnoletto, G.~Altavilla, T.~Aoki,
  E.~Brocato, E.~M. Corsini, A.~Di~Cianno, N.~Elias-Rosa, M.~Hamuy, K.~Enya,
  M.~Fiaschi, G.~Folatelli, S.~Desidera, A.~Harutyunyan, D.~A. Howell,
  A.~Kawka, Y.~Kobayashi, B.~Leibundgut, T.~Minezaki, H.~Navasardyan,
  K.~Nomoto, S.~Mattila, A.~Pietrinferni, G.~Pignata, G.~Raimondo, M.~Salvo,
  B.~P. Schmidt, J.~Sollerman, J.~Spyromilio, S.~Taubenberger, G.~Valentini,
  S.~Vennes, and Y.~Yoshii, ``Low luminosity type ii supernovae – ii.
  pointing towards moderate mass precursors,''
  \href{http://dx.doi.org/10.1093/mnras/stu156}{{\em Monthly Notices of the
  Royal Astronomical Society} {\bfseries 439} no.~3, (Feb., 2014) 2873–2892}.
  \url{http://dx.doi.org/10.1093/mnras/stu156}.

\bibitem{Fiorillo:2025yzf}
D.~F.~G. Fiorillo, T.~Pitik, and E.~Vitagliano, ``{Energy transfer by feebly
  interacting particles in supernovae: the trapping regime},''
  \href{http://arxiv.org/abs/2503.13653}{{\ttfamily arXiv:2503.13653
  [hep-ph]}}.

\bibitem{1979ICRC....5..135R}
J.~M. {Ryan}, L.~L. {Chupp}, D.~J. {Forrest}, M.~L. {Cherry}, I.~U. {Gleske},
  L.~{Rieger}, G.~{Kanbach}, K.~{Pinkau}, C.~{Reppin}, G.~{Share}, R.~L.
  {Kinzer}, W.~N. {Johnson}, and J.~D. {Kurfess}, ``{The Gamma Ray Spectrometer
  for the Solar Maximum Mission},'' in {\em International Cosmic Ray
  Conference}, vol.~5 of {\em International Cosmic Ray Conference}, p.~135.
\newblock Jan., 1979.

\bibitem{Oberauer:1993yr}
L.~Oberauer, C.~Hagner, G.~Raffelt, and E.~Rieger, ``{Supernova bounds on
  neutrino radiative decays},''
  \href{http://dx.doi.org/10.1016/0927-6505(93)90004-W}{{\em Astropart. Phys.}
  {\bfseries 1} (1993) 377--386}.

\bibitem{DeRocco:2019njg}
W.~DeRocco, P.~W. Graham, D.~Kasen, G.~Marques-Tavares, and S.~Rajendran,
  ``{Observable signatures of dark photons from supernovae},''
  \href{http://dx.doi.org/10.1007/JHEP02(2019)171}{{\em JHEP} {\bfseries 02}
  (2019) 171}, \href{http://arxiv.org/abs/1901.08596}{{\ttfamily
  arXiv:1901.08596 [hep-ph]}}.

\bibitem{Candon:2025ypl}
F.~R. Cand{\'o}n, D.~F.~G. Fiorillo, H.-T. Janka, B.~F.~A. van Baal, and
  E.~Vitagliano, ``{Small Progenitors, Large Couplings: Type Ic Supernova
  Constraints on Radiatively Decaying Particles},''
  \href{http://arxiv.org/abs/2509.18253}{{\ttfamily arXiv:2509.18253
  [hep-ph]}}.

\bibitem{Kozyreva:2024ksv}
A.~Kozyreva, A.~Caputo, P.~Baklanov, A.~Mironov, and H.-T. Janka, ``{SN
  2023ixf: An average-energy explosion with circumstellar medium and a
  precursor},'' \href{http://dx.doi.org/10.1051/0004-6361/202452758}{{\em
  Astron. Astrophys.} {\bfseries 694} (2025) A319},
  \href{http://arxiv.org/abs/2410.19939}{{\ttfamily arXiv:2410.19939
  [astro-ph.HE]}}.

\bibitem{Prantzos:2010wi}
N.~Prantzos {\em et~al.}, ``{The 511 keV emission from positron annihilation in
  the Galaxy},'' \href{http://dx.doi.org/10.1103/RevModPhys.83.1001}{{\em Rev.
  Mod. Phys.} {\bfseries 83} (2011) 1001--1056},
  \href{http://arxiv.org/abs/1009.4620}{{\ttfamily arXiv:1009.4620
  [astro-ph.HE]}}.

\bibitem{Siegert:2015knp}
T.~Siegert, R.~Diehl, G.~Khachatryan, M.~G.~H. Krause, F.~Guglielmetti,
  J.~Greiner, A.~W. Strong, and X.~Zhang, ``Gamma-ray spectroscopy of positron
  annihilation in the milky way,''
  \href{http://dx.doi.org/10.1051/0004-6361/201527510}{{\em Astron. Astrophys.}
  {\bfseries 586} (2016) A84},
  \href{http://arxiv.org/abs/1512.00325}{{\ttfamily arXiv:1512.00325
  [astro-ph.HE]}}.

\bibitem{Siegert:2019tus}
T.~Siegert, R.~M. Crocker, R.~Diehl, M.~G.~H. Krause, F.~H. Panther, M.~M.~M.
  Pleintinger, and C.~Weinberger, ``Constraints on positron annihilation
  kinematics in the inner galaxy,''
  \href{http://dx.doi.org/10.1051/0004-6361/201935408}{{\em Astron. Astrophys.}
  {\bfseries 627} (2019) A126},
  \href{http://arxiv.org/abs/1906.00498}{{\ttfamily arXiv:1906.00498
  [astro-ph.HE]}}.

\bibitem{Strong:2005fr}
A.~W. Strong, R.~Diehl, H.~Halloin, V.~Schoenfelder, L.~Bouchet, P.~Mandrou,
  F.~Lebrun, and R.~Terrier, ``Gamma-ray continuum emission from the inner
  galactic region as observed with integral/spi,''
  \href{http://dx.doi.org/10.1051/0004-6361:20053631}{{\em Astron. Astrophys.}
  {\bfseries 444} (2005) 495--503},
  \href{http://arxiv.org/abs/astro-ph/0509290}{{\ttfamily
  arXiv:astro-ph/0509290 [astro-ph]}}.

\bibitem{Bouchet:2010dj}
L.~Bouchet, J.-P. Roques, and E.~Jourdain, ``On the morphology of the
  electron-positron annihilation emission as seen by spi/integral,''
  \href{http://dx.doi.org/10.1088/0004-637X/720/2/1772}{{\em Astrophys. J.}
  {\bfseries 720} (2010) 1772--1780},
  \href{http://arxiv.org/abs/1007.4753}{{\ttfamily arXiv:1007.4753
  [astro-ph.HE]}}.

\bibitem{Beacom:2005qv}
J.~F. Beacom and H.~Yuksel, ``{Stringent constraint on galactic positron
  production},'' \href{http://dx.doi.org/10.1103/PhysRevLett.97.071102}{{\em
  Phys. Rev. Lett.} {\bfseries 97} (2006) 071102},
  \href{http://arxiv.org/abs/astro-ph/0512411}{{\ttfamily
  arXiv:astro-ph/0512411}}.

\bibitem{Calore:2021lih}
F.~Calore, P.~Carenza, M.~Giannotti, J.~Jaeckel, G.~Lucente, L.~Mastrototaro,
  and A.~Mirizzi, ``{511~keV line constraints on feebly interacting particles
  from supernovae},'' \href{http://dx.doi.org/10.1103/PhysRevD.105.063026}{{\em
  Phys. Rev. D} {\bfseries 105} no.~6, (2022) 063026},
  \href{http://arxiv.org/abs/2112.08382}{{\ttfamily arXiv:2112.08382
  [hep-ph]}}.

\bibitem{Li:2010kd}
W.~Li, R.~Chornock, J.~Leaman, A.~V. Filippenko, D.~Poznanski, X.~Wang,
  M.~Ganeshalingam, and F.~Mannucci, ``{Nearby Supernova Rates from the Lick
  Observatory Supernova Search. III. The Rate-Size Relation, and the Rates as a
  Function of Galaxy Hubble Type and Colour},''
  \href{http://dx.doi.org/10.1111/j.1365-2966.2011.18162.x}{{\em Mon. Not. Roy.
  Astron. Soc.} {\bfseries 412} (2011) 1473},
  \href{http://arxiv.org/abs/1006.4613}{{\ttfamily arXiv:1006.4613
  [astro-ph.SR]}}.

\bibitem{DelaTorreLuque:2024zsr}
P.~De~la Torre~Luque, S.~Balaji, P.~Carenza, and L.~Mastrototaro,
  ``{{\ensuremath{\gamma}} rays from in-flight positron annihilation as a probe
  of new physics},'' \href{http://dx.doi.org/10.1103/PhysRevD.111.L061303}{{\em
  Phys. Rev. D} {\bfseries 111} no.~6, (2025) L061303},
  \href{http://arxiv.org/abs/2405.08482}{{\ttfamily arXiv:2405.08482
  [hep-ph]}}.

\bibitem{Balaji:2025alr}
S.~Balaji, P.~Carenza, P.~De~la Torre~Luque, A.~Lella, and L.~Mastrototaro,
  ``{In-flight positron annihilation as a probe of feebly interacting
  particles},'' \href{http://dx.doi.org/10.1103/PhysRevD.111.083053}{{\em Phys.
  Rev. D} {\bfseries 111} no.~8, (2025) 083053},
  \href{http://arxiv.org/abs/2501.07725}{{\ttfamily arXiv:2501.07725
  [hep-ph]}}.

\bibitem{Piran:1999kx}
T.~Piran, ``{Gamma-ray bursts and the fireball model},''
  \href{http://dx.doi.org/10.1016/S0370-1573(98)00127-6}{{\em Phys. Rept.}
  {\bfseries 314} (1999) 575--667},
  \href{http://arxiv.org/abs/astro-ph/9810256}{{\ttfamily
  arXiv:astro-ph/9810256}}.

\bibitem{Kazanas:2014mca}
D.~Kazanas, R.~N. Mohapatra, S.~Nussinov, V.~L. Teplitz, and Y.~Zhang,
  ``{Supernova Bounds on the Dark Photon Using its Electromagnetic Decay},''
  \href{http://dx.doi.org/10.1016/j.nuclphysb.2014.11.009}{{\em Nucl. Phys. B}
  {\bfseries 890} (2014) 17--29},
  \href{http://arxiv.org/abs/1410.0221}{{\ttfamily arXiv:1410.0221 [hep-ph]}}.

\bibitem{Diamond:2023scc}
M.~Diamond, D.~F.~G. Fiorillo, G.~Marques-Tavares, and E.~Vitagliano,
  ``{Axion-sourced fireballs from supernovae},''
  \href{http://dx.doi.org/10.1103/PhysRevD.107.103029}{{\em Phys. Rev. D}
  {\bfseries 107} no.~10, (2023) 103029},
  \href{http://arxiv.org/abs/2303.11395}{{\ttfamily arXiv:2303.11395
  [hep-ph]}}. [Erratum: Phys.Rev.D 108, 049902 (2023)].

\bibitem{Diamond:2023cto}
M.~Diamond, D.~F.~G. Fiorillo, G.~Marques-Tavares, I.~Tamborra, and
  E.~Vitagliano, ``{Multimessenger Constraints on Radiatively Decaying Axions
  from GW170817},''
  \href{http://dx.doi.org/10.1103/PhysRevLett.132.101004}{{\em Phys. Rev.
  Lett.} {\bfseries 132} no.~10, (2024) 101004},
  \href{http://arxiv.org/abs/2305.10327}{{\ttfamily arXiv:2305.10327
  [hep-ph]}}.

\bibitem{Diamond:2021ekg}
M.~D. Diamond and G.~Marques-Tavares, ``{\ensuremath{\gamma}-Ray Flashes from
  Dark Photons in Neutron Star Mergers},''
  \href{http://dx.doi.org/10.1103/PhysRevLett.128.211101}{{\em Phys. Rev.
  Lett.} {\bfseries 128} no.~21, (2022) 211101},
  \href{http://arxiv.org/abs/2106.03879}{{\ttfamily arXiv:2106.03879
  [hep-ph]}}.

\bibitem{LIGOScientific:2017vwq}
{\bfseries LIGO Scientific, Virgo} Collaboration, B.~P. Abbott {\em et~al.},
  ``{GW170817: Observation of Gravitational Waves from a Binary Neutron Star
  Inspiral},'' \href{http://dx.doi.org/10.1103/PhysRevLett.119.161101}{{\em
  Phys. Rev. Lett.} {\bfseries 119} no.~16, (2017) 161101},
  \href{http://arxiv.org/abs/1710.05832}{{\ttfamily arXiv:1710.05832 [gr-qc]}}.

\bibitem{Murguia-Berthier:2020tfs}
A.~Murguia-Berthier, E.~Ramirez-Ruiz, F.~De~Colle, A.~Janiuk, S.~Rosswog, and
  W.~H. Lee, ``{The Fate of the Merger Remnant in GW170817 and its Imprint on
  the Jet Structure},'' \href{http://dx.doi.org/10.3847/1538-4357/abd08e}{{\em
  Astrophys. J.} {\bfseries 908} no.~2, (2021) 152},
  \href{http://arxiv.org/abs/2007.12245}{{\ttfamily arXiv:2007.12245
  [astro-ph.HE]}}.

\bibitem{Madau:2014bja}
P.~Madau and M.~Dickinson, ``{Cosmic Star Formation History},''
  \href{http://dx.doi.org/10.1146/annurev-astro-081811-125615}{{\em Ann. Rev.
  Astron. Astrophys.} {\bfseries 52} (2014) 415--486},
  \href{http://arxiv.org/abs/1403.0007}{{\ttfamily arXiv:1403.0007
  [astro-ph.CO]}}.

\bibitem{Beacom:2010kk}
J.~F. Beacom, ``{The Diffuse Supernova Neutrino Background},''
  \href{http://dx.doi.org/10.1146/annurev.nucl.010909.083331}{{\em Ann. Rev.
  Nucl. Part. Sci.} {\bfseries 60} (2010) 439--462},
  \href{http://arxiv.org/abs/1004.3311}{{\ttfamily arXiv:1004.3311
  [astro-ph.HE]}}.

\bibitem{Bethe:1985sox}
H.~A. Bethe and J.~R. Wilson, ``{Revival of a stalled supernova shock by
  neutrino heating},'' \href{http://dx.doi.org/10.1086/163343}{{\em Astrophys.
  J.} {\bfseries 295} (1985) 14--23}.

\bibitem{Janka:2000bt}
H.~T. Janka, ``{Conditions for shock revival by neutrino heating in core
  collapse supernovae},''
  \href{http://dx.doi.org/10.1051/0004-6361:20010012}{{\em Astron. Astrophys.}
  {\bfseries 368} (2001) 527},
  \href{http://arxiv.org/abs/astro-ph/0008432}{{\ttfamily
  arXiv:astro-ph/0008432}}.

\bibitem{Essig:2010xa}
R.~Essig, P.~Schuster, and N.~Toro, ``Probing dark forces and light hidden
  sectors at low-energy e+e- colliders,''
  \href{http://dx.doi.org/10.1103/PhysRevD.80.015003}{{\em Phys. Rev. D}
  {\bfseries 80} (2010) 015003},
  \href{http://arxiv.org/abs/1001.2557}{{\ttfamily arXiv:1001.2557 [hep-ph]}}.

\bibitem{HPS:2018xkw}
C.~Moreno and others (HPS~Collaboration), ``Search for a dark photon in
  electroproduced e+e− pairs,''
  \href{http://dx.doi.org/10.1103/PhysRevLett.121.041802}{{\em Phys. Rev.
  Lett.} {\bfseries 121} (2018) 041802},
  \href{http://arxiv.org/abs/1807.11530}{{\ttfamily arXiv:1807.11530
  [hep-ex]}}.

\bibitem{Bjorken:1988as}
J.~D. Bjorken, S.~Ecklund, W.~R. Nelson, A.~Abashian, C.~Church, B.~Lu, L.~W.
  Mo, T.~A. Nunamaker, and P.~Rassmann, ``Search for neutral metastable
  penetrating particles produced in the slac beam dump,''
  \href{http://dx.doi.org/10.1103/PhysRevD.38.3375}{{\em Phys. Rev. D}
  {\bfseries 38} (1988) 3375}.

\bibitem{Riordan:1987aw}
E.~M. e.~a. Riordan, ``A search for short lived axions in an electron beam dump
  experiment,'' \href{http://dx.doi.org/10.1103/PhysRevLett.59.755}{{\em Phys.
  Rev. Lett.} {\bfseries 59} (1987) 755}.

\bibitem{CMS:2023hwl}
{\bfseries CMS} Collaboration, A.~Hayrapetyan {\em et~al.}, ``{Search for
  direct production of GeV-scale resonances decaying to a pair of muons in
  proton-proton collisions at $ \sqrt{s} $ = 13 TeV},''
  \href{http://dx.doi.org/10.1007/JHEP12(2023)070}{{\em JHEP} {\bfseries 12}
  (2023) 070}, \href{http://arxiv.org/abs/2309.16003}{{\ttfamily
  arXiv:2309.16003 [hep-ex]}}.

\bibitem{CMS:2021sch}
{\bfseries CMS} Collaboration, A.~Tumasyan {\em et~al.}, ``{Search for
  long-lived particles decaying into muon pairs in proton-proton collisions at
  $ \sqrt{s} $ = 13 TeV collected with a dedicated high-rate data stream},''
  \href{http://dx.doi.org/10.1007/JHEP04(2022)062}{{\em JHEP} {\bfseries 04}
  (2022) 062}, \href{http://arxiv.org/abs/2112.13769}{{\ttfamily
  arXiv:2112.13769 [hep-ex]}}.

\bibitem{ATLAS:2023cjw}
{\bfseries ATLAS} Collaboration, G.~Aad {\em et~al.}, ``{Search for light
  long-lived neutral particles from Higgs boson decays via vector-boson-fusion
  production from pp collisions at $\sqrt{s}=13$ TeV with the ATLAS
  detector},'' \href{http://dx.doi.org/10.1140/epjc/s10052-024-12902-7}{{\em
  Eur. Phys. J. C} {\bfseries 84} no.~7, (2024) 719},
  \href{http://arxiv.org/abs/2311.18298}{{\ttfamily arXiv:2311.18298
  [hep-ex]}}.

\bibitem{Ilten:2015hya}
P.~Ilten, J.~Thaler, M.~Williams, and W.~Xue, ``Dark photons from charm mesons
  at lhcb,'' \href{http://dx.doi.org/10.1103/PhysRevD.92.115017}{{\em Phys.
  Rev. D} {\bfseries 92} no.~11, (2015) 115017},
  \href{http://arxiv.org/abs/1509.06765}{{\ttfamily arXiv:1509.06765
  [hep-ex]}}.

\bibitem{Ilten:2016tkc}
P.~Ilten, Y.~Soreq, J.~Thaler, M.~Williams, and W.~Xue, ``Proposed inclusive
  dark photon search at lhcb,''
  \href{http://dx.doi.org/10.1103/PhysRevLett.116.251803}{{\em Phys. Rev.
  Lett.} {\bfseries 116} no.~25, (2016) 251803},
  \href{http://arxiv.org/abs/1603.08926}{{\ttfamily arXiv:1603.08926
  [hep-ph]}}.

\bibitem{Jaegle:2015sxl}
I.~B. Jaegle, ``Search for the dark photon and the dark higgs boson at belle,''
  \href{http://dx.doi.org/10.1103/PhysRevLett.114.211801}{{\em Phys. Rev.
  Lett.} {\bfseries 114} no.~21, (2015) 211801},
  \href{http://arxiv.org/abs/1502.00084}{{\ttfamily arXiv:1502.00084
  [hep-ex]}}.

\bibitem{KLOE-2:2015nli}
{\bfseries KLOE-2} Collaboration, A.~Anastasi {\em et~al.}, ``Search for dark
  higgsstrahlung in $e^+e^- \to \mu^+\mu^-$ and missing energy events with the
  kloe experiment,''
  \href{http://dx.doi.org/10.1016/j.physletb.2015.06.015}{{\em Phys. Lett. B}
  {\bfseries 747} (2015) 365--372},
  \href{http://arxiv.org/abs/1501.06795}{{\ttfamily arXiv:1501.06795
  [hep-ex]}}.

\bibitem{Belle-II:2022jyy}
{\bfseries Belle-II} Collaboration, F.~Abudinen {\em et~al.}, ``Search for a
  dark photon and an invisible dark higgs boson in $\mu^+\mu^-$ and missing
  energy final states with the belle ii experiment,''
  \href{http://dx.doi.org/10.1103/PhysRevLett.130.071804}{{\em Phys. Rev.
  Lett.} {\bfseries 130} no.~7, (2023) 071804},
  \href{http://arxiv.org/abs/2207.00509}{{\ttfamily arXiv:2207.00509
  [hep-ex]}}.

\bibitem{Chang:2018rso}
J.~H. Chang, R.~Essig, and S.~D. McDermott, ``{Supernova 1987A Constraints on
  Sub-GeV Dark Sectors, Millicharged Particles, the QCD Axion, and an
  Axion-like Particle},'' \href{http://dx.doi.org/10.1007/JHEP09(2018)051}{{\em
  JHEP} {\bfseries 09} (2018) 051},
  \href{http://arxiv.org/abs/1803.00993}{{\ttfamily arXiv:1803.00993
  [hep-ph]}}.

\bibitem{Altherr:1992jg}
T.~Altherr, E.~Petitgirard, and T.~de~Rio~Gaztelurrutia, ``{Photon propagation
  in dense media},'' \href{http://dx.doi.org/10.1016/0927-6505(93)90015-6}{{\em
  Astropart. Phys.} {\bfseries 1} (1993) 289--296},
  \href{http://arxiv.org/abs/hep-ph/9212264}{{\ttfamily arXiv:hep-ph/9212264}}.

\bibitem{Braaten:1993jw}
E.~Braaten and D.~Segel, ``{Neutrino energy loss from the plasma process at all
  temperatures and densities},''
  \href{http://dx.doi.org/10.1103/PhysRevD.48.1478}{{\em Phys. Rev. D}
  {\bfseries 48} (1993) 1478--1491},
  \href{http://arxiv.org/abs/hep-ph/9302213}{{\ttfamily arXiv:hep-ph/9302213}}.

\bibitem{Hall:2009bx}
L.~J. Hall, K.~Jedamzik, J.~March-Russell, and S.~M. West, ``{Freeze-In
  Production of FIMP Dark Matter},''
  \href{http://dx.doi.org/10.1007/JHEP03(2010)080}{{\em JHEP} {\bfseries 03}
  (2010) 080}, \href{http://arxiv.org/abs/0911.1120}{{\ttfamily arXiv:0911.1120
  [hep-ph]}}.

\bibitem{BaBar:2012bdw}
{\bfseries BaBar} Collaboration, J.~P. Lees {\em et~al.}, ``{Precise
  Measurement of the $e^+ e^- \to \pi^+\pi^- (\gamma)$ Cross Section with the
  Initial-State Radiation Method at BABAR},''
  \href{http://dx.doi.org/10.1103/PhysRevD.86.032013}{{\em Phys. Rev. D}
  {\bfseries 86} (2012) 032013},
  \href{http://arxiv.org/abs/1205.2228}{{\ttfamily arXiv:1205.2228 [hep-ex]}}.

\bibitem{BaBar:2013jqz}
{\bfseries BaBar} Collaboration, J.~P. Lees {\em et~al.}, ``{Precision
  measurement of the $e^+e^- \to K^+K^-(\gamma)$ cross section with the
  initial-state radiation method at BABAR},''
  \href{http://dx.doi.org/10.1103/PhysRevD.88.032013}{{\em Phys. Rev. D}
  {\bfseries 88} no.~3, (2013) 032013},
  \href{http://arxiv.org/abs/1306.3600}{{\ttfamily arXiv:1306.3600 [hep-ex]}}.

\bibitem{Fradette:2017sdd}
A.~Fradette and M.~Pospelov, ``{BBN for the LHC: constraints on lifetimes of
  the Higgs portal scalars},''
  \href{http://dx.doi.org/10.1103/PhysRevD.96.075033}{{\em Phys. Rev. D}
  {\bfseries 96} no.~7, (2017) 075033},
  \href{http://arxiv.org/abs/1706.01920}{{\ttfamily arXiv:1706.01920
  [hep-ph]}}.

\bibitem{Mukhanov:2003xs}
V.~F. Mukhanov, ``{Nucleosynthesis without a computer},''
  \href{http://dx.doi.org/10.1023/B:IJTP.0000048169.69609.77}{{\em Int. J.
  Theor. Phys.} {\bfseries 43} (2004) 669--693},
  \href{http://arxiv.org/abs/astro-ph/0303073}{{\ttfamily
  arXiv:astro-ph/0303073}}.

\end{thebibliography}\endgroup
\bibliographystyle{utphys}
\end{document}